\documentclass[11pt]{article}

\usepackage{amsfonts}
\usepackage{amsmath}
\usepackage{amssymb}
\usepackage{amsthm}
\usepackage{xcolor}
\usepackage[T1]{fontenc}
\usepackage[scaled]{helvet}
\usepackage{rotating}
\usepackage{multirow}
\usepackage{natbib}
\usepackage{appendix}
\usepackage{enumitem}
\usepackage{epstopdf}
\usepackage{bm}
\usepackage{multirow}
\usepackage{float}
\usepackage{multicol}
\usepackage{subfig}
\usepackage{xr}
\usepackage{pdfpages}
\makeatletter
\@addtoreset{equation}{section}
\renewcommand{\theequation}{\thesection.\arabic{equation}}
\makeatother

\newcommand{\mub}{\bm{\mu}}
\newcommand{\xb}{\bm{x}}

\newcommand{\kappab}{\bm{\kappa}}
\newcommand{\lambdab}{\bm{\lambda}}
\newcommand{\varthetab}{\bm{\vartheta}}

\newcommand{\R}{\mathbb R}

\begin{document}

\title{Sine-skewed toroidal distributions and their application in protein bioinformatics}

\author{Jose Ameijeiras-Alonso$^{1,*}$
and Christophe Ley$^{2,\dagger}$
}
\date{%
    $^*$KU Leuven and 
    $^\dagger$Ghent University
}
\footnotetext[1]{Supported by the FWO research project G.0826.15N (Flemish Science Foundation), GOA/12/014 project (Research Fund KU Leuven), Project MTM2016-76969-P from the Spanish State Research Agency (AEI) co-funded by the European Regional Development Fund (ERDF) and the Competitive Reference Groups 2017-2020 (ED431C 2017/38) from the Xunta de Galicia through the ERDF.}
\footnotetext[2]{Supported by the FWO Krediet aan Navorsers grant with reference number 1510391N.}
\maketitle

\begin{abstract}
In the bioinformatics field, there has been a growing interest in modelling dihedral angles of amino acids by viewing them as data on the torus. This has motivated, over the past years, new proposals of distributions on the bivariate torus. The main drawback of most of these models is that the related densities are (pointwise) symmetric, despite the fact that the data usually present asymmetric patterns. This motivates the need to find a new way of constructing asymmetric toroidal distributions starting from a symmetric distribution. We tackle this problem in this paper by introducing the sine-skewed toroidal distributions. The general properties of the new models are derived. Based on the initial symmetric model, explicit expressions for the shape parameters are obtained, a simple algorithm for generating random numbers is provided, and asymptotic results for the maximum likelihood estimators are established. An important feature of our construction is that no normalizing constant needs to be calculated, leading to  more flexible distributions without increasing the complexity of the models. The  benefit of employing these new sine-skewed distributions is shown on the basis of protein data, where, in general, the new models outperform their symmetric antecedents.
\end{abstract}

\noindent%

\section{Introduction}

Toroidal data are observations taking values on the unit torus, that is, the product of unit circles, the typical example being the torus in three dimensions formed by the product of two unit circles. This also explains the alternative terminology of circular-circular or bivariate circular data in $\R^3$. Toroidal data appear  in numerous domains. Typical examples are wind directions measured at two distinct moments of the day \citep[see, e.g.,][]{JW77,KSS08,K09}, earthquake data consisting of pre-earthquake direction of steepest descent and the direction of lateral ground movement \citep{R97,circulas} or peak systolic blood pressure times, converted to angles, during two separate time periods \citep{FL83,DM02}. The interest in modelling especially circular-circular data has been strongly growing over the past years, mainly thanks to important applications in bioinformatics. In particular, it has been noted that dihedral angles of amino acids can  be much better modelled by viewing them as data on the torus, which has led to crucial contributions in the protein structure prediction problem, see for instance \cite{mardia2007}, \cite{FD07}, \cite{BMTFKH08}, \cite{HBPFJH10}, \cite{GSMH19} or the monograph \cite{HMF12} and Chapters 1 and 4 of \cite{LV19}.

One of the most famous toroidal distributions is the \emph{bivariate von Mises density} of \cite{mardia1975}, which arises as a maximum entropy distribution with von Mises marginals. Several more parameter-parsimonious submodels have since been proposed in the literature, e.g. by \cite{R88}, \cite{singh2002}, \cite{mardia2007} and \cite{KMT08}. The extensions to higher dimensions of the so-called Sine model by \cite{singh2002} and Cosine model by \cite{mardia2007} can be found in \cite{mardia2008} and \cite{MP05}, respectively. Details about these distributions are given in Section~\ref{sec:cases}. Another famous class of toroidal distributions that allows specifying the marginal distributions has been put forward by \cite{WJ80}. Their construction resembles copulas in multivariate Euclidean spaces, which is why \cite{circulas} coin the term ``circulas'' in their general study of these models. A particularly salient special case is the bivariate wrapped Cauchy distribution of \cite{kato2015}, which has wrapped Cauchy distributions both as marginal and conditional distributions. We refer the reader to Sections 2.4 and 2.5 of \cite{ley2017} for an overview of further toroidal distributions.

A common trait of the large majority of these distributions is that they are symmetric around the location on the torus. However, several datasets are not symmetric, as noted for instance by \cite{SZ11} in the context of orthologous genes shared by circular prokaryotic genomes.  Quoting \cite{Mardia13}  in the context of protein data: ``\emph{Another unexplored area is constructing plausible bivariate skew distributions, i.e. the field is still evolving, but it should be noted that for bioinformatics we generally need computationally efficient methods}''. Thus there is a strong  demand for asymmetric distributions on the torus, and we will palliate that need in this paper by introducing a general construction that  skews any symmetric distribution. This thus allows us to exploit precisely the large amount of existing symmetric distributions and is a tool that can easily be used in conjunction with any future new construction of symmetric distributions.

Our proposal, which can be found in~\eqref{sine_skewed}, inscribes itself in the vein of symmetry modulation \`a la Azzalini, also known as the perturbation approach. The idea consists in shifting the probability mass into a certain direction indicated by an additional parameter, called skewness parameter, and this shift is obtained by multiplying a symmetric density with a so-called skewing function. Initially designed for the real line, resulting in the famous scalar skew-normal distribution \citep{A85}, this approach has been extended to the multivariate setting \citep{AD96, WBG04}, to the unit circle \citep{UJ09, abe2011} and to the unit hypersphere \citep{LV17}. We also refer the interested reader to \cite{JRA16} for a general setting of symmetry modulation. 

Besides their wide usage, our sine-skewed toroidal distributions enjoy the following advantages: ease of interpretation, absence of the need to calculate normalizing constants (which can be very hard on the torus), simple random number generation, versatile shapes, and simple parameter estimation. We will discuss all these properties as well as special cases in the subsequent sections. 

The present paper is organized as follows. In Section~\ref{sec:proposal} we introduce our mechanism to skew any symmetric toroidal distribution and provide the main properties of the resulting distributions. Special cases of sine-skewed toroidal distributions are described in Section~\ref{sec:cases}, while inferential aspects are considered in Section~\ref{sec:Inf}. In Section~\ref{sec:real} we show the superior fit provided by our sine-skewed distributions in comparison to their symmetric antecedents when modelling real data from bioinformatics, hereby showing the benefits from using our potentially asymmetric models in this field. Final conclusions are drawn in Section~\ref{sec:conclu}, while additional results are given in the online Supplementary Material.

\section{Basic formulation and properties}\label{sec:proposal}

\subsection{The general expression of sine-skewed densities}

The objective of this subsection is to provide a general formulation of our approach to obtain skewed distributions on the torus. Consider a (pointwise) symmetric density $\xb\mapsto f(\xb-\bm{\mu};\varthetab), \xb\in\mathbb{T}^d:=[-\pi,\pi)^d,$ around a location $\bm{\mu} \in\mathbb{T}^d$, i.e., a toroidal density satisfying $f(\bm{\mu}+\bm{x}; \bm{\vartheta})=f(\bm{\mu}-\bm{x}; \bm{\vartheta})$, where $\varthetab$ denotes the set of all parameters involved in the density other than location. In general, this concerns  concentration parameters $\bm{\kappa}\in(\mathbb{R}^+)^d$, but also a correlation parameter $r \in \mathbb{R}$ (or a dependence matrix) and/or a $d$-dimensional peakedness/kurtosis vector $\bm{\eta}\in \mathbb{R}^{d}$ can be present in $\varthetab$, depending on the distribution.  

Starting from the base density $f(\xb-\bm{\mu};\varthetab)$, our approach consists in transforming it into the density 
\begin{equation}\label{sine_skewed}
\xb\mapsto g(\bm{x}-\mub;\bm{\vartheta},\bm{\lambda}):=f(\bm{x}-\bm{\mu};\bm{\vartheta}) \left(1+\sum_{s=1}^{d} \lambda_s \sin (x_s-\mu_s)\right),
\end{equation}
where $\lambdab\in[-1,1]^d$ plays the role of skewness parameter and satisfies $\sum_{s=1}^{d} |\lambda_s| \leq 1$. It is clear that $\xb\mapsto g(\bm{x}-\mub;\bm{\vartheta},\bm{\lambda})$ is always positive and integrates to 1 over $\mathbb{T}^d$, see  the next subsection for more details. One can readily see that this density is asymmetric (probability mass shifted in the direction of $\lambdab$) unless $\lambdab=\bm{0}$ in which case we retrieve the base symmetric density $f(\bm{x}-\bm{\mu};\bm{\vartheta})$. One of the main advantages of this formulation is that a skewness parameter can be included  without altering the normalizing constant. When $d=1$, we retrieve the sine-skewed distributions on the circle from \cite{UJ09} and \cite{abe2011}. In the next subsections, we highlight the main properties of this model and how to generate random data.

\subsection{Main properties}\label{general_properties}

In this subsection, we study some basic properties of the toroidal sine-skewed distributions with associated density~\eqref{sine_skewed}. First, the cumulative distribution function is analyzed, where the main advantage of the proposed distribution is pointed out: the proposed distributions provide a method of skewing any toroidal distribution without altering the normalizing constant. Second, we investigate when the marginals also follow  a (sub-toroidal) sine-skewed distribution. Finally, the shape parameters of our new models are analyzed by providing the expressions of trigonometric moments.

\paragraph{Cumulative distribution function and normalizing constant.} Without loss of generality, we can set $\bm{\mu}=\bm{0}$. Denoting by $F(\cdot;\bm{\vartheta})$ the cumulative distribution function (cdf) of the toroidal base symmetric distribution, the cdf of its sine-skewed version, $G(\bm{x};\bm{\vartheta},\bm{\lambda})$, in a point $\bm{x}\in \mathbb{T}^d$, is obtained as
\begin{eqnarray*}
G(\bm{x};\bm{\vartheta},\bm{\lambda})&=& F(\bm{x};\bm{\vartheta}) + \sum_{s=1}^d  \lambda_s \int_{-\pi}^{x_1} \ldots \int_{-\pi}^{x_d} \sin(t_s) f(\bm{t};\bm{\vartheta}) \mbox{d} \bm{t} \\
G(\bm{\pi};\bm{\vartheta},\bm{\lambda})-G(\bm{-\pi};\bm{\vartheta},\bm{\lambda})&=& F(\bm{\pi};\bm{\vartheta})-F(\bm{-\pi};\bm{\vartheta}) = 1.
\end{eqnarray*} 
Consequently, the normalizing constant does not need to be altered, which is a particularly nice property given the difficulty to calculate integrals on the multidimensional torus.

\paragraph{Marginals.} Let $\bm{X}=[X_1,\ldots,X_d]^T$ follow a $d$-dimensional toroidal sine-skewed distribution and let $\bm{m}$ be a subset of $\{1,\ldots,d\}$ and $\bm{m}^c$ its complementary. Studying the behaviour of the marginals (of any dimension lower than $d$) leads us to consider the following three scenarios:
\begin{itemize}
\item[$\bullet$] If the initial symmetric density $f$ is pointwise symmetric in ${\bm x}_{\bm{m}^c}:=\{x_s: s\in\bm{m}^c\}$ alone (while fixing the other components), then the distribution of $\bm{X}_{\bm{m}}=[X_s]_{s\in\bm{m}}^T$ is sine-skewed. For example, if the marginals of the base density are independent, the distribution of $\bm{X}_{\bm{m}}$ is sine-skewed. 
\item[$\bullet$] If $\lambda_s=0$ for all $s\in\bm{m}^c$, then $\bm{X}_{\bm{m}}$ is sine-skewed. 
\item[$\bullet$] Otherwise, in general, it is not true that the marginals are also sine-skewed, and skewed distributions for $\bm{X}_{\bm m}$ can be obtained even if $\lambda_s=0$ for all $s\in\bm{m}$ (see Figure~\ref{figsswc_multimodal}, right panel).
\end{itemize}
A formal proof of these statements can be found in Section~\ref{proofs} of the Supplementary Material.

\paragraph{Trigonometric moments and shape parameters.} Assume again, without loss of generality and for the sake of simplicity, that $\bm{\mu}=\bm{0}$. The characteristic function of the $d$-toroidal density is given by the sequence of numbers $\{\phi_{\bm{p}}: p_s=0,\pm 1, \pm 2, \ldots; 1\leq s \leq d\}$, where $\phi_{\bm{p}}=\mathbb{E}[\exp(i\bm{p}^T\bm{X})]=\mathbb{E}[\cos(\bm{p}^T\bm{X})]+i\mathbb{E}[\sin(\bm{p}^T\bm{X})]=:\alpha_{\bm{p}}+i\beta_{\bm{p}}$. If $\alpha^0_{\bm{p}}$  denotes the cosine  moment of order $\bm{p}$ of the base symmetric distribution (note that its sine moments are zero), then, for its sine-skewed version, these moments are given by
\begin{eqnarray*}
\alpha_{\bm{p}}&=& \alpha^0_{\bm{p}} \\
\beta_{\bm{p}}&=& \frac{1}{2}\sum_{s=1}^{d} \lambda_s \left(\alpha^0_{[p_1,\ldots,p_{s-1},(p_s-1),p_{s+1},\ldots,p_d]} - \alpha^0_{[p_1,\ldots,p_{s-1},(p_s+1),p_{s+1},\ldots,p_d]} \right).\\
\end{eqnarray*}
The elements of the toroidal mean, which we denote by $\bm{\mu}_1=[\mu_{1;s}]^T_{s\in\{1,\ldots,d\}}$, are obtained as
\begin{equation*}
{\mu}_{1;s}=\mbox{atan2}(\beta_{\bm{e}_s},\alpha_{\bm{e}_s})=\mbox{atan2}\left(\frac{1}{2} \left(\lambda_s(1-\alpha^0_{2\bm{e}_s})+ \sum_{p=1, p\neq s}^{d} \lambda_p (\alpha^0_{\bm{e}_s-\bm{e}_p}-\alpha^0_{\bm{e}_s+\bm{e}_p})\right) ,\alpha^0_{\bm{e}_s}\right)
\end{equation*}
where $\bm{e}_s$ stands for the vector with all components zero except in the position $s$ which is equal to 1, and atan2($y$,$x$) computes the principal value of the argument function applied to the complex number $x+i y$. Note that this definition guarantees that, at $\lambdab=\pmb 0$, the mean coincides with the mean of $f$. In an analogous way, the elements of the concentration vector $\bm{\rho}_1$ and variance  $\bm{V}$ can be computed: \begin{eqnarray*}
\rho_{1;s}&=&\sqrt{\alpha_{\bm{e}_s}^2 + \beta_{\bm{e}_s}^2}= \sqrt{(\alpha^0_{\bm{e}_s})^2+\frac{1}{4} \left(\lambda_s(1-\alpha^0_{2\bm{e}_s})+ \sum_{p=1, p\neq s}^{d} \lambda_p (\alpha^0_{\bm{e}_s-\bm{e}_p}-\alpha^0_{\bm{e}_s+\bm{e}_p})\right)^2},\\
V_s&=& 1-\rho_{1;s}.
\end{eqnarray*}
A toroidal equivalent of the circular skewness and kurtosis parameters \citep[see, e.g.,][Subsection 3.4]{mardia2000} can be obtained via the cosine and sine moments about the mean direction $\bm{\mu}_1$, that is, $\bar{\alpha}_{\bm{p}}=\mathbb{E}[\cos(\bm{p}^T(\bm{X}-\bm{\mu}_1)] $ and $\bar{\beta}_{\bm{p}}=\mathbb{E}[\sin(\bm{p}^T(\bm{X}-\bm{\mu}_1))]$. In that case, the skewness parameter $\bm{\mathfrak{s}}$ is given by
\begin{eqnarray*}
\bar{\beta}_{2\bm{e}_s}&=& \frac{\lambda_s(\alpha^0_{\bm{e}_s}-\alpha^0_{3\bm{e}_s})+ \sum_{p=1, p\neq s}^{d} \lambda_p (\alpha^0_{2\bm{e}_s-\bm{e}_p}-\alpha^0_{2\bm{e}_s+\bm{e}_p})}{2\rho_{1;s}^2} \\
&\times&\left((\alpha^0_{\bm{e}_s})^2-\frac{1}{4} \left(\lambda_s(1-\alpha^0_{2\bm{e}_s})+ \sum_{p=1, p\neq s}^{d} \lambda_p (\alpha^0_{\bm{e}_s-\bm{e}_p}-\alpha^0_{\bm{e}_s+\bm{e}_p})\right)^2\right)\\
& -&\frac{\alpha^0_{\bm{e}_s}\alpha^0_{2\bm{e}_s}}{\rho_{1;s}^2} \left(\lambda_s(1-\alpha^0_{2\bm{e}_s})+ \sum_{p=1, p\neq s}^{d} \lambda_p (\alpha^0_{\bm{e}_s-\bm{e}_p}-\alpha^0_{\bm{e}_s+\bm{e}_p})\right),\\
\mathfrak{s}_s&=& \bar{\beta}_{2\bm{e}_s}/ V_s^{3/2}.
\end{eqnarray*}
From the value of $\bar{\alpha}_{2\bm{e}_s}$, the kurtosis parameter, $\bm{\mathfrak{k}}$, can be obtained as

\begin{eqnarray*}
\bar{\alpha}_{2\bm{e}_s}&=& \frac{\alpha^0_{2\bm{e}_s}}{\rho_{1;s}^2}\left((\alpha^0_{\bm{e}_s})^2-\frac{1}{4} \left(\lambda_s(1-\alpha^0_{2\bm{e}_s})+ \sum_{p=1, p\neq s}^{d} \lambda_p (\alpha^0_{\bm{e}_s-\bm{e}_p}-\alpha^0_{\bm{e}_s+\bm{e}_p})\right)^2\right)\\
& +&\frac{\alpha^0_{\bm{e}_s}}{2\rho_{1;s}^2} \left(\lambda_s(1-\alpha^0_{2\bm{e}_s})+ \sum_{p=1, p\neq s}^{d} \lambda_p (\alpha^0_{\bm{e}_s-\bm{e}_p}-\alpha^0_{\bm{e}_s+\bm{e}_p})\right)\\
&\times& \left(\lambda_s(\alpha^0_{\bm{e}_s}-\alpha^0_{3\bm{e}_s})+ \sum_{p=1, p\neq s}^{d} \lambda_p (\alpha^0_{2\bm{e}_s-\bm{e}_p}-\alpha^0_{2\bm{e}_s+\bm{e}_p})\right), \\
\mathfrak{k}_s&=& (\bar{\alpha}_{2\bm{e}_s}-\rho_{1;s}^2)/ V_s^{2}.
\end{eqnarray*}
 We refer the interested reader to Section~\ref{proofs} of the Supplementary Material for an overview of useful identities leading up to the expressions of $\bar{\alpha}_{2\bm{e}_s}$ and $\bar{\beta}_{2\bm{e}_s}$.
 
 It is clear from the preceding formulas that, quite nicely, whenever trigonometric moments from the base symmetric distribution are available, one can readily compute their sine-skewed counterparts  and even express the toroidal variance, skewness, and kurtosis.

\subsection{Generating mechanism}

Random number generation from our toroidal sine-skewed distributions is straightforward, provided that we can generate data from the base symmetric density $f(\cdot-\mub;\bm{\vartheta})$. For each $i\in\{1,\ldots,n\}$, where $n$ is the sample size, the following algorithm shows how to generate data from our model:
\begin{enumerate}
\item Simulate a random number $\bm{Y}_i$ from the distribution associated to the base density $f(\cdot-\mub;\bm{\vartheta})$.
\item Generate, independently, a random number $U_i$ from the uniform distribution on $[0,1]$.
\item The new data point $\bm{X}_i$ from the sine-skewed density will then be equal to
\begin{equation*} 
\bm{X}_i =
\begin{cases} 
\bm{Y}_i& \mbox{if } U_i \leq  \left(1+\sum_{s=1}^{d} \lambda_s \sin (Y_{i;s}-\mu_s)\right)/2  \\
-\bm{Y}_i+2\bm{\mu} & \mbox{if }  U_i > \left(1+\sum_{s=1}^{d} \lambda_s \sin (Y_{i;s}-\mu_s)\right)/2. 
\end{cases}
\end{equation*} 
\end{enumerate} 
After generating the random numbers, if needed, the module $2\pi$ can be employed to guarantee that each $\bm{X}_i\in \mathbb{T}^d$. A proof that the algorithm above provides the desired sine-skewed data is given in Section~\ref{proofs} of the Supplementary Material.

\section{Special cases of sine-skewed toroidal distributions}\label{sec:cases}

As mentioned in the Introduction, a strength of our approach is that it can precisely make use of the various existing symmetric toroidal distributions from the literature as we can turn them all into their sine-skewed versions. In this section, we will present  some of these well-known distributions and discuss their resulting sine-skewed equivalent.  Except for the sine-skewed uniform distribution, all these models will have $d(d+5)/2$ parameters corresponding to location $\mub$ (dimension $d$), concentration $\kappab$ (dimension $d$), skewness $\lambdab$ (dimension $d$) and dependence $\bm R$ (dimension $d(d-1)/2$). Their particular cases when $\bm{\kappa}=\bm{0}$ will correspond to the uniform distribution, those where  $\bm{R}$ is the zero matrix  will be related to the independent case, and $\bm{\lambda}=\bm{0}$ will lead back to the symmetric case.

\subsection{Sine-skewed uniform distribution}

The particular case where the base density $f$ is the uniform toroidal distribution leads to the sine-skewed uniform with density
\begin{equation*}
g_{\tiny{\mbox{SU}}}(\bm{x}-\bm{\mu};\bm{\lambda})=\frac{1}{(2\pi)^d} \left(1+\sum_{s=1}^{d} \lambda_s \sin (x_s-\mu_s)\right).
\end{equation*}
This density can be seen as a toroidal extension of the cardioid distribution on the circle, extension that is unimodal and symmetric around $\bm{\mu}+({\pi/2},\pi/2,\ldots,\pi/2)^T$, with concentration $\bm{\lambda}/2$. The marginals of this density are independent and the $s$-th univariate marginal corresponds to the cardioid density with mean $\mu_s+\pi/2$ and concentration $\lambda_s/2$ if the parametrization of \citet[][Equation~3.5.47]{mardia2000} is employed.

\subsection{Sine-skewed  Sine distribution}

\begin{figure}
 \centering
\begin{tabular}{cccc}
\subfloat{
    \includegraphics[height=0.16\textheight]{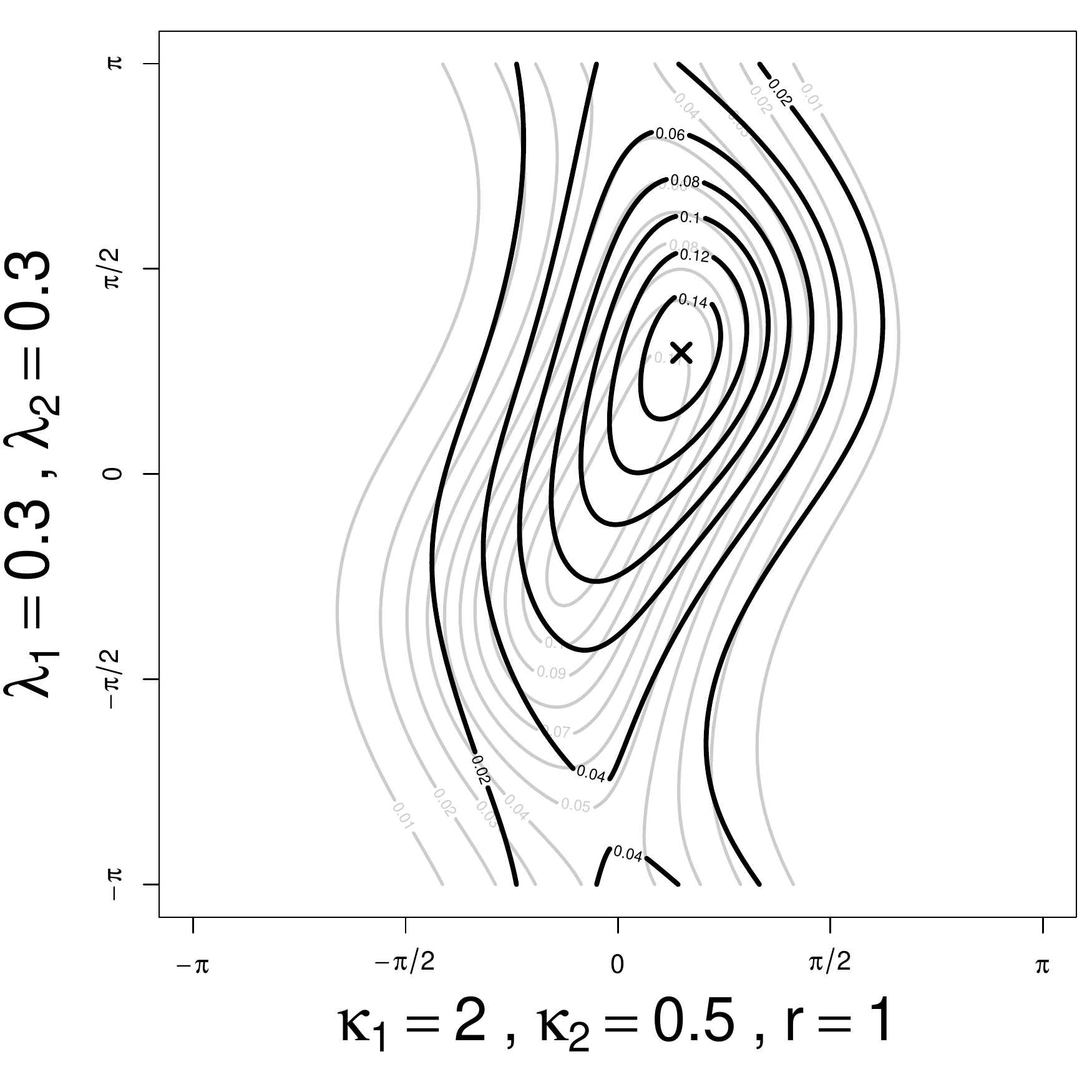}}
\subfloat{
    \includegraphics[height=0.16\textheight]{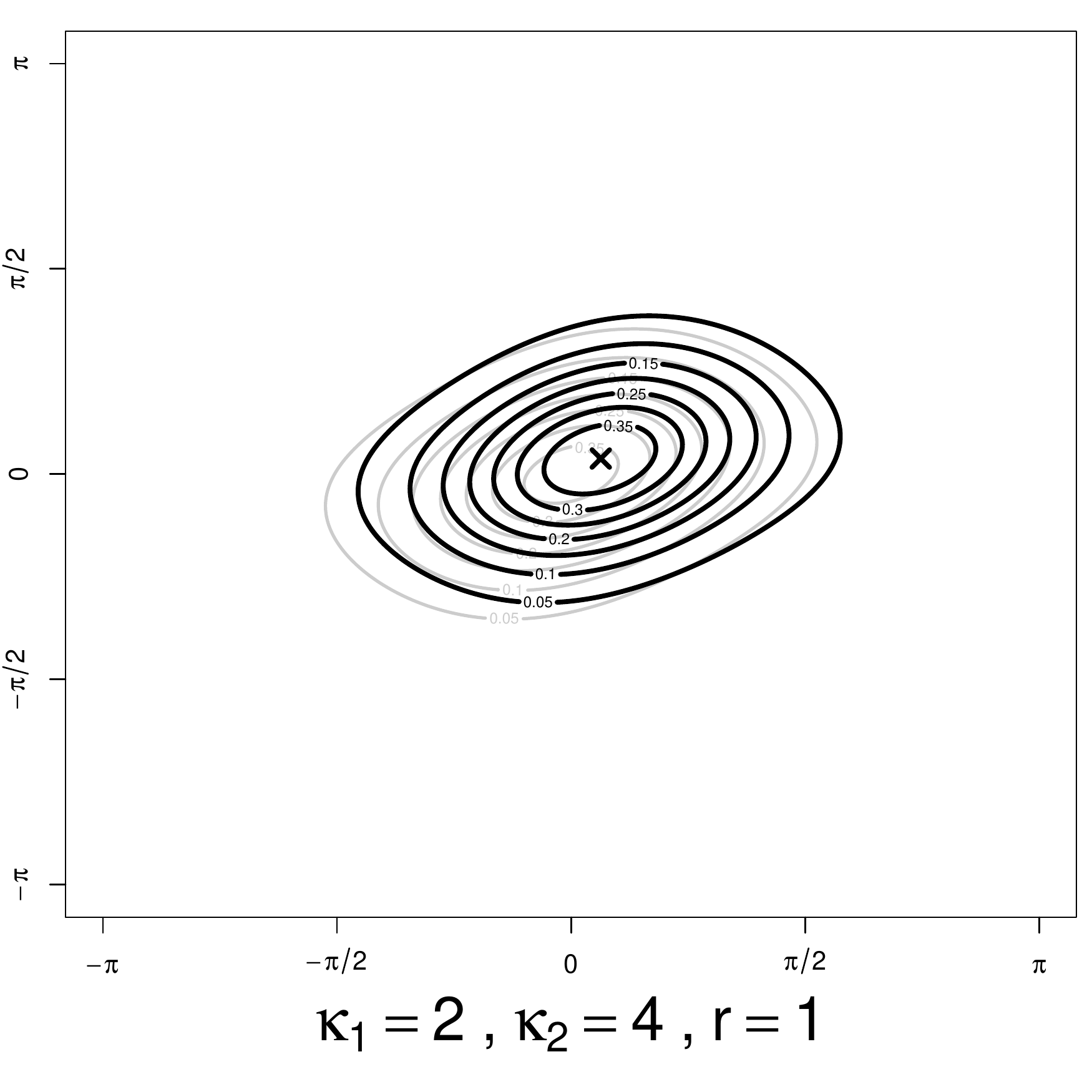}}
		\subfloat{
    \includegraphics[height=0.16\textheight]{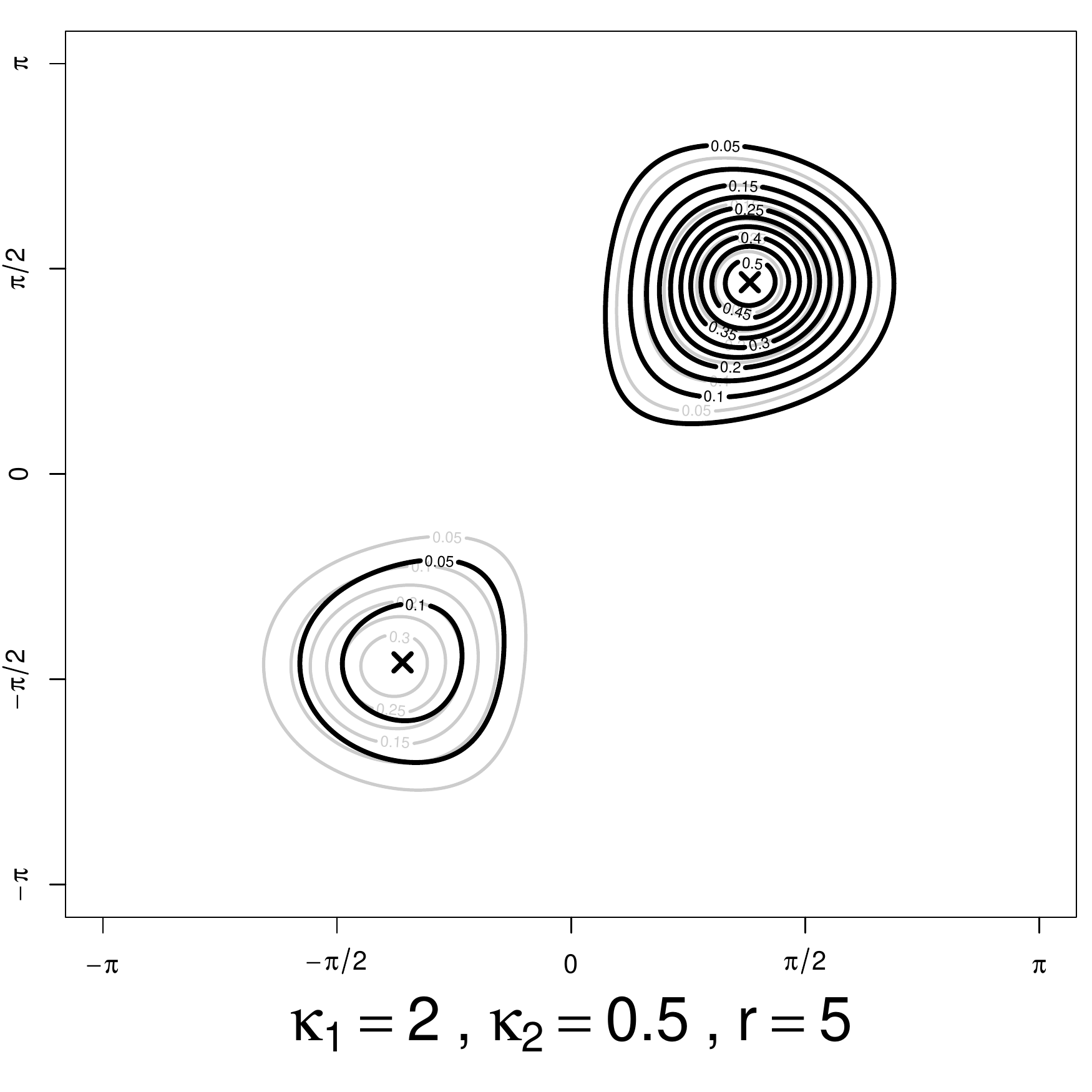}}
		\subfloat{
    \includegraphics[height=0.16\textheight]{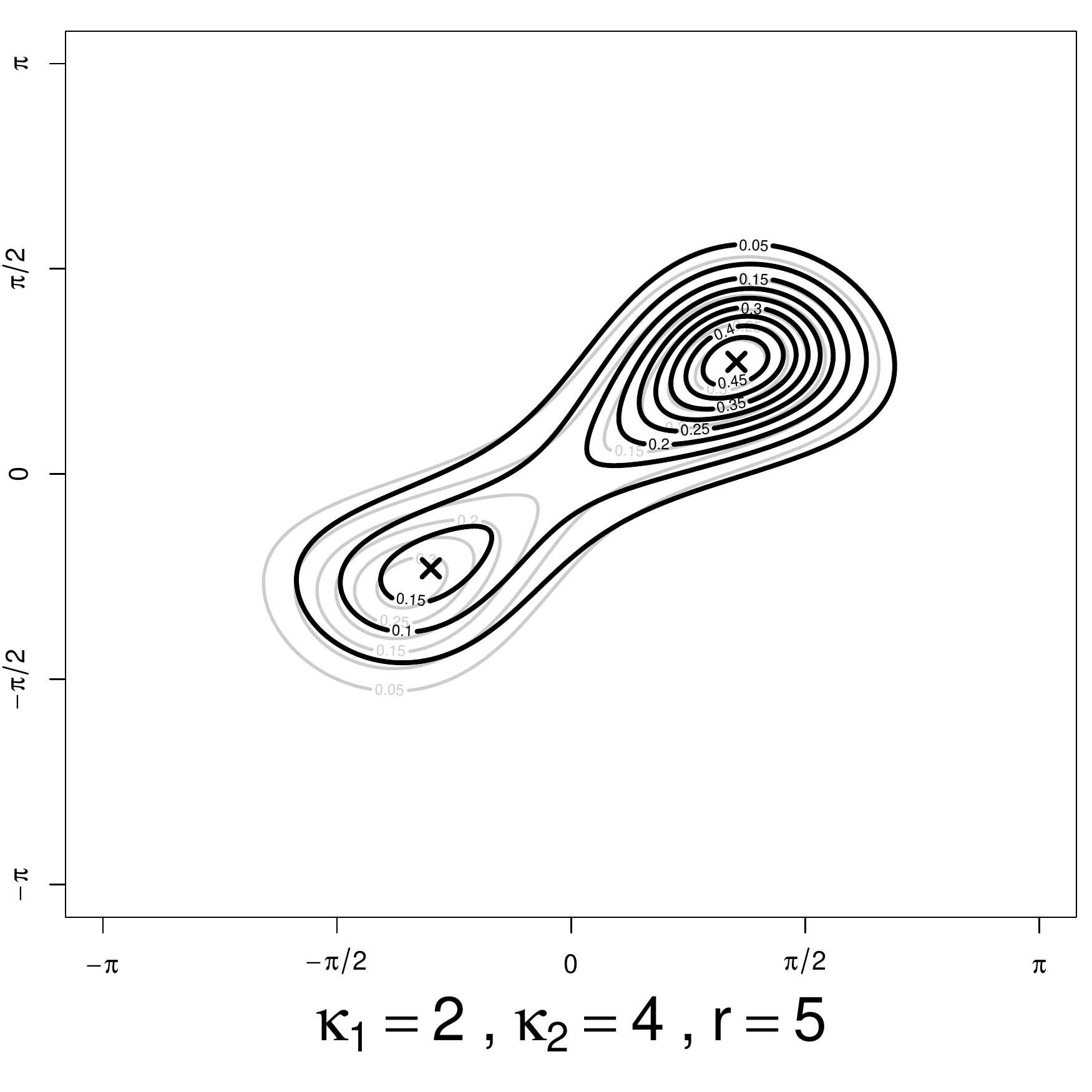}}\\
\subfloat{
    \includegraphics[height=0.16\textheight]{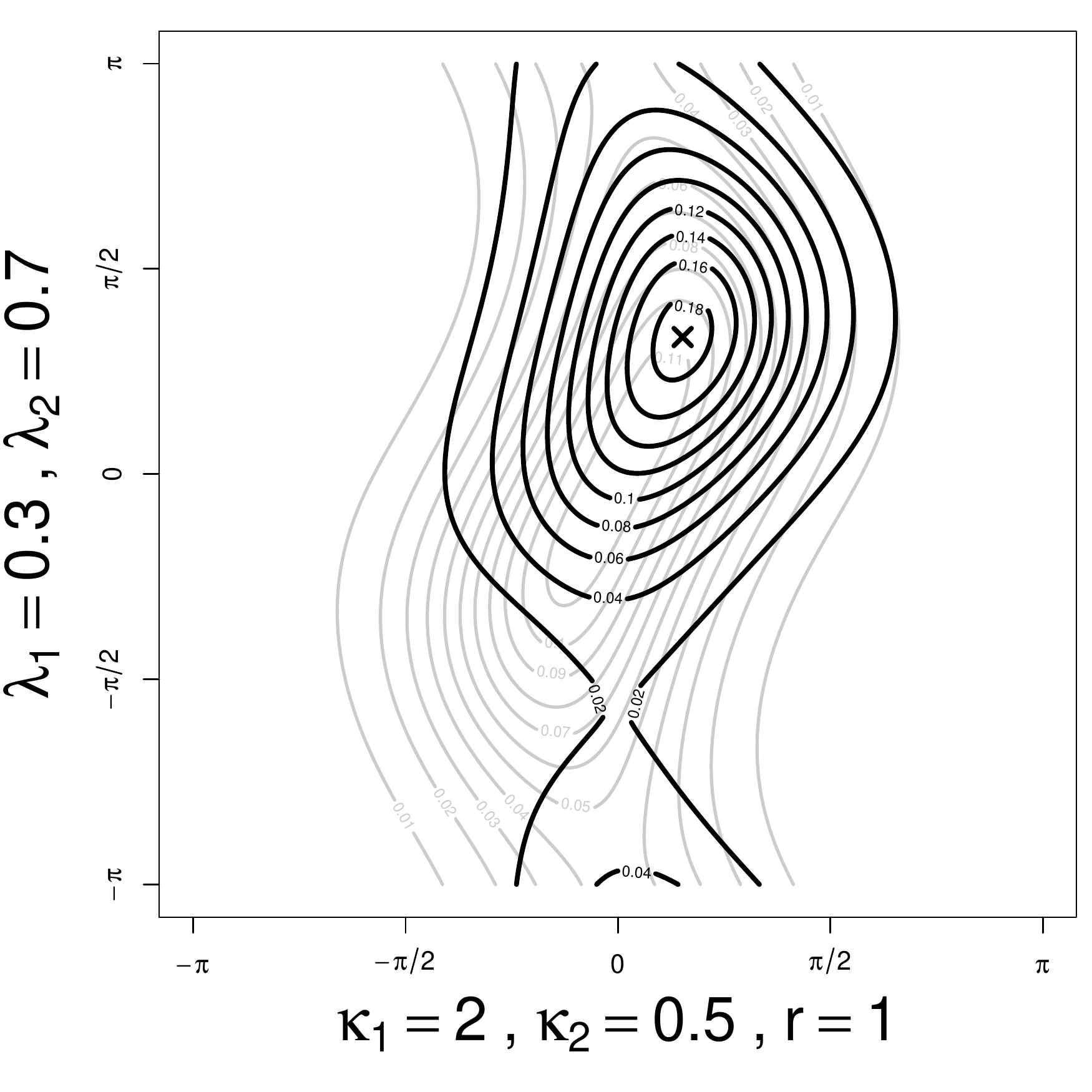}}
		\subfloat{
    \includegraphics[height=0.16\textheight]{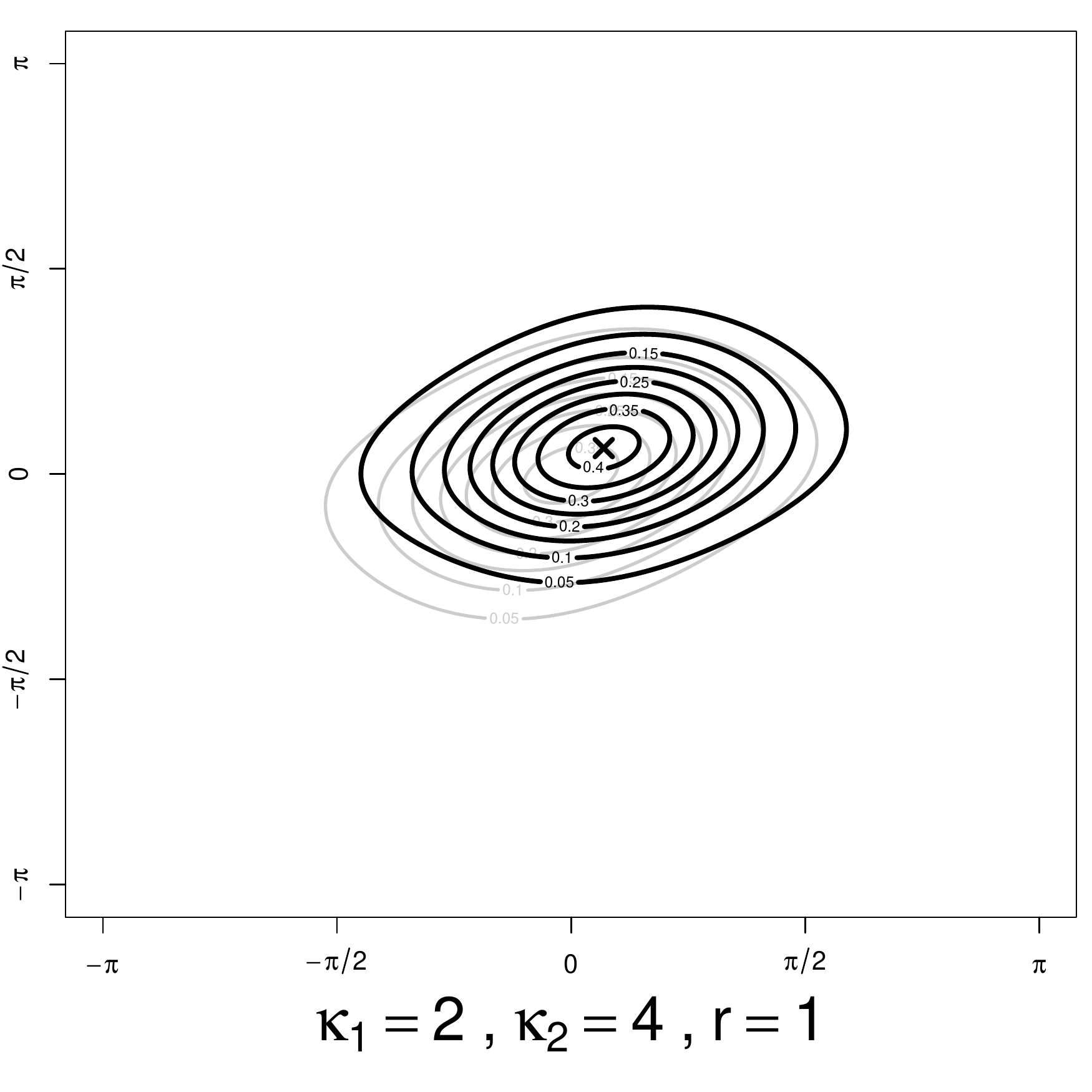}}
		\subfloat{
    \includegraphics[height=0.16\textheight]{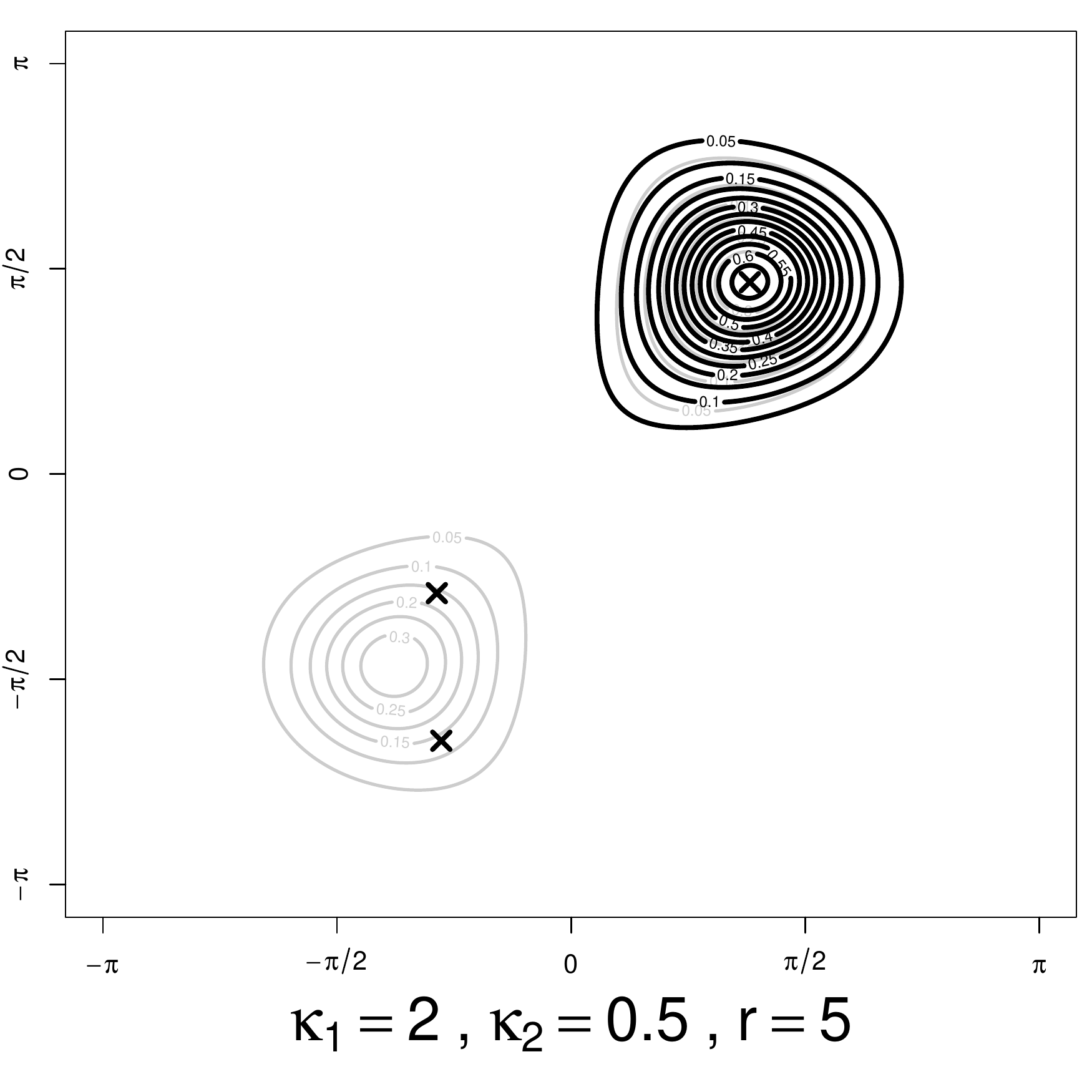}}
			\subfloat{
    \includegraphics[height=0.16\textheight]{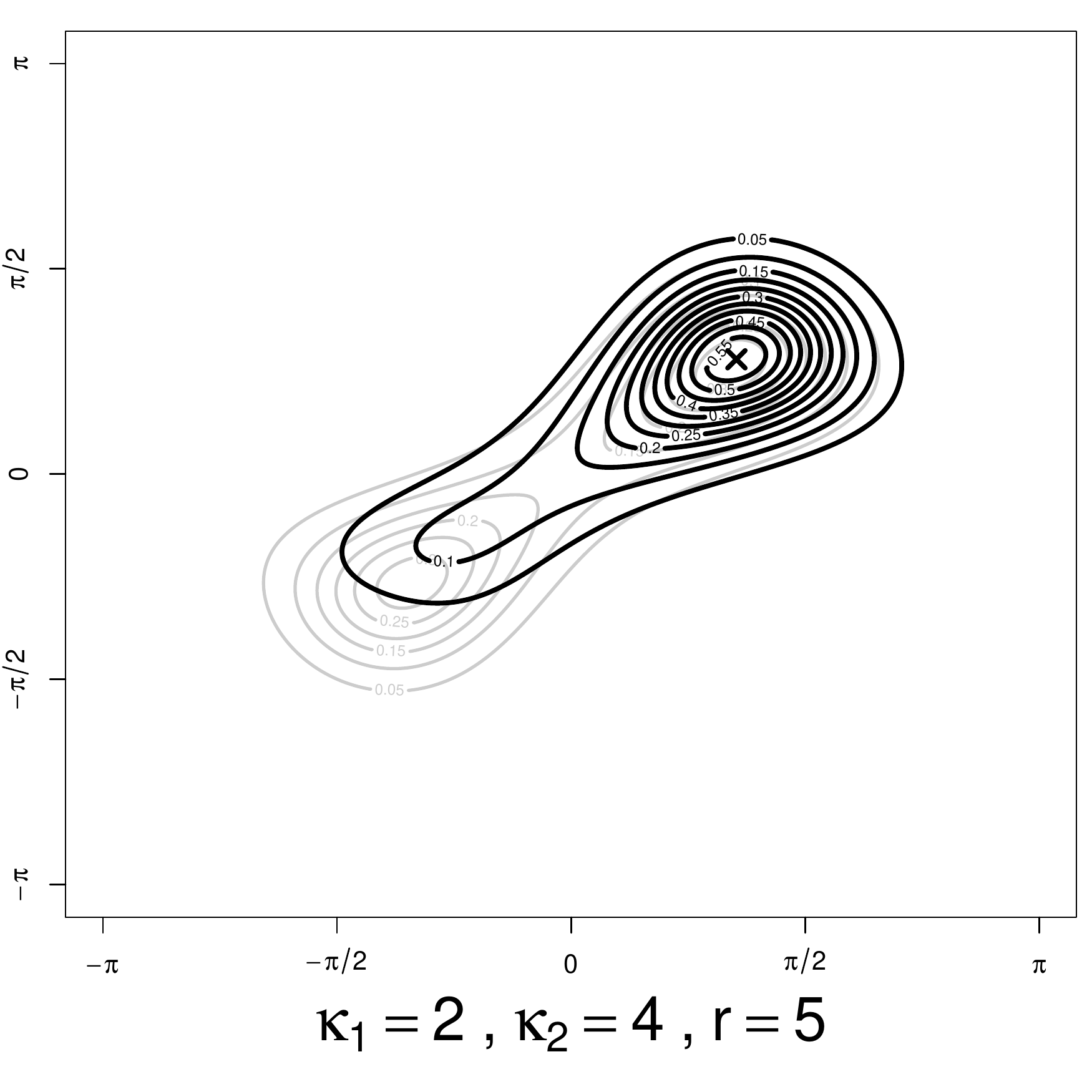}}\\
		\end{tabular}
		\vspace{0.7cm}
 \caption{Bivariate sine-skewed  Sine density functions (with $\bm{\mu}=\bm{0}$), where the gray contour indicates the symmetric bivariate Sine density ($\lambda_1=\lambda_2=0$) and the dark contour  the sine-skewed version obtained for the skewness parameters indicated at the left. The other parameters are indicated in the labels. \\From left to right: effect of the concentration and dependence parameters. From top to bottom: effect of the skewness parameters. The crosses identify the mode locations.}
 \label{figsssine}
\end{figure}

From the original bivariate von Mises density introduced by \cite{mardia1975}, different particular cases were proposed. The main issue with the original model is that it was ``overparametrized'' \citep[see][Subsection 2.4.2]{ley2017} with eight parameters, two for location, two for concentration and four for circular-circular dependence, which creates difficulties of interpretation. One of the simplest particular cases is the \textit{Sine model} of \cite{singh2002} with only five parameters, the location $\bm{\mu}\in \mathbb{T}^2$, the concentration $\bm{\kappa}\in(\mathbb{R}^+)^2$ and the correlation $r\in \mathbb{R}$. Using the Sine distribution as the base density,  the bivariate sine-skewed  Sine distribution has the density
\begin{eqnarray*}
g_{\tiny{\mbox{SS}}}(\bm{x}-\bm{\mu};\bm{\kappa},r,\bm{\lambda})&=& C_{\bm{\kappa},r}^{-1} \exp (\kappa_1 \cos(x_1-\mu_1) + \kappa_2 \cos(x_2-\mu_2) + r \sin(x_1-\mu_1)\sin(x_2-\mu_2)) \\
&\times&(1+ \lambda_1 \sin (x_1-\mu_1)+ \lambda_2 \sin (x_2-\mu_2) ) , \\
C_{\bm{\kappa},r} &=& 4\pi^2 \sum_{i=0}^{\infty} \binom{2i}{i}  \left( \frac{r^2}{4\kappa_1\kappa_2} \right)^i I_i (\kappa_1) I_i (\kappa_2),
\end{eqnarray*}
where $I_i (\kappa)$ denotes the modified Bessel function of order $i$ evaluated at the value $\kappa$. Besides the original five parameters, two skewness parameters $\lambda_1,\lambda_2\in[-1,1]$, with $|\lambda_1|+|\lambda_2|\leq 1$, are introduced for this model. Focusing on the case $\kappa_1 \kappa_2\neq 0$,  this model is symmetric around $\bm{\mu}$ when $\bm{\lambda}=\bm{0}$ and, in that case, unimodality holds if $\kappa_1 \kappa_2>r^2$, while the density is bimodal if $\kappa_1 \kappa_2<r^2$ \citep[see][Theorem 3]{mardia2007}. A representation of this model for different parameter configurations is shown in Figure~\ref{figsssine}. We can see from the third and fourth column of that figure that, starting from a bimodal Sine density and depending on the skewness parameters, new modes can be added (leading to a trimodal density) or disappear (yielding a unimodal density). {An equivalent observation is made in the  circular sine-skewed von Mises case by \cite{abe2011}, where for some configurations of the concentration and skewness parameters bimodal densities are obtained.}

{Let us now calculate the density of the first marginal. To this end, we denote by $a(x_1)$ and $b(x_1)$ the quantities such that $\kappa_2=a(x_1)\cos(b(x_1))$ and $r \sin(x_1-\mu_1)=a(x_1)\sin(b(x_1))$. Then, using integration results involving the already mentioned modified Bessel function of order 0, $I_0$, as well as the modified Bessel function of order 1, $I_1$, we get }
\begin{eqnarray}
g_{1;\tiny{\mbox{SS}}}(x_1-\mu_1;{\mu_2},\bm{\kappa},r,\bm{\lambda}) 
&=&  \frac{1}{C_{\bm{\kappa},r}} \exp(\kappa_1 \cos(x_1-\mu_1)) \left[ 2\pi I_0(a(x_1)) (1+\lambda_1 \sin(x_1)) \right. \nonumber \\
&+& \left. \lambda_2 \frac{I_1(a(x_1))}{I_0(a(x_1))} \cos(\mu_2+b(x_1)) \right], \label{marginal_sine}\\
a(x_1)^2 &=& \kappa_2^2+r^2 \sin^2(x_1-\mu_1), \nonumber \\
\tan(b(x_1)) &=& ({r}/{\kappa_2}) \sin(x_1-\mu_1). \nonumber 
\end{eqnarray} 
{The second marginal can be obtained in a similar way. Regarding the first marginal, as expected from the results in Subsection~\ref{general_properties}, if $\lambda_2=0$, then the density $g_{1;\tiny{\mbox{SS}}}$ is a sine-skewed version of a symmetric density, but, in general (unless $r=0$), not a sine-skewed von Mises. If $\lambda_1=0$ and $\lambda_2\neq0$, in general (if $r\neq 0$), the marginal density $g_{1;\tiny{\mbox{SS}}}$ is not (reflectively) symmetric. }

An extension of the Sine model to the $d$-variate setting was provided by \cite{mardia2008}. The resulting  $d$-variate sine-skewed Sine density is given by
\begin{eqnarray*}
g_{\tiny{\mbox{SS}}}(\bm{x}-\bm{\mu};\bm{\kappa},\bm{R},\bm{\lambda})&=& \frac{1}{ C_{\bm{\kappa},\bm{R}}} \exp (\bm{\kappa}^T c(\bm{x},\bm{\mu})  + s(\bm{x},\bm{\mu})^T \bm{R} s(\bm{x},\bm{\mu}) ) \left(1+\bm{\lambda}^T s(\bm{x},\bm{\mu}) \right) ,\\
c(\bm{x},\bm{\mu}) &=& (\cos(x_1-\mu_1), \ldots, \cos(x_d-\mu_d))^T, \\
s(\bm{x},\bm{\mu}) &=& (\sin(x_1-\mu_1), \ldots, \sin(x_d-\mu_d))^T, \\
 R_{i,j} &=& R_{j,i} , \quad -\infty<R_{i,j} < \infty ,\quad R_{i,i} = 0 ,\quad i,j\in\{1,\ldots,d\},
\end{eqnarray*} 
where $\bm{\mu}\in \mathbb{T}^d$, $\bm{\kappa}\in(\mathbb{R}^+)^d$, $\bm{\lambda}\in[-1,1]^d$ with $\sum_{s=1}^{d} |\lambda_s| \leq 1$ and where, in general, the analytic expression of the normalizing constant $C_{\bm{\kappa},\bm{R}}$ is unknown when $d>2$.

\subsection{Sine-skewed  Cosine distribution}

\begin{figure}[t]
 \centering
\begin{tabular}{cccc}
\subfloat{
    \includegraphics[height=0.16\textheight]{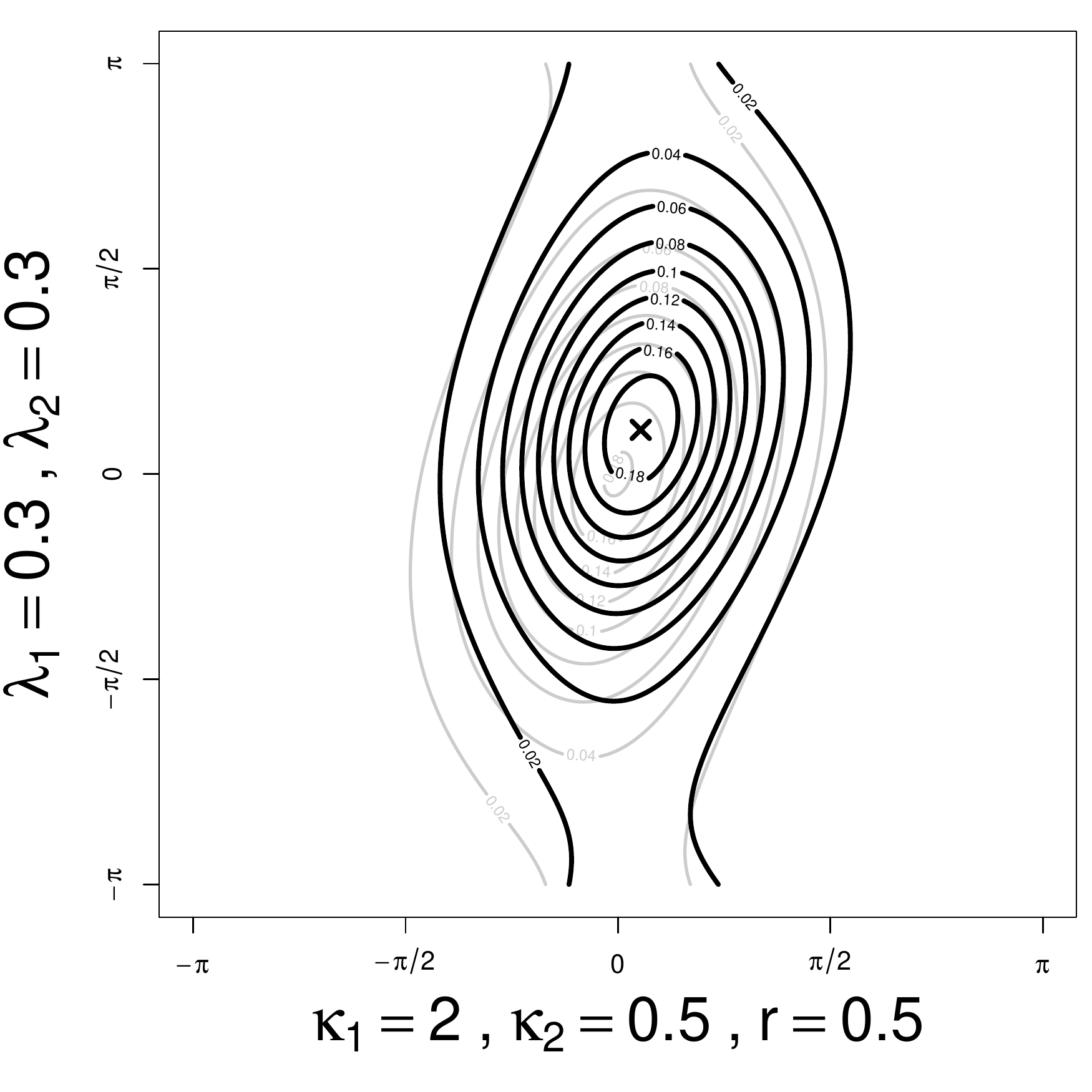}}
\subfloat{
    \includegraphics[height=0.16\textheight]{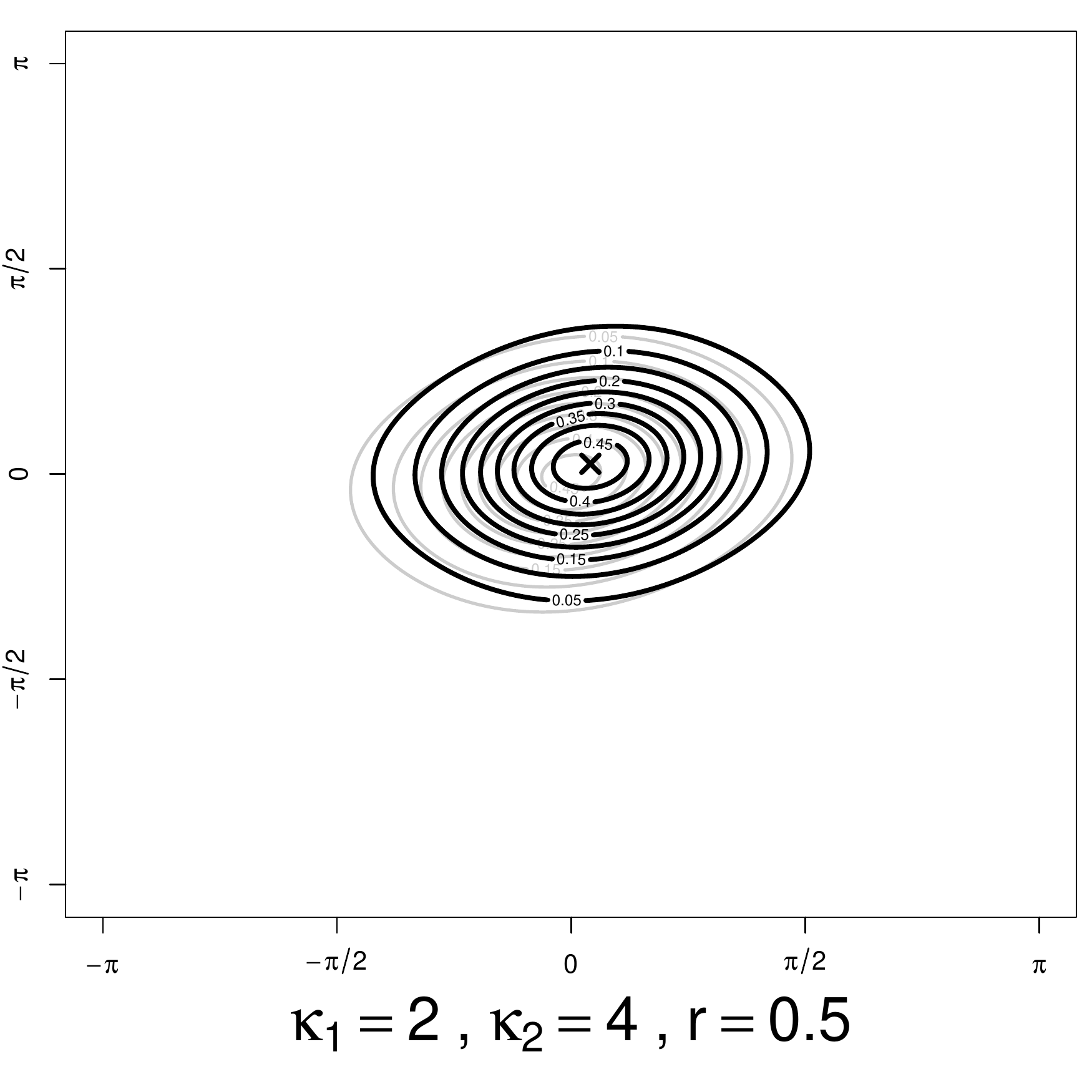}}
		\subfloat{
    \includegraphics[height=0.16\textheight]{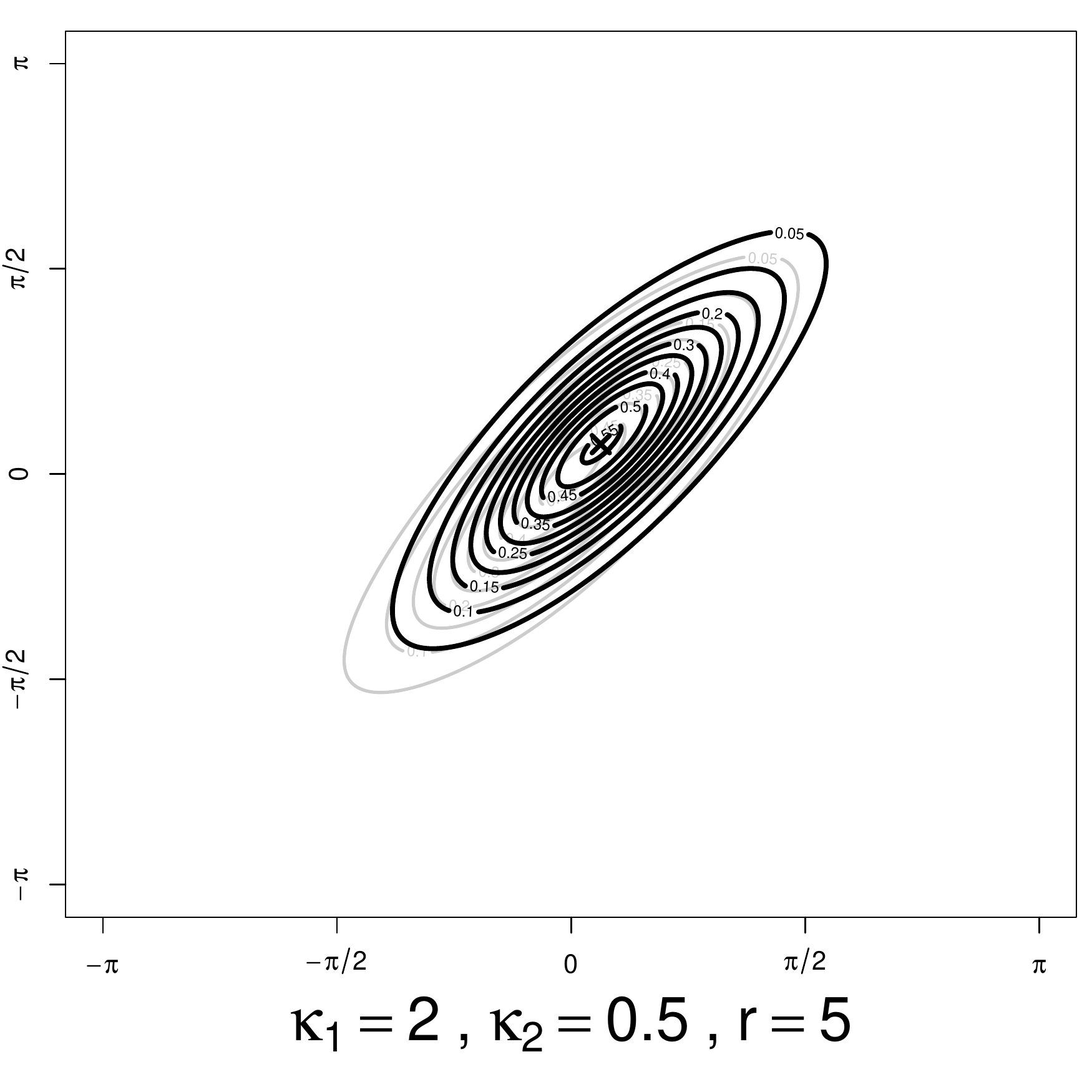}}
		\subfloat{
    \includegraphics[height=0.16\textheight]{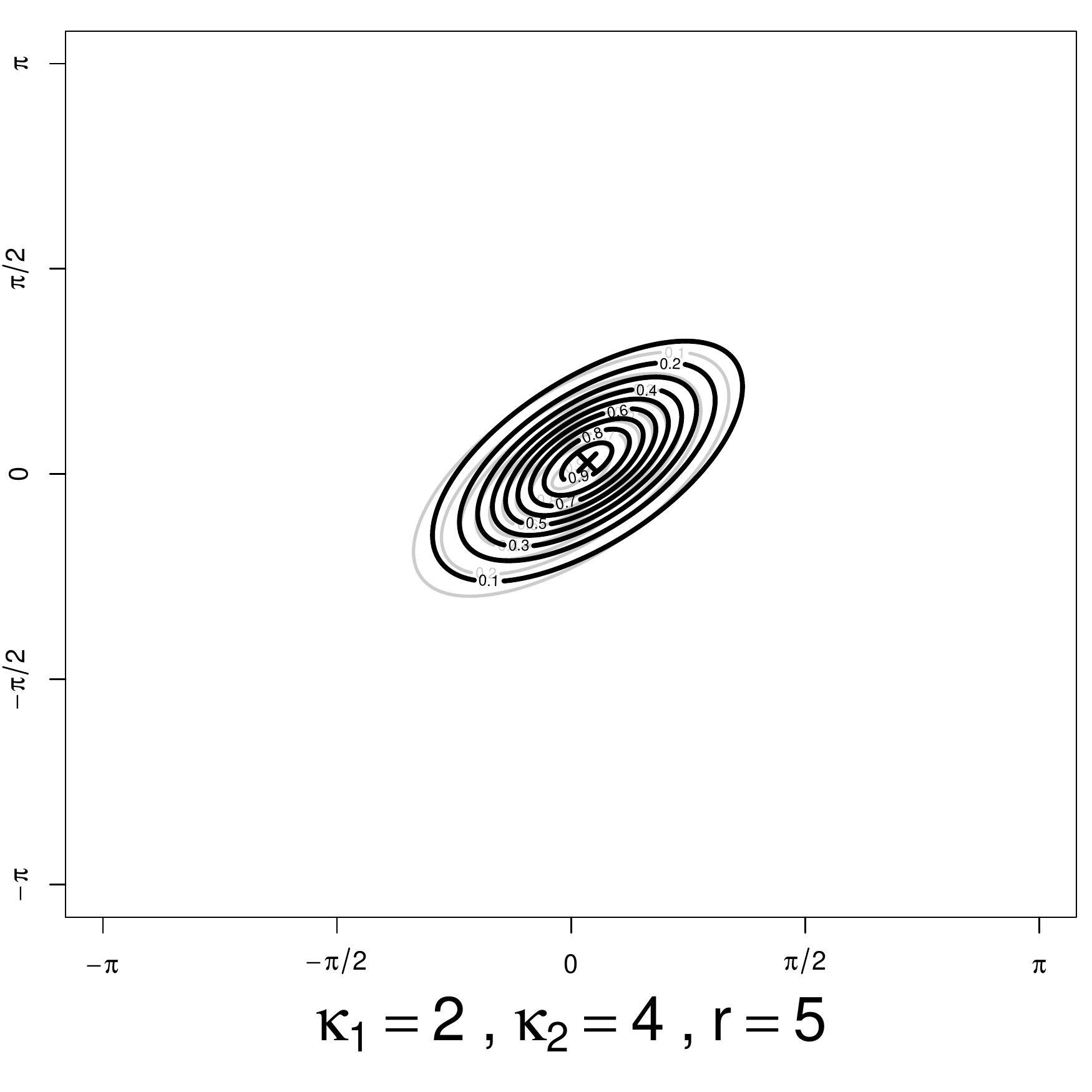}}\\
\subfloat{
    \includegraphics[height=0.16\textheight]{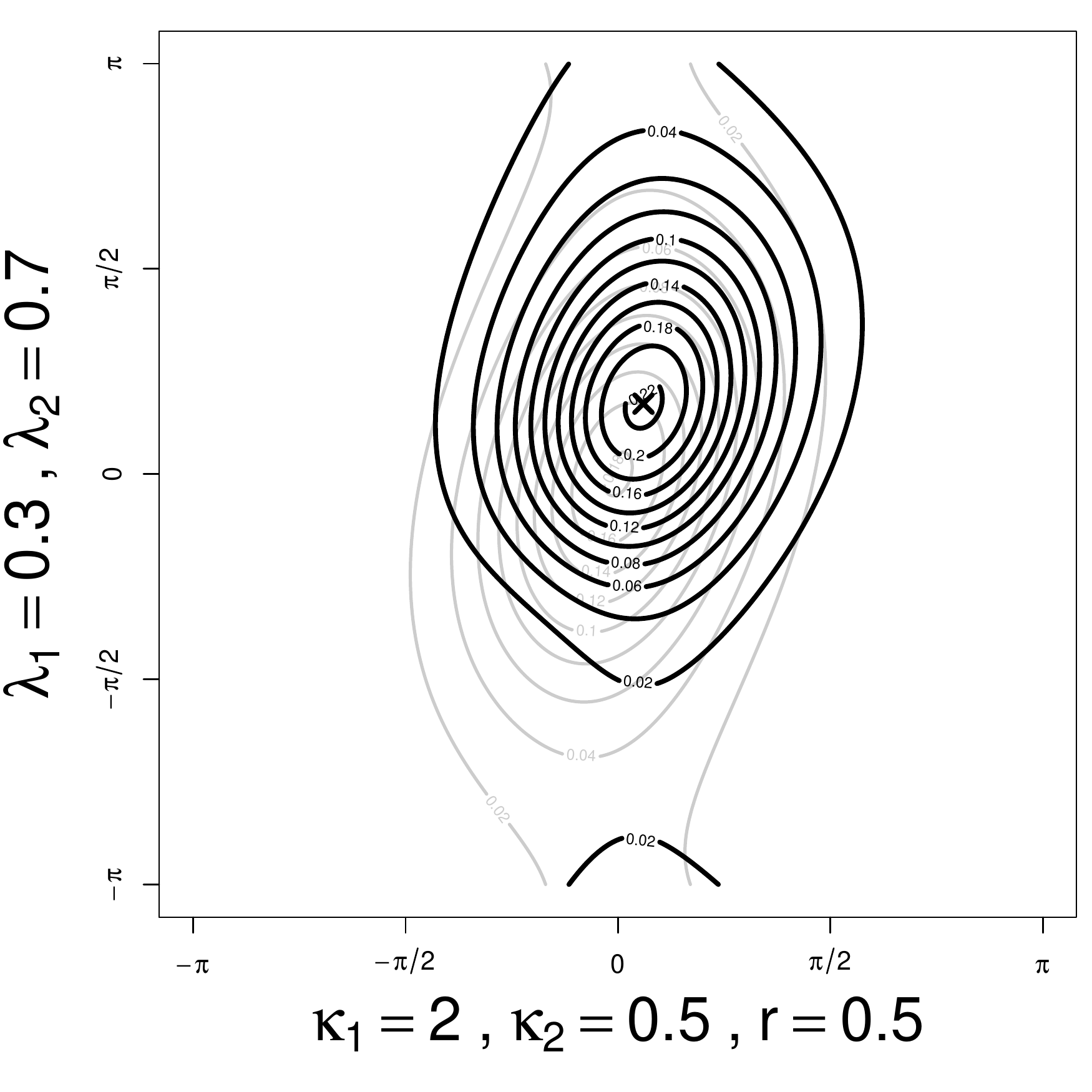}}
		\subfloat{
    \includegraphics[height=0.16\textheight]{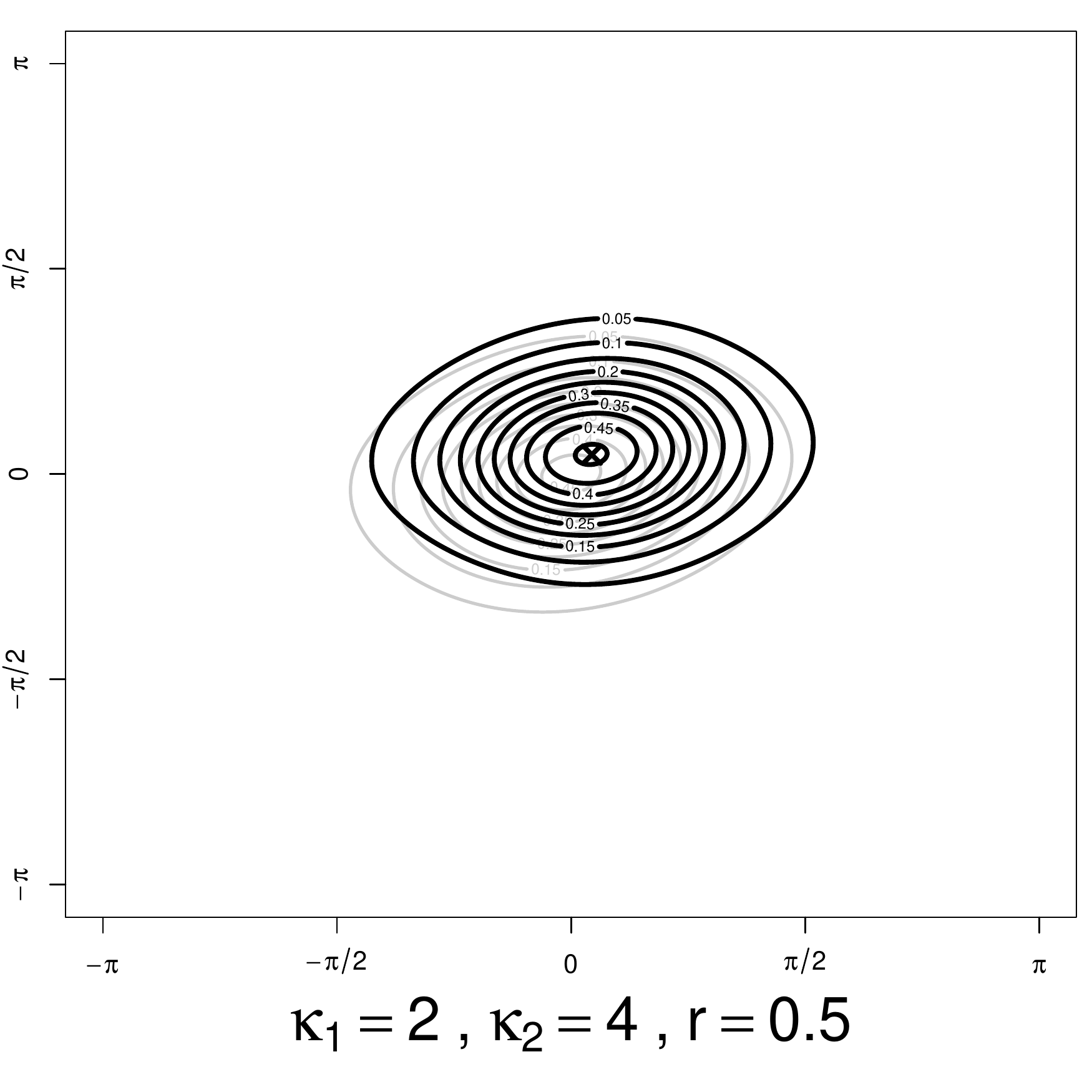}}
		\subfloat{
    \includegraphics[height=0.16\textheight]{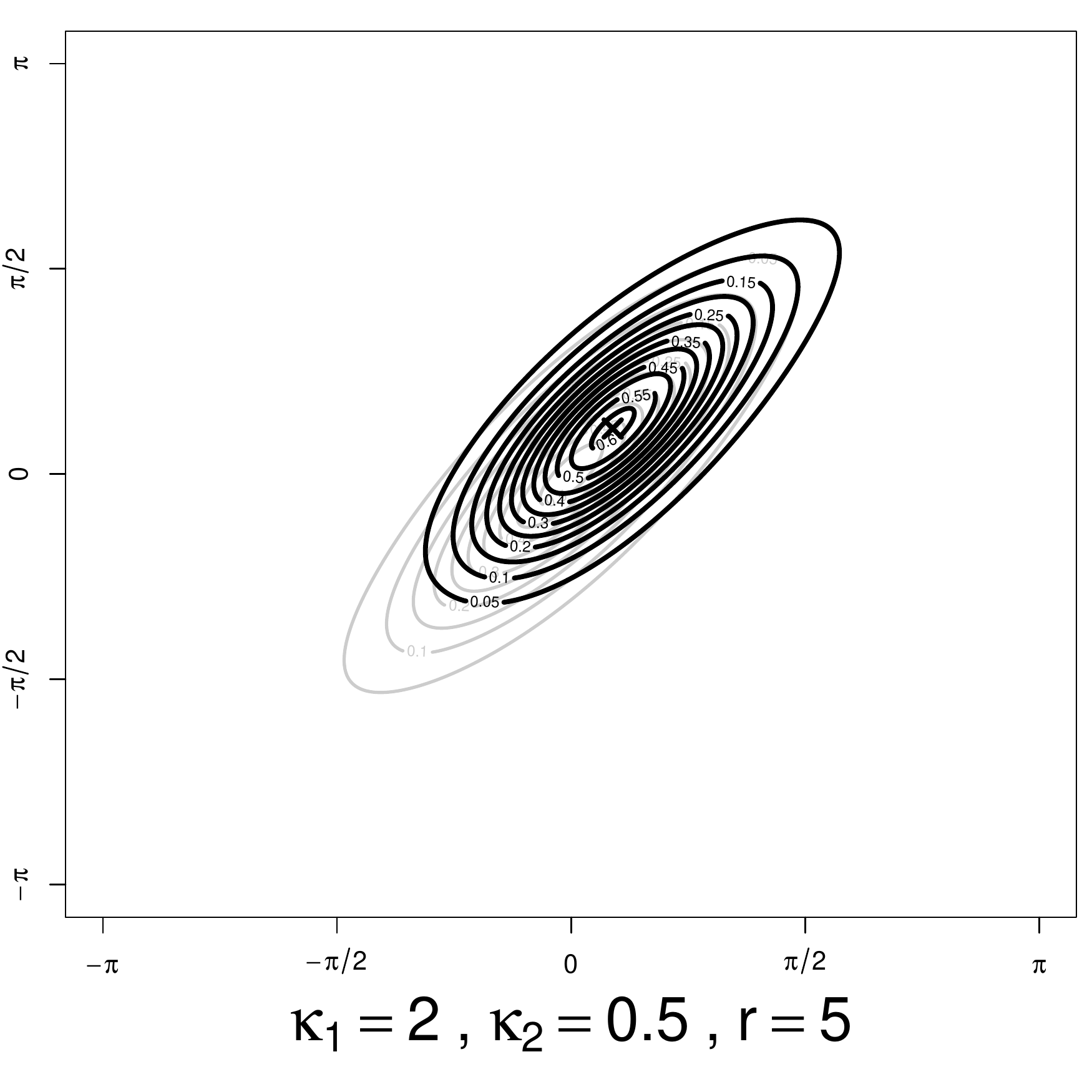}}
			\subfloat{
    \includegraphics[height=0.16\textheight]{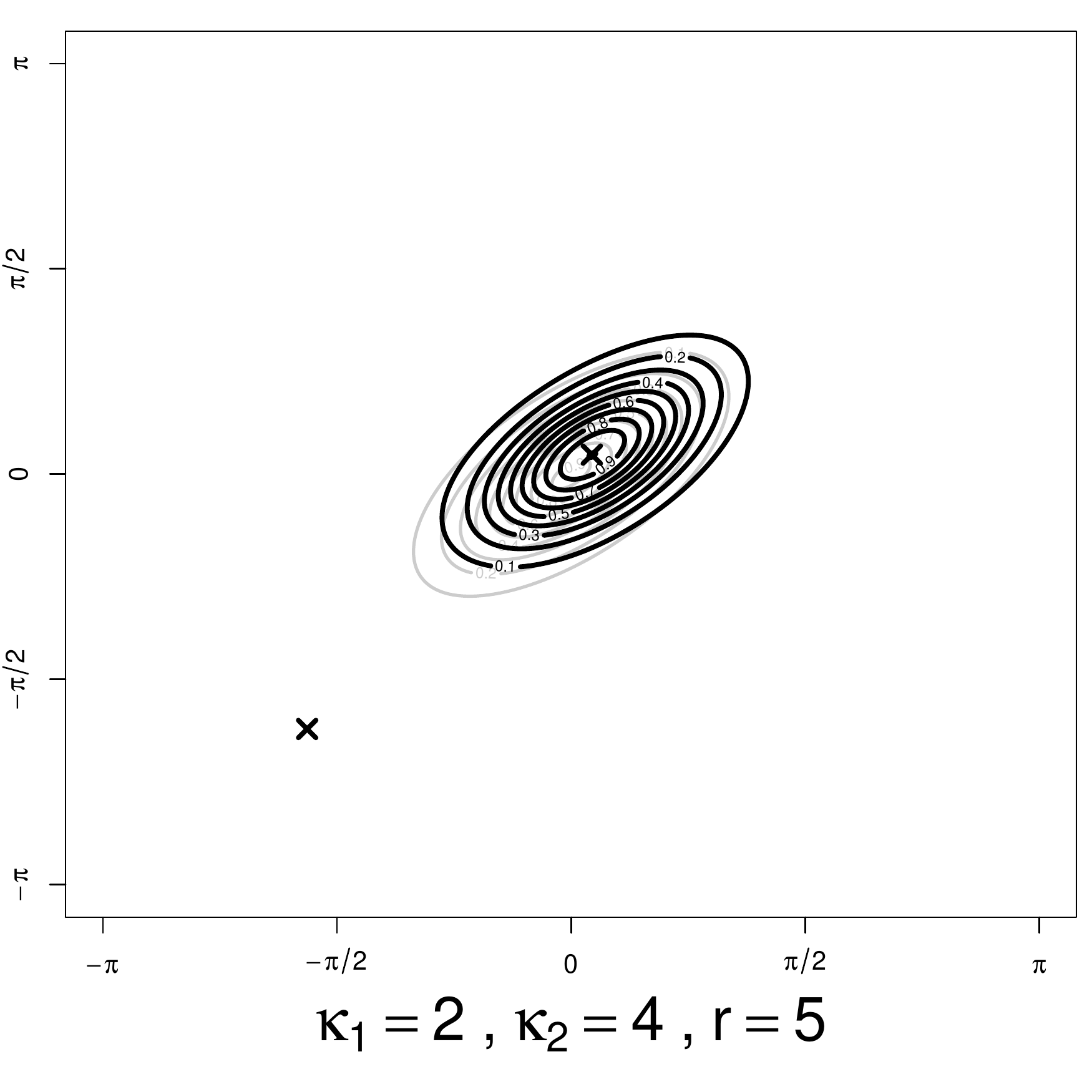}}\\
		\end{tabular}
		\vspace{0.7cm}
 \caption{Bivariate sine-skewed  Cosine density functions (with $\bm{\mu}=\bm{0}$), where the gray contour indicates the symmetric bivariate Cosine density ($\lambda_1=\lambda_2=0$) and the dark contour  the sine-skewed version obtained for the skewness parameters indicated at the left. The other parameters are indicated in the labels. \\From left to right: effect of the concentration and dependence parameters. From top to bottom: effect of the skewness parameters. The crosses identify the mode locations.}
 \label{figsscosine}
\end{figure}

Another popular submodel of the bivariate density of \cite{mardia1975} is the so-called \textit{Cosine model} of \cite{mardia2007}. A comparison between both models (Sine and Cosine) can be found in \cite{mardia2007}. Again five parameters are needed for the original density, and two are introduced as skewness parameters $\lambda_1,\lambda_2\in[-1,1]$ subject to $|\lambda_1|+|\lambda_2|\leq 1$. The sine-skewed Cosine density then reads
\begin{eqnarray*}
g_{\tiny{\mbox{SC}}}(\bm{x}-\bm{\mu};\bm{\kappa},r,\bm{\lambda})&=& C_{\bm{\kappa},r}^{-1} \exp (\kappa_1 \cos(x_1-\mu_1) + \kappa_2 \cos(x_2-\mu_2) + r \cos(x_1-\mu_1-x_2+\mu_2)) \\
&\times&(1+ \lambda_1 \sin (x_1-\mu_1)+ \lambda_2 \sin (x_2-\mu_2) ) \mbox{, with} \\
C_{\bm{\kappa},r} &=& 4\pi^2 \left( I_0 (\kappa_1) I_0 (\kappa_2)  I_0 (r) + 2 \sum_{i=1}^{\infty} I_i (\kappa_1) I_i (\kappa_2)  I_i (r)\right).
\end{eqnarray*}
For the original symmetric case $\bm{\lambda}=\bm{0}$, unimodality will depend on the values of $\bm{\kappa}$ and $r$, holding when $-r<\kappa_1\kappa_2/(\kappa_1+\kappa_2)$ \citep[see][Theorem 2]{mardia2007}. In this case, since the concentration parameters are always positive, unimodality will hold if also $r$ is positive. Plots of its sine-skewed version are provided in Figure~\ref{figsscosine}. As shown on the bottom-left panel, even when $r>0$, unimodality may not hold for the sine-skewed Cosine distribution. 

{Regarding the marginals, the same density as in~(\ref{marginal_sine}) is obtained, up to the  value of the normalizing constant $C_{\bm{\kappa},r}$ and the values of $a(x_1)$ and $b(x_1)$ which, in this case, are the quantities such that $\kappa_2+r \cos(x_1-\mu_1)=a(x_1)\cos(b(x_1))$ and $r \sin(x_1-\mu_1)=a(x_1)\sin(b(x_1))$. Thus, similar comments as for the sine-skewed Sine distribution apply.}

{From the $d$-dimensional von Mises density proposed by \cite{MP05}, as a particular case, the generalization of the Cosine distribution can be derived. In that case, the resulting $d$-variate sine-skewed Cosine density is
\begin{eqnarray*}
g_{\tiny{\mbox{SC}}}(\bm{x}-\bm{\mu};\bm{\kappa},\bm{R},\bm{\lambda})&=& \frac{1}{C_{\bm{\kappa},\bm{R}}} \exp (\bm{\kappa}^T c(\bm{x},\bm{\mu})  + s(\bm{x},\bm{\mu})^T \bm{R} s(\bm{x},\bm{\mu}) +c(\bm{x},\bm{\mu})^T \bm{R} c(\bm{x},\bm{\mu}) ) \\
&\times& \left(1+\bm{\lambda}^T s(\bm{x},\bm{\mu}) \right) ,\\
 R_{i,j} &=& R_{j,i} , \quad -\infty<R_{i,j} < \infty ,\quad R_{i,i} = 0 ,\quad i,j\in\{1,\ldots,d\},
\end{eqnarray*} 
where $\bm{\mu}\in \mathbb{T}^d$, $\bm{\kappa}\in(\mathbb{R}^+)^d$ and $\bm{\lambda}\in[-1,1]^d$ with $\sum_{s=1}^{d} |\lambda_s| \leq 1$.
}

\subsection{Bivariate sine-skewed  wrapped Cauchy distribution}

\begin{figure}
 \centering
\begin{tabular}{cccc}
\subfloat{
    \includegraphics[height=0.16\textheight]{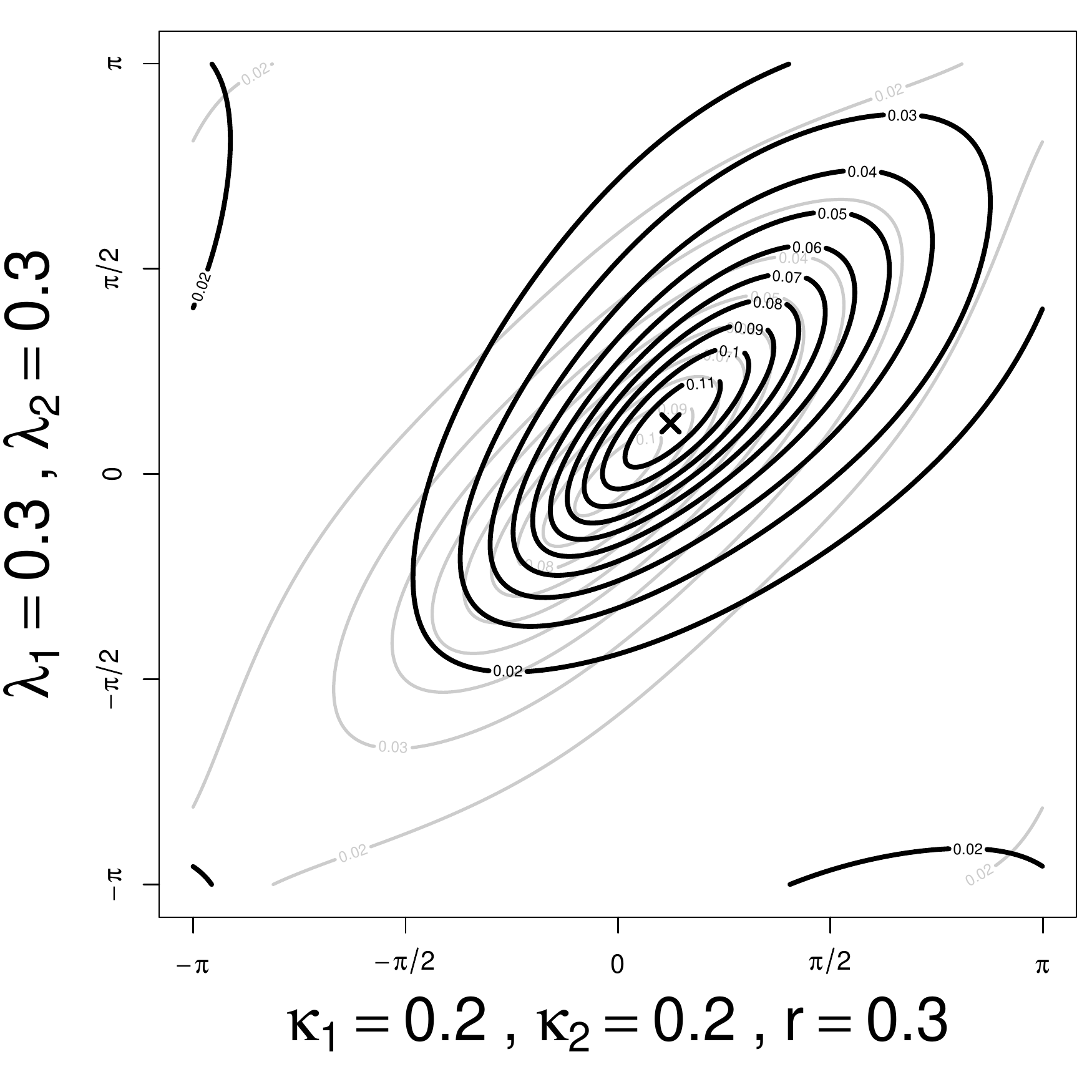}}
\subfloat{
    \includegraphics[height=0.16\textheight]{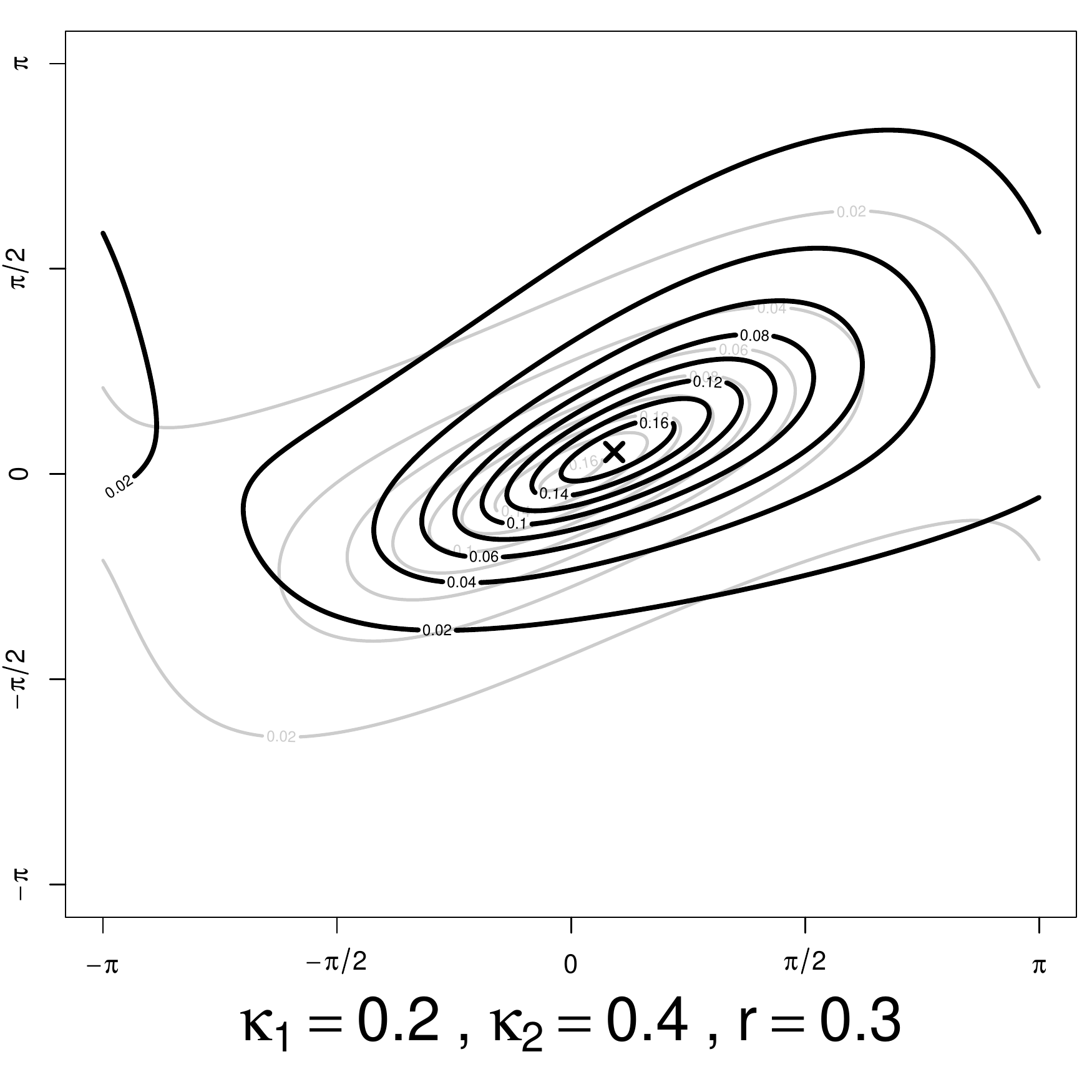}}
		\subfloat{
    \includegraphics[height=0.16\textheight]{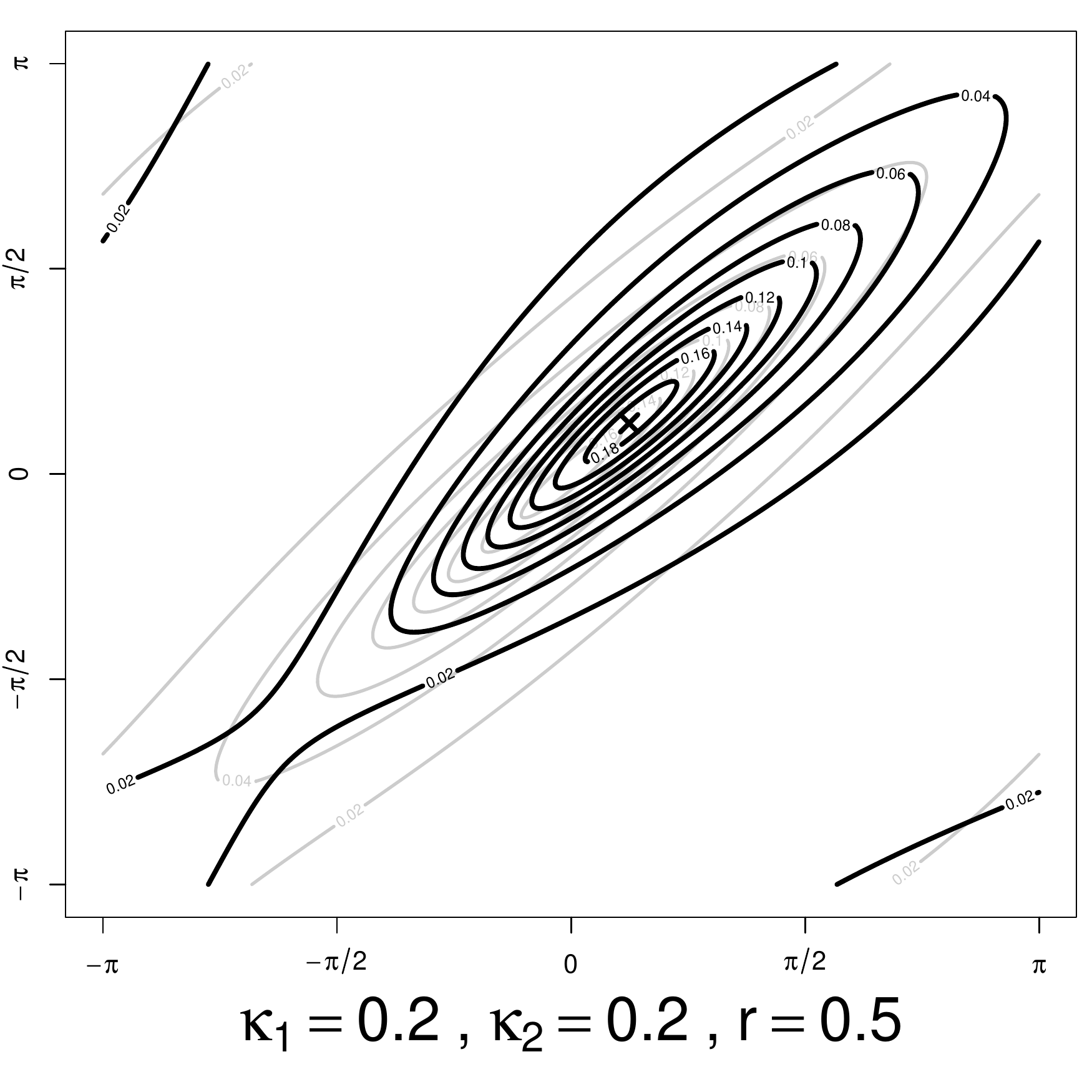}}
		\subfloat{
    \includegraphics[height=0.16\textheight]{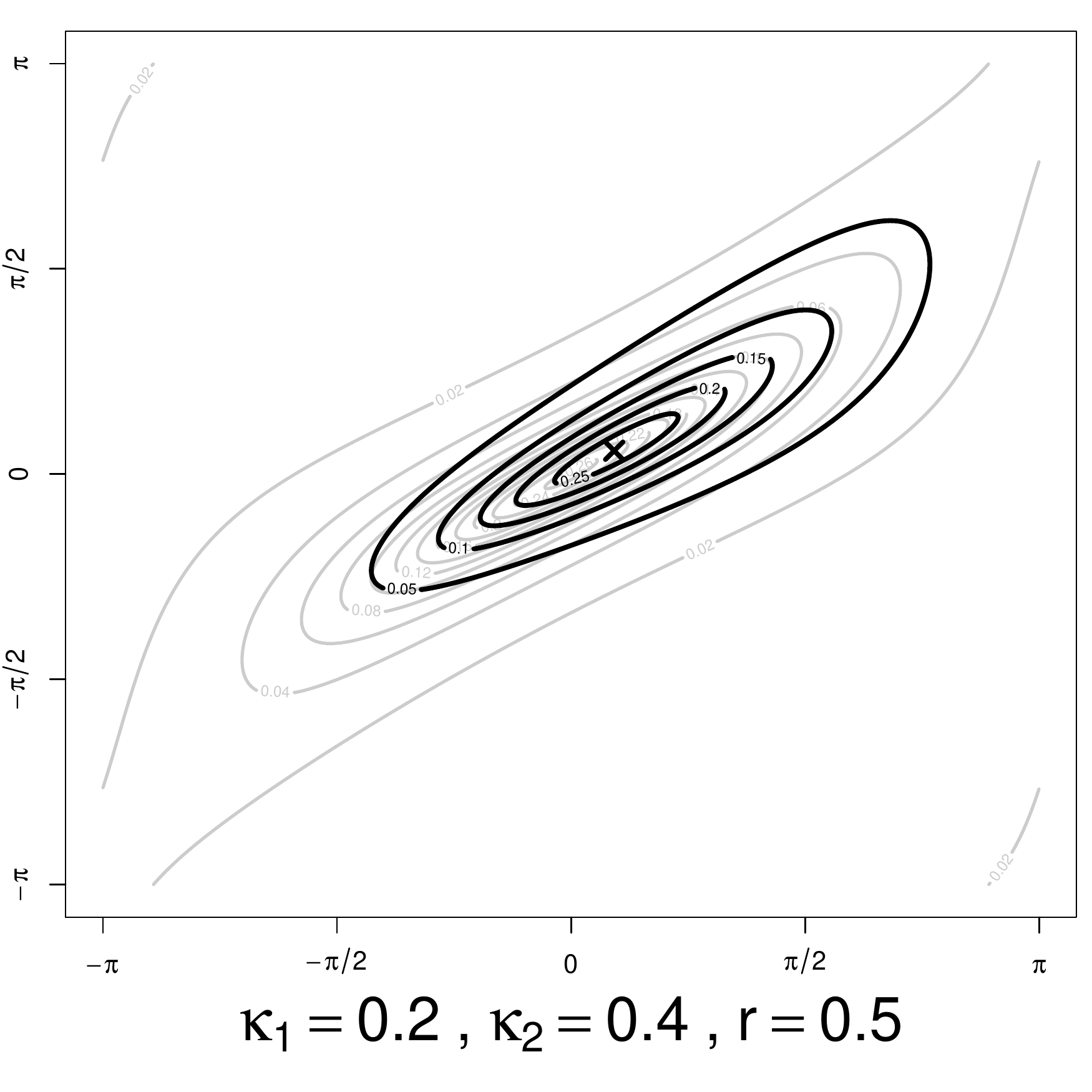}}\\
\subfloat{
    \includegraphics[height=0.16\textheight]{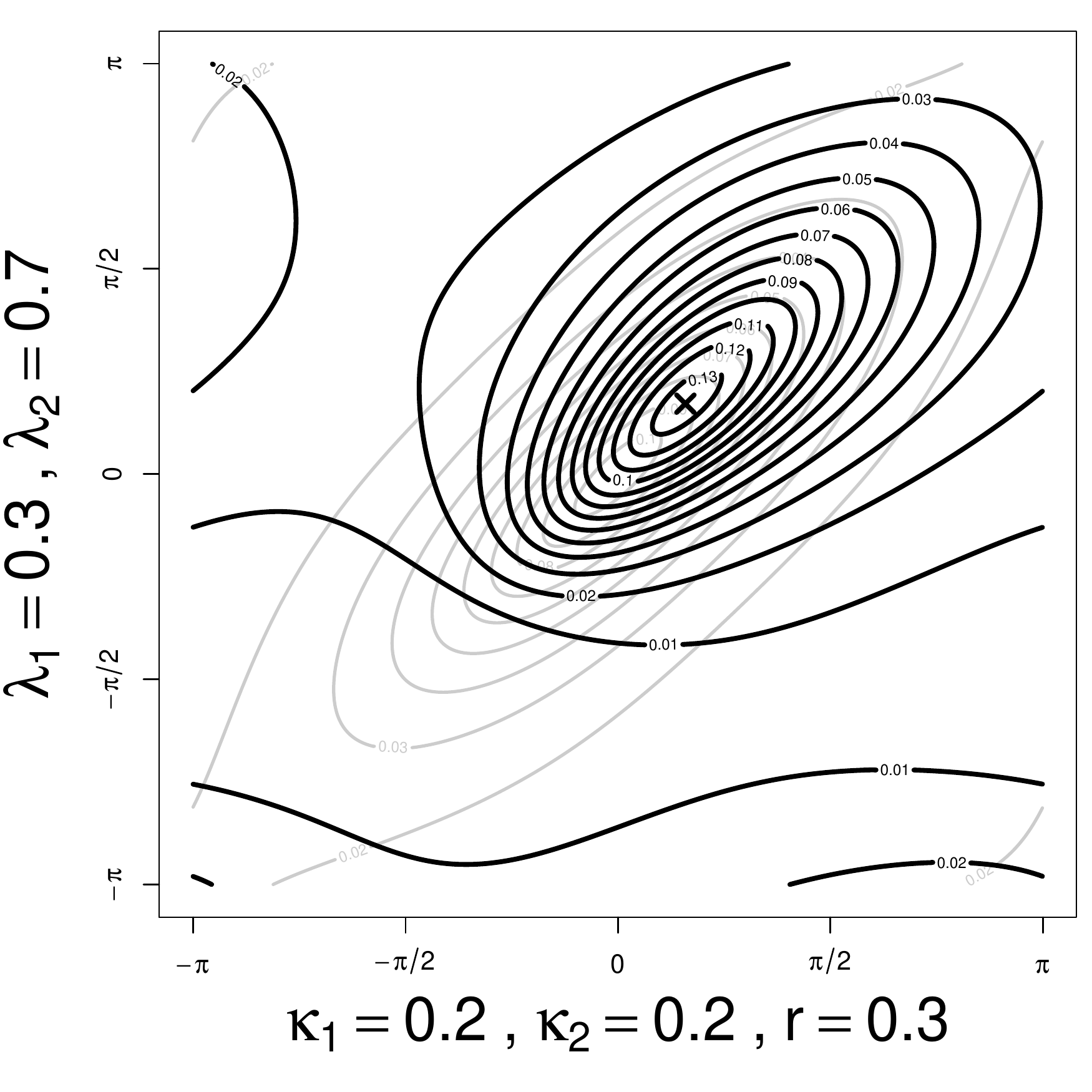}}
		\subfloat{
    \includegraphics[height=0.16\textheight]{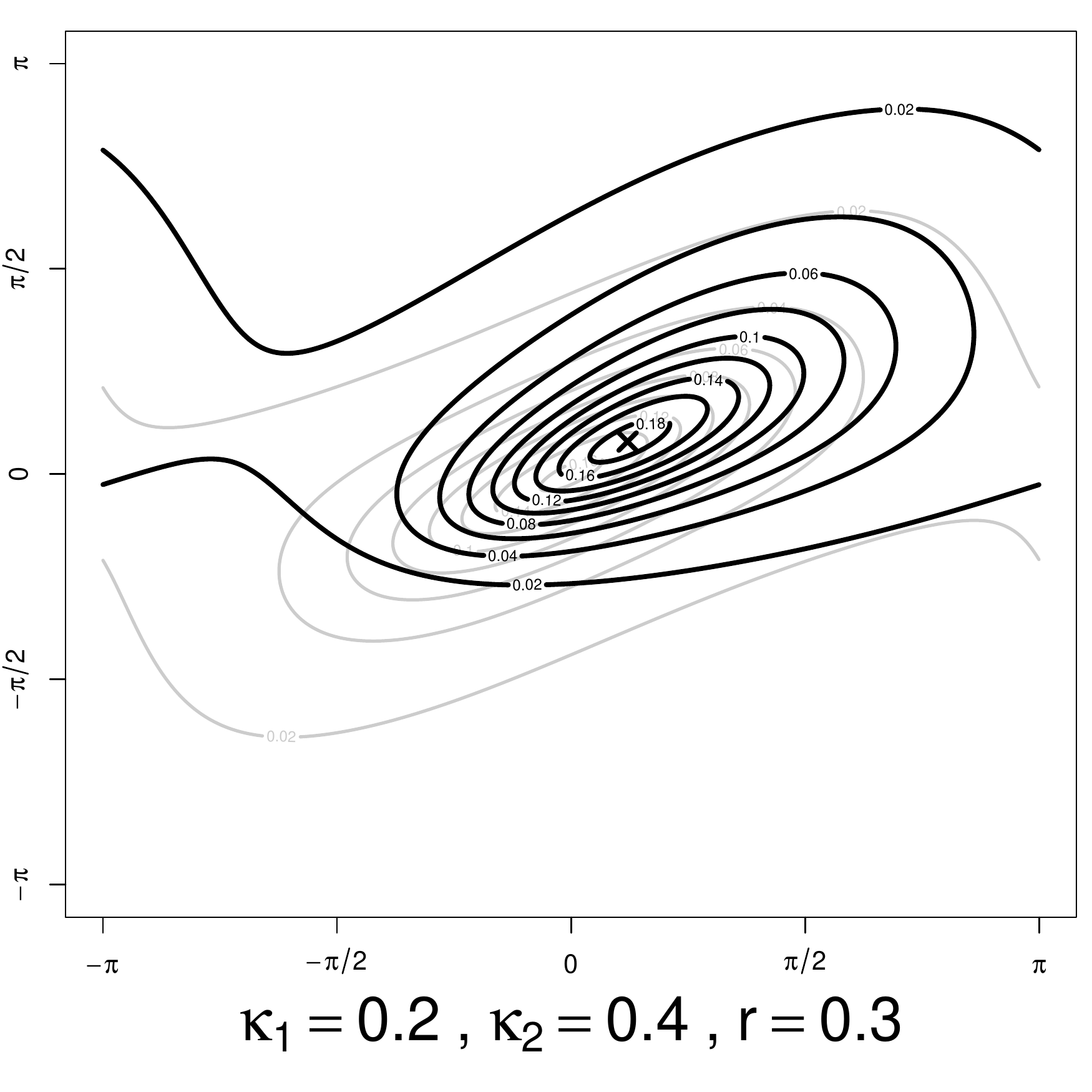}}
		\subfloat{
    \includegraphics[height=0.16\textheight]{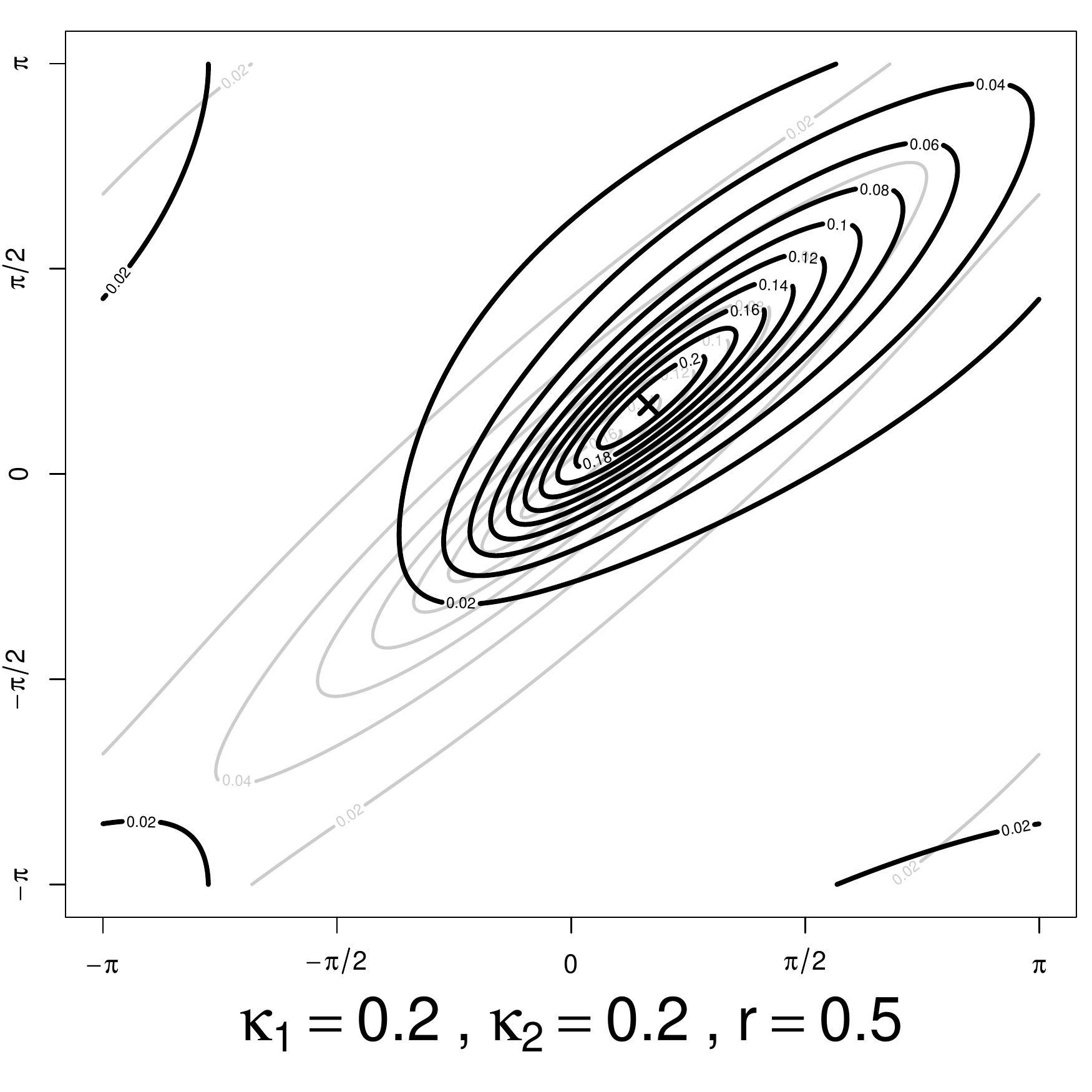}}
			\subfloat{
    \includegraphics[height=0.16\textheight]{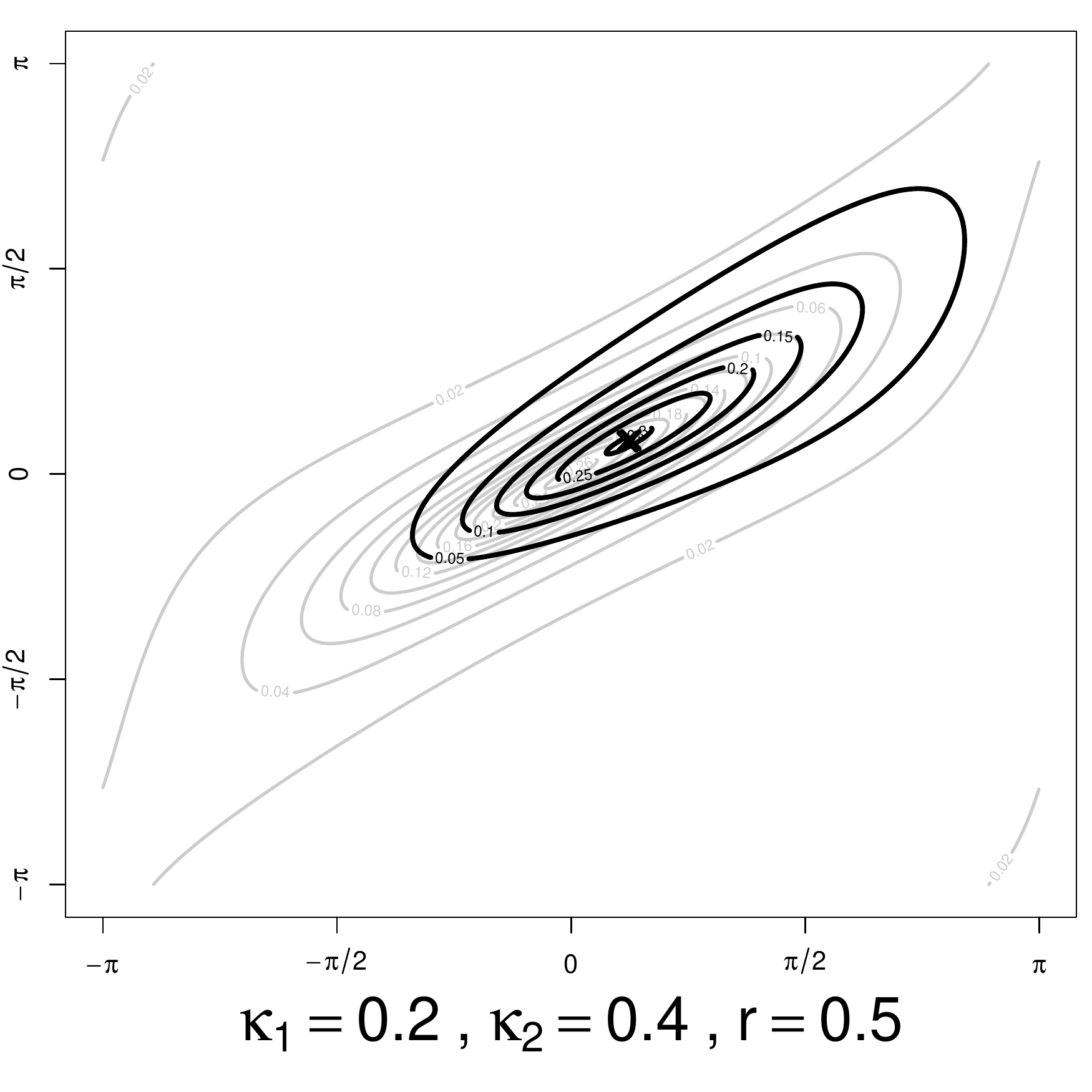}}\\
		\end{tabular}
		\vspace{0.7cm}
 \caption{Bivariate sine-skewed  wrapped Cauchy density functions (with $\bm{\mu}=\bm{0}$), where the gray contour indicates the symmetric bivariate wrapped Cauchy density ($\lambda_1=\lambda_2=0$) and the dark contour  the sine-skewed version obtained for the skewness parameters indicated at the left. The other parameters are indicated in the labels. \\From left to right: effect of the concentration and dependence parameters. From top to bottom: effect of the skewness parameters. The crosses identify the mode locations.}
 \label{figsswc}
\end{figure}

While Sine and Cosine models focus on having conditional von Mises densities, inherited logically from the bivariate von Mises distribution, \cite{kato2015} provide a way of constructing toroidal densities focusing on the wrapped Cauchy distribution. Starting from their density, the sine-skewed bivariate wrapped Cauchy distribution corresponds to
\begin{eqnarray}
g_{\tiny{\mbox{SWC}}}(\bm{x}-\bm{\mu};\bm{\kappa},r,\bm{\lambda})&=& (1-r^2)(1-\kappa_1^2)(1-\kappa_2^2)[4\pi^2 (c_0-c_1 \cos(x_1-\mu_1) - c_2 \cos(x_2-\mu_2)  \nonumber   \\
&-&  c_3 \cos(x_1-\mu_1)\cos(x_2-\mu_2) - c_4 \sin(x_1-\mu_1)\sin(x_2-\mu_2))]^{-1}  \nonumber   \\
&\times& (1+ \lambda_1 \sin (x_1-\mu_1)+ \lambda_2 \sin (x_2-\mu_2) ),  \label{sswC} \\ 
c_0&=& (1+r^2)(1+\kappa_1^2)(1+\kappa_2^2)-8|r|\kappa_1\kappa_2 , \nonumber   \\
c_1&=& 2(1+r^2)\kappa_1(1+\kappa_2^2)-4|r|(1+\kappa_1^2)\kappa_2 ,  \nonumber   \\
c_2&=& 2(1+r^2)(1+\kappa_1^2)\kappa_2-4|r|\kappa_1(1+\kappa_2^2) , \nonumber   \\
c_3&=& -4(1+r^2)\kappa_1\kappa_2+2|r|(1+\kappa_1^2)(1+\kappa_2^2) , \nonumber   \\
c_4&=& 2r (1-\kappa_1^2)(1-\kappa_2^2), \nonumber   
\end{eqnarray}
with  location parameter $\bm{\mu}\in \mathbb{T}^2$, concentration parameter $\bm{\kappa}\in[0,1)^2$, skewness parameter $\bm{\lambda}\in[-1,1]^2$ subject to with $|\lambda_1|+|\lambda_2|\leq 1$, and  dependence parameter $-1<r < 1$. The contour plots of the density~(\ref{sswC}) for different values of these parameters are shown in Figure~\ref{figsswc}.

\begin{figure}
\begin{tabular}{cccc}
 \centering
\subfloat{
    \includegraphics[height=0.16\textheight]{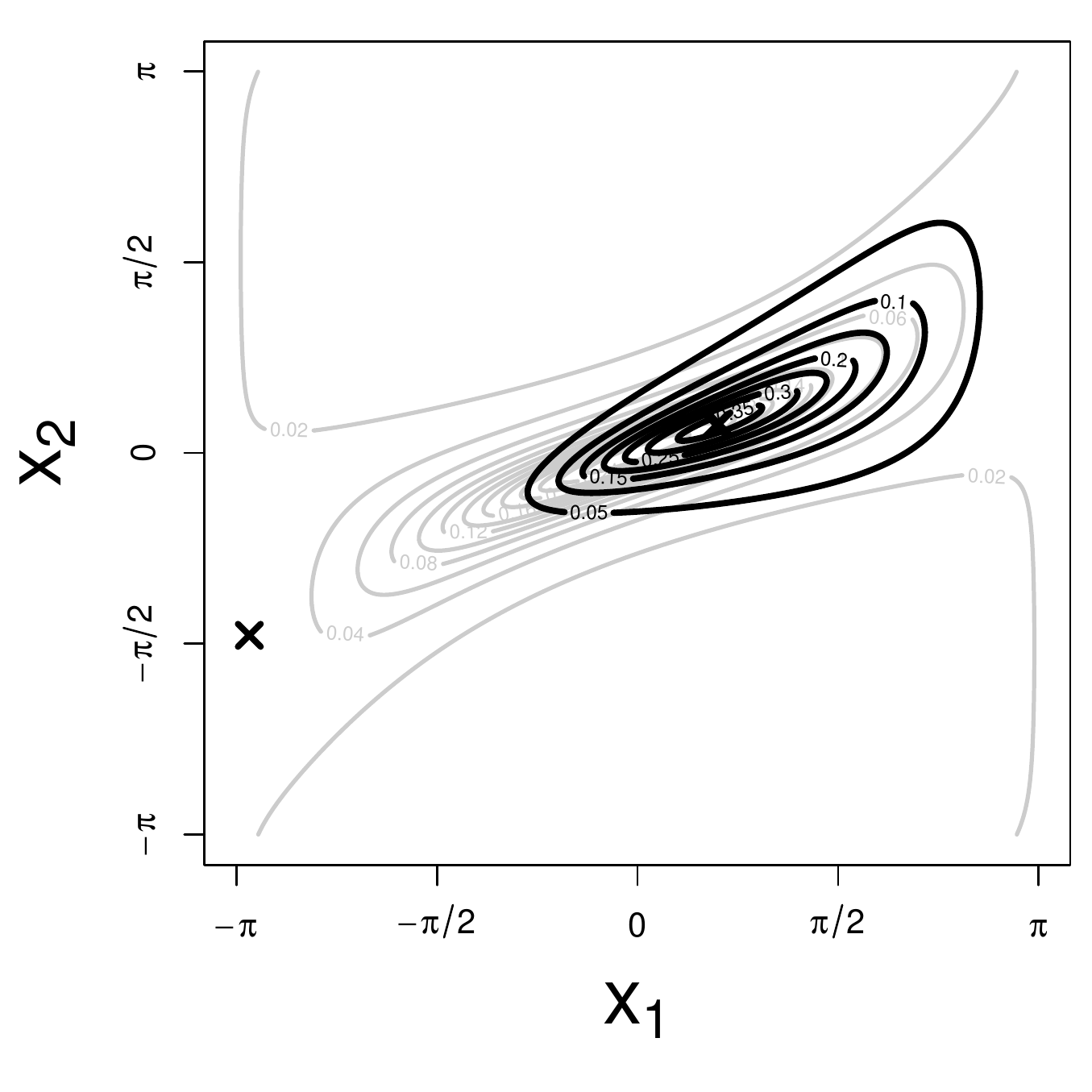}}
\subfloat{
    \includegraphics[height=0.16\textheight]{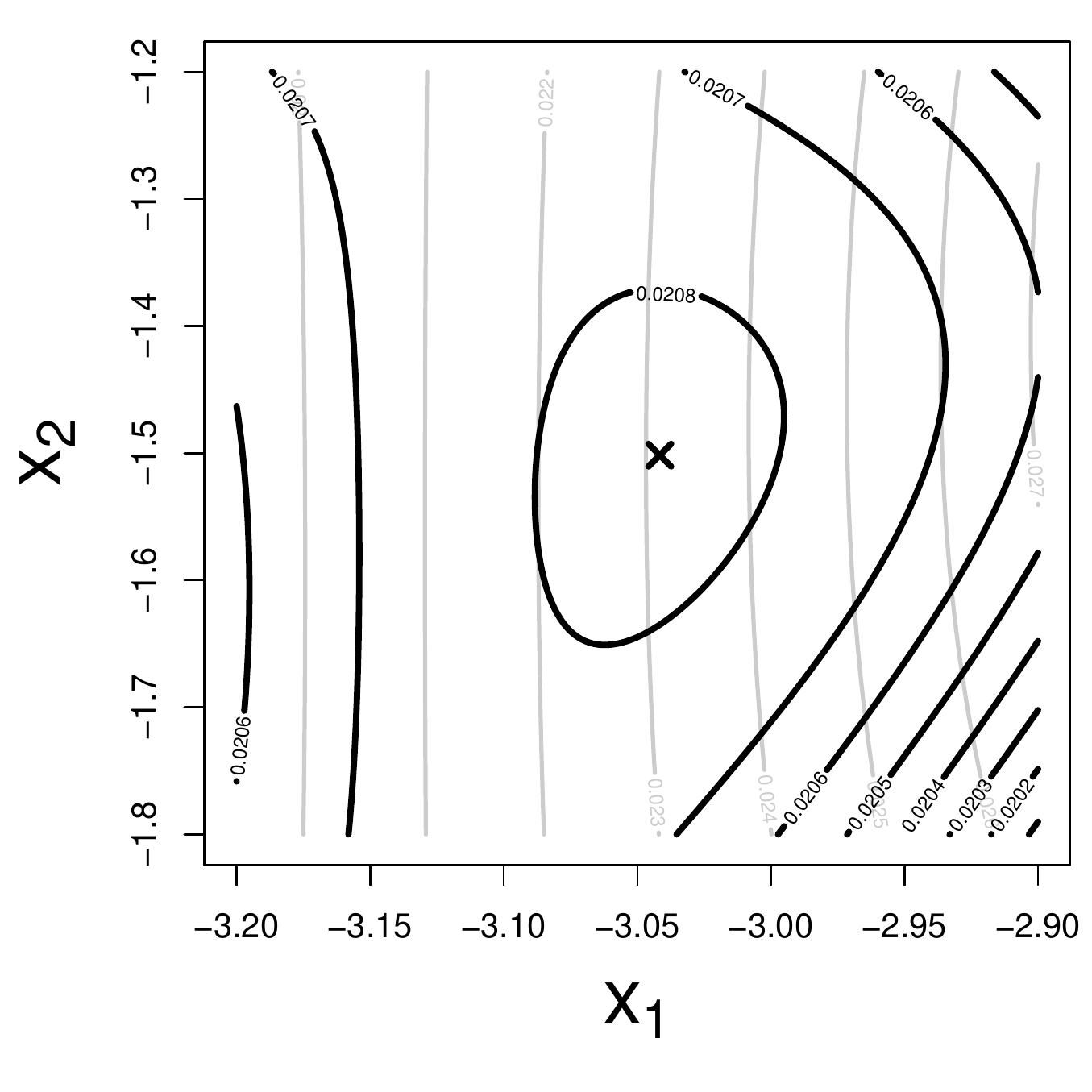}}
		\subfloat{
    \includegraphics[height=0.16\textheight]{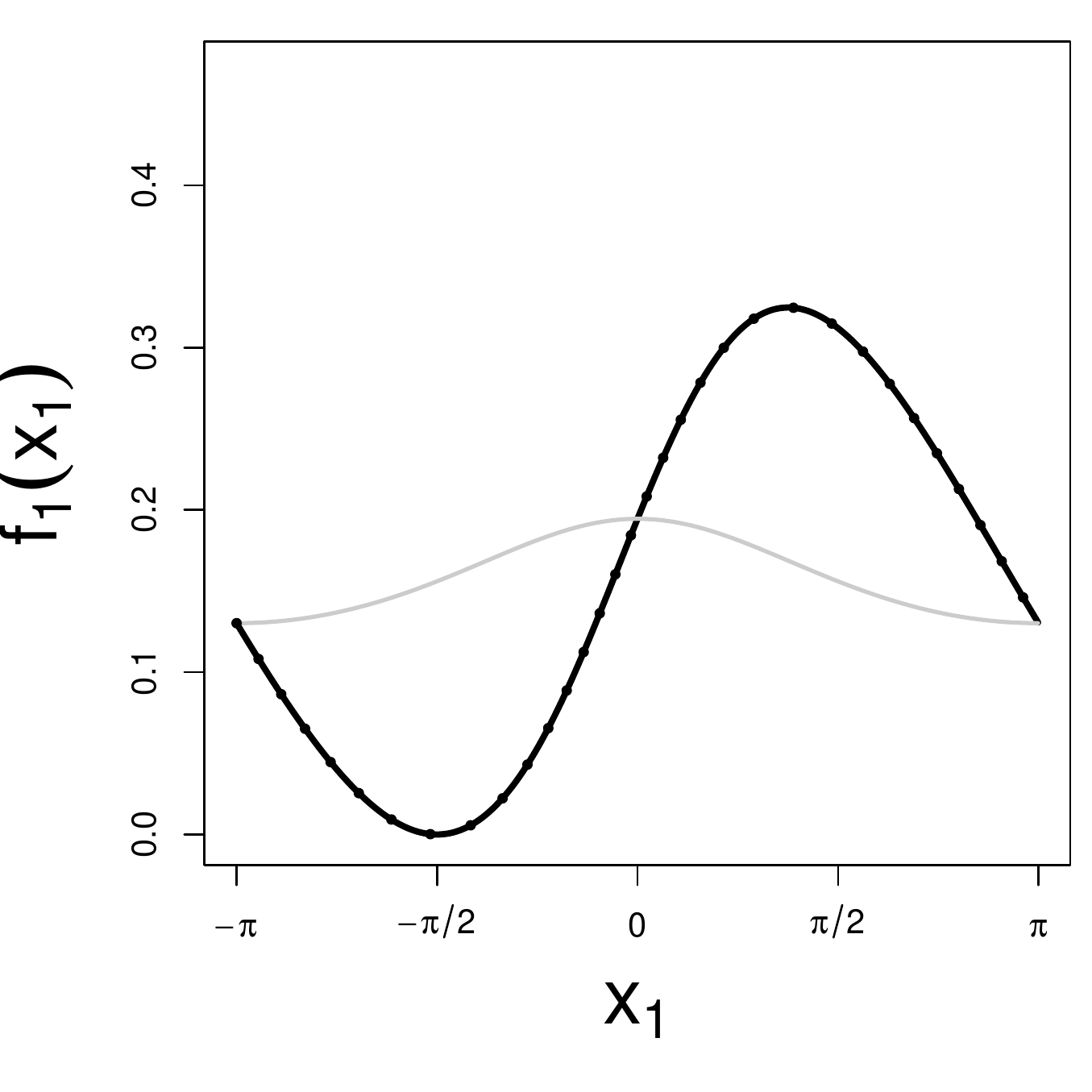}}
		\subfloat{
    \includegraphics[height=0.16\textheight]{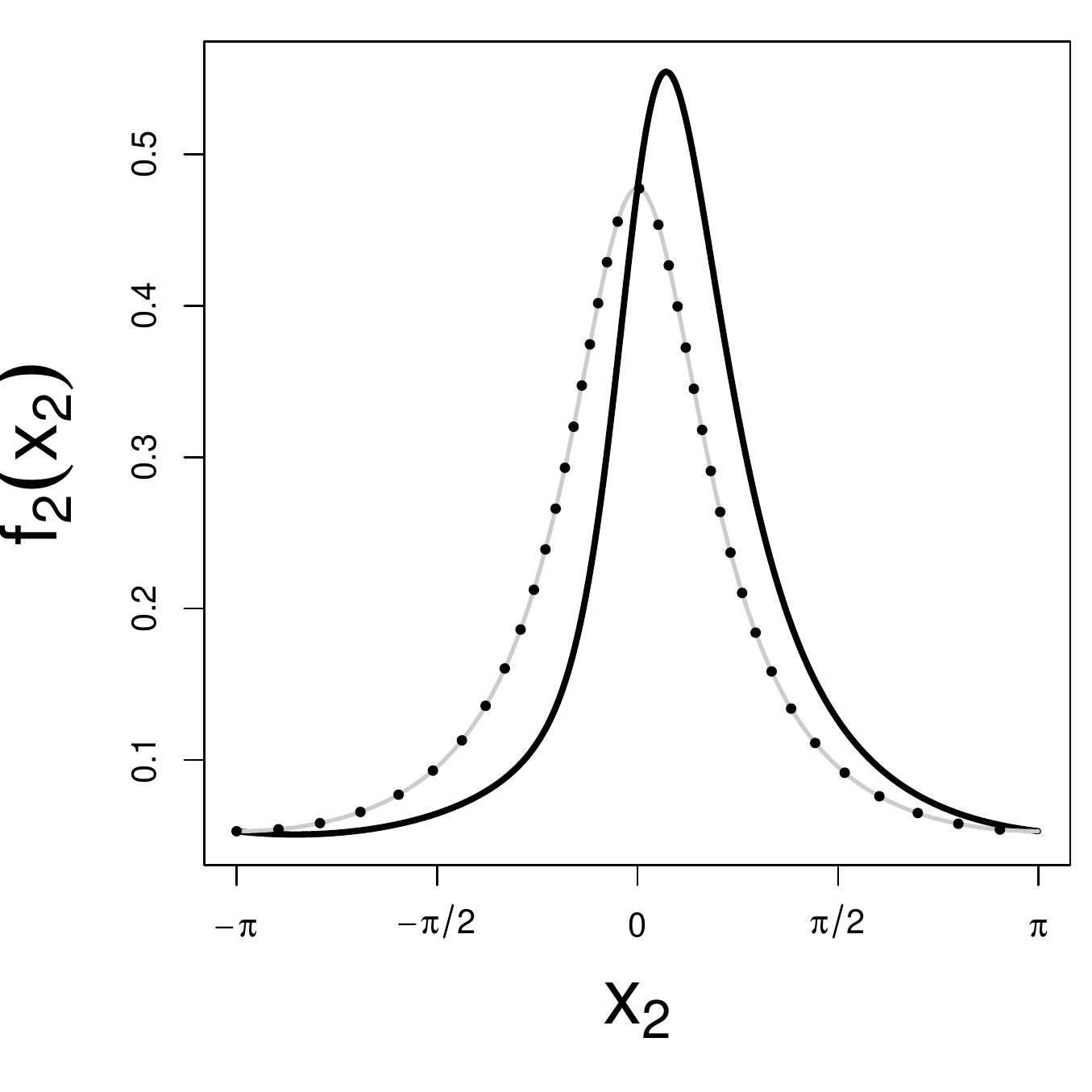}}
		\end{tabular}
		\vspace{0.7cm}
 \caption{Bivariate sine-skewed wrapped Cauchy density functions with parameters $\bm{\mu}=\bm{0},\kappa_1=0.1,\kappa_2=0.5$ and $r=0.5$, where the dark continuous lines correspond to $\lambda_1=1$ and $\lambda_2=0$ and the gray continuous lines indicate the bivariate wrapped Cauchy density ($\lambda_1=\lambda_2=0$). The crosses identify the mode locations, and in the  two right-most settings the dotted lines stand for the univariate sine-skewed densities with parameters $\kappa_s$ and $\lambda_s$.\\
 Left: entire support $\mathbb{T}^2$. Center-left: close-up to show the secondary mode. Center-right: first marginal. Right: second marginal.}
 \label{figsswc_multimodal}
\end{figure}

The marginal distributions of the model  by \cite{kato2015} are univariate wrapped Cauchy distributions.  From the different (non-uniform) base symmetric densities investigated by \cite{abe2011} for the univariate case, the wrapped Cauchy is the only one providing always unimodal sine-skewed alternatives. In \cite{kato2015}, it is mentioned that the bivariate wrapped Cauchy is always unimodal when $\kappa_1,\kappa_2>0$. This persistent unimodality feature is not inherited in the bivariate sine-skewed toroidal setting. Despite the fact that, in most of the investigated cases,  unimodality was obtained, this property cannot be guaranteed to hold unless both marginals are independent. A counter-example to unimodality corresponds to the following choice of parameter values: $\kappa_1=0.1,\kappa_2=0.5,r=0.5, \lambda_1=1$ and $\lambda_2=0$ (see Figure~\ref{figsswc_multimodal}, left panels). Figure~\ref{figsswc_multimodal} (right panels) informs us about the behavior of the marginals. While $\lambda_2=0$ guarantees that the first marginal will be equivalent to the sine-skewed version of the circular wrapped Cauchy density with parameters $\kappa_1$ and $\lambda_1$, the same is not true for the second case as $\lambda_1\neq 0$. In this case, note also that even if $\lambda_2=0$, a skewed univariate marginal in the second dimension is obtained.

\section{Inference}\label{sec:Inf}

Let $\bm{\mathcal{X}}=(\bm{X}_1,\ldots,\bm{X}_n)$, with $\bm{X}_i=({X}_{i;1},\ldots,{X}_{i;d})^T$, denote a random sample of $d$-toroidal data generated by the distribution associated to the sine-skewed toroidal density~(\ref{sine_skewed}). The objective of this section is to explain how to estimate the various parameters, derive the asymptotic properties of the estimates and build a test for symmetry against sine-skewedness. 

If the trigonometric moments of the base symmetric density are known, the method of moments estimators can easily be obtained by equating the expressions of the shape parameters (see Subsection~\ref{general_properties}) with their sample counterparts. The main issue of this approach is that, even for the univariate case, the solutions of those equations need not necessarily exist \citep[see][Subsection~4.3]{abe2011}. For that reason, in the remainder of this section, we will focus on the maximum likelihood estimators.

\subsection{Maximum likelihood estimation}\label{mle_section}

Given a base density $f(\xb-\mub;\bm{\vartheta})$,  the log-likelihood function of its sine-skewed counterpart is given by
\begin{eqnarray}\label{log-like}
\ell(\bm{\mu},\bm{\vartheta},\bm{\lambda})=\sum_{i=1}^n \log \left(1+\sum_{s=1}^{d} \lambda_s \sin (X_{i;s}-\mu_s)\right)+\sum_{i=1}^n \log f(\bm{X}_i-\bm{\mu};\bm{\vartheta}),
\end{eqnarray}
whose maximization with respect to $\bm{\mu},\bm{\vartheta},\bm{\lambda}$ yields the maximum likelihood estimates (MLEs).  Closed-form expressions in general do not exist for the MLEs, and we have to revert to numerical methods able to deal with a nonlinear optimization problem since we need to satisfy the constraint $\sum_{s=1}^{d} |\lambda_s| \leq 1$. We use the solver proposed by \cite{ye1987} implemented in the \texttt{Rsolnp} package \citep{Ghalanos2015} of the statistical software $\mathtt{R}$, and suggest to use different initial points to avoid having a local maximum as a solution.

\subsection{Asymptotic behavior of the maximum likelihood estimators}

Denote by $(\bm{\mu}_0,\bm{\vartheta}_0,\bm{\lambda}_0)$ the true parameter values and suppose that they lie in the interior of their domain. Then, from Equation~(\ref{log-like}), we can see that if the base density satisfies the assumptions for asymptotic normality results on the MLE of $\bm{\mu},\bm{\vartheta}$ \citep[see][Theorem 5.39 or the classical conditions in Subsection 5.6]{vaart2000}, then its sine-skewed counterpart will also satisfy them. We thus have
\begin{equation*}
\sqrt{n}\left((\hat{\bm{\mu}}^T,\hat{\bm{\vartheta}}^T,\hat{\bm{\lambda}}^T)^T - (\bm{\mu}_0^T,\bm{\vartheta}_0,^T\bm{\lambda}_0^T)^T \right) \overset{d}{\rightarrow} N\left(\bm{0},\mathfrak{I}^{-1}\right), 
\end{equation*}
where $\overset{d}{\rightarrow}$ represents convergence in distribution and $\mathfrak{I}$ denotes the Fisher Information matrix. The elements of that symmetric matrix $\mathfrak{I}$ are equal to 
 \begin{eqnarray*}
\mathfrak{i}_{\mu_j,\mu_k}&=& \mathfrak{i}_{\mu_j,\mu_k}^0 +
\int_{\mathbb{T}^d}{\frac{\lambda_j \lambda_k\cos(x_j) \cos(x_k)}{1+\sum_{s=1}^{d} \lambda_s \sin (x_s)} f(\bm{x};\bm{\vartheta})} d \bm{x}, \\
\mathfrak{i}_{\mu_j,\lambda_k}&=& \int_{\mathbb{T}^d}{\frac{\lambda_j \cos(x_j) \sin(x_k)}{1+\sum_{s=1}^{d} \lambda_s \sin (x_s)} f(\bm{x};\bm{\vartheta})} d \bm{x}, \\
\mathfrak{i}_{\mu_j,\vartheta_k}&=& \mathfrak{i}_{\mu_j,\vartheta_k}^0 + \sum_{s=1}^{d} \lambda_s \int_{\mathbb{T}^d}  \frac{\sin (x_s)}{f(\bm{x};\bm{\vartheta})} \frac{\partial}{\partial x_j} f(\bm{x};\bm{\vartheta}) \frac{\partial}{\partial \vartheta_k} f(\bm{x};\bm{\vartheta}) d \bm{x},\\
\mathfrak{i}_{\lambda_j,\lambda_k}&=& \int_{\mathbb{T}^d}{\frac{\sin(x_j) \sin(x_k)}{1+\sum_{s=1}^{d} \lambda_s \sin (x_s)} f(\bm{x};\bm{\vartheta})} d \bm{x},\\
\mathfrak{i}_{\lambda_j,\vartheta_k}&=& 0,\\
\mathfrak{i}_{\vartheta_j,\vartheta_k}&=& \mathfrak{i}_{\vartheta_j,\vartheta_k}^0 + \sum_{s=1}^{d} \lambda_s \int_{\mathbb{T}^d}  \frac{\sin (x_s)}{f(\bm{x};\bm{\vartheta})} \frac{\partial}{\partial \vartheta_j} f(\bm{x};\bm{\vartheta}) \frac{\partial}{\partial \vartheta_k} f(\bm{x};\bm{\vartheta}) d \bm{x},\\
 \end{eqnarray*}
where the elements of $\mathfrak{I}$ with a superscript 0 denote the values of the corresponding elements of the Fisher Information matrix for the base density. 

\subsection{Testing for symmetry}\label{test_sk}

A question of natural interest in our investigation is whether the sine-skewed models indeed improve significantly on their symmetric antecedents for a given sample $\bm{\mathcal{X}}$. This boils down to testing for the nested symmetry inside a sine-skewed toroidal distribution, which can be formulated as the hypothesis testing problem
\begin{equation*}
H_0: \bm{\lambda}=\bm{0} \quad \mbox{versus} \quad H_1:\bm{\lambda}\neq \bm{0}.
\end{equation*}
A likelihood ratio test is a straightforward and  efficient procedure to tackle this problem. For each parameter $\bm{\psi}\in\{\mub,\varthetab,\lambdab\}$, we denote as before by $\hat{\bm{\psi}}$ the unconstrained maximum likelihood estimate and by $\hat{\bm{\psi}}_0$  the maximum likelihood estimate under the null hypothesis of symmetry (i.e., the MLEs from the base symmetric distribution). The likelihood ratio test then rejects the null hypothesis at asymptotic level $\alpha$ whenever the test statistic $-2(\ell(\hat{\bm{\mu}}_0,\hat{\bm{\vartheta}}_0,\bm{0})-\ell(\hat{\bm{\mu}},\hat{\bm{\vartheta}},\hat{\bm{\lambda}}))$ exceeds $\chi^2_{d;1-\alpha}$, the $\alpha$-upper quantile of a chi-square distribution with $d$ degrees of freedom.

\section{Application on protein data}\label{sec:real}

Modeling the three-dimensional structure of proteins has become a topic of special interest in the protein bioinformatics field. Proteins are constructed from a linear sequence of monomers, called \textit{amino acids}. The folded structure of a protein is determined by its amino acid sequence. The relative orientation of subsequent amino acids can be described by the pairs of dihedral angles $\phi$ and $\psi$, which correspond to the rotations of the ${\rm N}-{\rm C}_\alpha$ and ${\rm C}_\alpha-{\rm C}'$ bonds in the backbone of the protein. A more detailed biochemical background for analyzing protein structures can be found in \cite{mardia2018}. In this section, our objective is to use the proposed bivariate sine-skewed toroidal distributions to model the pairs of dihedral angles on the different types of amino acids. 

The analyzed dataset was obtained from the Top500 database, a selection of 500 files from the Protein Data Bank that are high resolution ($1.8 \mathring{A}$ or better), low homology, and high quality (see \url{http://kinemage.biochem.duke.edu/databases/top500.php} for details). In the employed data, all proteins with chain breaks and PDB files that caused errors upon parsing were removed, resulting in data on 402 proteins. Also, the amino acids with a C- or N-terminus are not employed as they do not have one of the corresponding rotation-angles. Then, for each type of amino acid, since there may be a dependency between two amino acids that are part of the same protein, just one pair of rotation-angle was selected for each protein.

\begin{table}
\centering
\scalebox{0.74}{
\begin{tabular}{|l|l|l|cccccccc|ccc|}
  \hline
  & Model & Comp & $\hat{\mu}_{1}$ & $\hat{\mu}_{2}$ & $\hat{\kappa}_{1}$ & $\hat{\kappa}_{2}$ & $\hat{r}$ & $\hat{\lambda}_1$ & $\hat{\lambda}_2$ & $\hat{p}$ & LL & AIC & BIC \\ 
  \hline
\hline
\parbox[t]{4mm}{\multirow{12}{*}{\rotatebox[origin=c]{90}{S ($n=396$)}}} & \multirow{2}{*}{S} & 1 & -1.8 & 2.493 & 2.215 & 2.116 & -0.094 & -- & -- & 0.607 & \multirow{2}{*}{-780.5} & \multirow{2}{*}{1583.1} & \multirow{2}{*}{1626.9} \\ 
   &  & 2 & -1.225 & -0.486 & 38.466 & 29.498 & -23.664 & -- & -- & 0.393 &  &  &  \\ 
   \cline{2-14} & \multirow{2}{*}{SS} & 1 & -1.338 & -0.389 & 17.805 & 19.977 & -13.535 & 0.168 & -0.582 & 0.442 & \multirow{2}{*}{-725.1*} & \multirow{2}{*}{1480.1} & \multirow{2}{*}{1539.8} \\ 
      &  & 2 & 2.903 & 1.222 & 0.016 & 0.986 & 4.446 & 0.014 & 0.985 & 0.558 &  &  &  \\ 
	  \cline{2-14} & \multirow{2}{*}{C} & 1 & -1.879 & 2.529 & 1.578 & 1.502 & 0.843 & -- & -- & 0.589 & \multirow{2}{*}{-814.1} & \multirow{2}{*}{1650.1} & \multirow{2}{*}{1693.9} \\ 
   &  & 2 & -1.224 & -0.486 & 23.795 & 14.28 & 0 & -- & -- & 0.411 &  &  &  \\ 
   \cline{2-14} & \multirow{2}{*}{SC} & 1 & -1.393 & -0.474 & 8.333 & 8.555 & 0 & 0.776 & 0.182 & 0.45 & \multirow{2}{*}{-811.6} & \multirow{2}{*}{1653.1} & \multirow{2}{*}{1712.8} \\ 
   &  & 2 & -2.384 & 2.335 & 0.268 & 3.393 & 1.801 & 0.945 & -0.018 & 0.55 &  &  &  \\ 
   \cline{2-14} & \multirow{2}{*}{WC} & 1 & -1.787 & 2.637 & 0.604 & 0.788 & -0.036 & -- & -- & 0.539 & \multirow{2}{*}{-764} & \multirow{2}{*}{1550} & \multirow{2}{*}{1593.8} \\ 
   &  & 2 & -1.17 & -0.567 & 0.865 & 0.789 & -0.44 & -- & -- & 0.461 &  &  &  \\ 
   \cline{2-14} & \multirow{2}{*}{SWC} & 1 & -1.311 & 2.631 & 0.639 & 0.801 & -0.129 & -1 & 0 & 0.524 & \multirow{2}{*}{-717*} & \multirow{2}{*}{1464} & \multirow{2}{*}{1523.7} \\ 
   &  & 2 & -1.144 & -0.638 & 0.844 & 0.78 & -0.51 & 0.039 & 0.961 & 0.476 &  &  &  \\ 
   \hline
\hline
\parbox[t]{4mm}{\multirow{12}{*}{\rotatebox[origin=c]{90}{G ($n=399$)}}} & \multirow{2}{*}{S} & 1 & 2.875 & 1.546 & 1.091 & 0.66 & 5.899 & -- & -- & 0.6 & \multirow{2}{*}{-1047.1} & \multirow{2}{*}{2116.3} & \multirow{2}{*}{2160.2} \\ 
   &  & 2 & 0.269 & -1.795 & 0.462 & 1.959 & -6.248 & -- & -- & 0.4 &  &  &  \\ 
   \cline{2-14} & \multirow{2}{*}{SS} & 1 & 0.277 & -1.784 & 0.401 & 1.936 & -6.071 & -0.386 & -0.234 & 0.403 & \multirow{2}{*}{-1041.1*} & \multirow{2}{*}{2112.2} & \multirow{2}{*}{2172.1} \\ 
   &  & 2 & 2.895 & 1.53 & 1.086 & 0.702 & 5.988 & -0.548 & 0.452 & 0.597 &  &  &  \\ 
	 \cline{2-14} & \multirow{2}{*}{C} & 1 & 0.52 & -0.155 & 0.001 & 4.937 & 1.024 & -- & -- & 0.522 & \multirow{2}{*}{-1233} & \multirow{2}{*}{2488} & \multirow{2}{*}{2531.8} \\ 
   &  & 2 & -2.733 & 3.111 & 0.631 & 4.086 & 0 & -- & -- & 0.478 &  &  &  \\ 
   \cline{2-14} & \multirow{2}{*}{SC} & 1 & 2.598 & 2.748 & 0.001 & 0.631 & 0.568 & 0.493 & 0.506 & 0.712 & \multirow{2}{*}{-1220.5*} & \multirow{2}{*}{2471} & \multirow{2}{*}{2530.8} \\ 
   &  & 2 & 1.495 & -0.004 & 10.463 & 6.567 & 0 & 0.005 & 0.993 & 0.288 &  &  &  \\ 
	   \cline{2-14} & \multirow{2}{*}{WC} & 1 & -1.42 & 0.504 & 0.603 & 0 & -0.676 & -- & -- & 0.636 & \multirow{2}{*}{-1110.7} & \multirow{2}{*}{2243.5} & \multirow{2}{*}{2287.3} \\ 
   &  & 2 & 1.443 & 0.039 & 0.819 & 0.741 & -0.542 & -- & -- & 0.364 &  &  &  \\ 
   \cline{2-14} & \multirow{2}{*}{SWC} & 1 & 1.385 & 0.121 & 0.816 & 0.723 & -0.536 & 1 & 0 & 0.383 & \multirow{2}{*}{-1079*} & \multirow{2}{*}{2188.1} & \multirow{2}{*}{2247.9} \\ 
   &  & 2 & -1.328 & 0.371 & 0.647 & 0 & -0.665 & -0.786 & -0.214 & 0.617 &  &  &  \\ 
   \hline
\hline
\parbox[t]{4mm}{\multirow{12}{*}{\rotatebox[origin=c]{90}{P ($n=390$)}}} &  \multirow{2}{*}{S} & 1 & -1.169 & 2.557 & 32.752 & 9.399 & -4.165 & -- & -- & 0.603 & \multirow{2}{*}{-173.4} & \multirow{2}{*}{368.9} & \multirow{2}{*}{412.5} \\ 
   &  & 2 & -1.133 & -0.415 & 73.17 & 46.121 & -51.988 & -- & -- & 0.397 &  &  &  \\ 
   \cline{2-14} & \multirow{2}{*}{SS} & 1 & -1.686 & -2.256 & 33.22 & 0.79 & -20.318 & 0 & -1 & 0.607 & \multirow{2}{*}{-150.1*} & \multirow{2}{*}{330.1} & \multirow{2}{*}{389.6} \\ 
   &  & 2 & -1.165 & -0.364 & 81.556 & 49.781 & -57.897 & 0 & -1 & 0.393 &  &  &  \\ 
	  \cline{2-14} & \multirow{2}{*}{C} & 1 & -1.111 & -0.44 & 15.2 & 13.869 & 0.005 & -- & -- & 0.421 & \multirow{2}{*}{-285.5} & \multirow{2}{*}{593.1} & \multirow{2}{*}{636.7} \\ 
   &  & 2 & -1.17 & 2.612 & 29.292 & 7.04 & 0 & -- & -- & 0.579 &  &  &  \\ 
   \cline{2-14} & \multirow{2}{*}{SC} & 1 & -1.112 & -0.441 & 15.201 & 13.869 & 0 & 0.689 & 0.136 & 0.421 & \multirow{2}{*}{-282.2} & \multirow{2}{*}{594.5} & \multirow{2}{*}{654} \\ 
   &  & 2 & -1.17 & 2.613 & 29.292 & 7.04 & 0.001 & 0.147 & -0.561 & 0.579 &  &  &  \\ 
	   \cline{2-14} & \multirow{2}{*}{WC} & 1 & -1.143 & 2.589 & 0.884 & 0.857 & -0.416 & -- & -- & 0.595 & \multirow{2}{*}{-191.1} & \multirow{2}{*}{404.2} & \multirow{2}{*}{447.8} \\ 
   &  & 2 & -1.067 & -0.494 & 0.881 & 0.822 & -0.583 & -- & -- & 0.405 &  &  &  \\ 
   \cline{2-14} & \multirow{2}{*}{SWC} & 1 & -1.115 & 2.568 & 0.883 & 0.857 & -0.425 & -1 & 0 & 0.594 & \multirow{2}{*}{-178.8*} & \multirow{2}{*}{387.5} & \multirow{2}{*}{447} \\ 
   &  & 2 & -1.039 & -0.548 & 0.882 & 0.819 & -0.586 & 0 & 1 & 0.406 &  &  &  \\ 
   \hline
\end{tabular}
}
\caption{Parameter estimation for each component, log-likelihood (LL), AIC and BIC of the mixture of Sine (S), Cosine (C) and wrapped Cauchy (WC) distributions and their sine-skewed versions (respectively, SS, SC, SWC). In the LL value of the sine-skewed models, an asterisk (*) indicates when the null hypothesis of symmetry is rejected against the sine-skewed version  at the level $\alpha=0.01$. Each block corresponds to one amino acid type, from top to bottom: serine (S), glycine (G) and proline (P).} 
\label{tab_prot}
\end{table}

\begin{table}
\centering
\scalebox{0.98}{
\begin{tabular}{|l|c|c|c|c|c|c|c|c|c|c|}
  \hline
Amino acid & A & C & D & E & F & G & H & I & K & L \\
Model & SWC & SWC & SS & SWC & SS & SS/S & SS & SWC & SWC & SWC \\
 \hline	
Amino acid & M & N & P & Q & R & S & T & V & W & Y \\
Model & SWC & SS & SS & SWC & SWC & SWC & SS/S & SWC & SWC & SS \\
 \hline
\end{tabular}
}
\caption{Model providing the best fit on the basis of the  AIC/BIC for each amino acid type. For G and T, a distinct model is obtained according to AIC and BIC.} 
\label{tab_prot2}
\end{table}

From the total of 20 amino acid types, in this section, we focus on three of them as all of them present a similar pattern except for the glycine (G symbol) and proline (P symbol). The amino acid type chosen to represent the rest of them is the serine (S symbol). The full analysis of the 20 different amino acid types is provided in Section~\ref{extra_protein} of the Supplementary Material. The scatter (planar) plot of the toroidal protein data, also known as Ramachandran plot \citep{ramachandran1963} in the protein bioinformatics field, is represented in Figure~\ref{fig_prot} for each amino acid type (see also the figures from Section~\ref{extra_protein} of the Supplementary Material). The Ramachandran plot suggests that at least a bimodal distribution is needed. For that reason, the following mixture of two toroidal densities was employed:
\begin{equation}\label{mixture}
g_M(\bm{x};\mub_1,\mub_2,\bm{\vartheta}_1,\bm{\vartheta}_2,\bm{\lambda}_1,\bm{\lambda}_2,p)= p g_1(\bm{x}-\mub_1;\bm{\vartheta}_1,\bm{\lambda}_1) + (1-p) g_2(\bm{x}-\mub_2;\bm{\vartheta}_2,\bm{\lambda}_2),
\end{equation} 
where $p\in[0,1]$ and $g_j$, with $j\in\{1,2\}$, are two sine-skewed densities of the form~\eqref{sine_skewed}. {The parameter estimates are obtained via the optimization approach described in Subsection~\ref{mle_section}, after  having adapted the optimization algorithm such that it takes into account the constraints of both mixture  components and $0\leq p \leq 1$.} As base symmetric densities, we considered the three non-uniform distributions described in Section~\ref{sec:cases}: the bivariate Sine, Cosine and wrapped Cauchy distributions.  For comparative purposes, we also computed the MLEs of the mixtures of these symmetric distributions, corresponding to $\lambdab_1=\lambdab_2={\pmb 0}$ in~\eqref{mixture}. In Table~\ref{tab_prot} we provide, for every amino acid type, the sample size, the MLEs, the maximum value of the log-likelihood function (LL), the Akaike Information Criterion (AIC) and the Bayesian Information Criterion (BIC) for each model. The test of symmetry described in Subsection~\ref{test_sk} is  applied at  significance level $\alpha=0.01$ and the rejection of the null hypothesis is indicated in Table~\ref{tab_prot} with an asterisk in the LL value. As we can see, in most cases the sine-skewed version improves on the original symmetric density, and for each amino acid type a sine-skewed toroidal distribution provides the best fit in terms of  AIC: for S it is the sine-skewed wrapped Cauchy model, for G and P the sine-skewed Sine model. Regarding the BIC, for S and P, it is the same models, while for G the base Sine model has the lowest value. Table~\ref{tab_prot2} includes a summary indicating which is the best model in terms of AIC and BIC for all amino acid types, hence also those not treated here but in the Supplementary Material. We  also see that in most cases the null hypothesis of symmetry is rejected at a significance level $\alpha=0.01$. Thus, it is clear that the proposed sine-skewing transformation of this paper strongly improves the goodness-of-fit  without overfitting the data.

The graphical representation of the fitted models is provided in Figure~\ref{fig_prot}, and the Supplementary Material contains the figures corresponding to the remaining amino acid types.  A common trait of all figures is that we can see that the best-fitting mixture of sine-skewed toroidal distributions corresponds to a very good fit of the data points in the Ramachandran plot.

\begin{figure}
 \centering
\begin{tabular}{cccc}
\subfloat{
    \includegraphics[height=0.15\textheight]{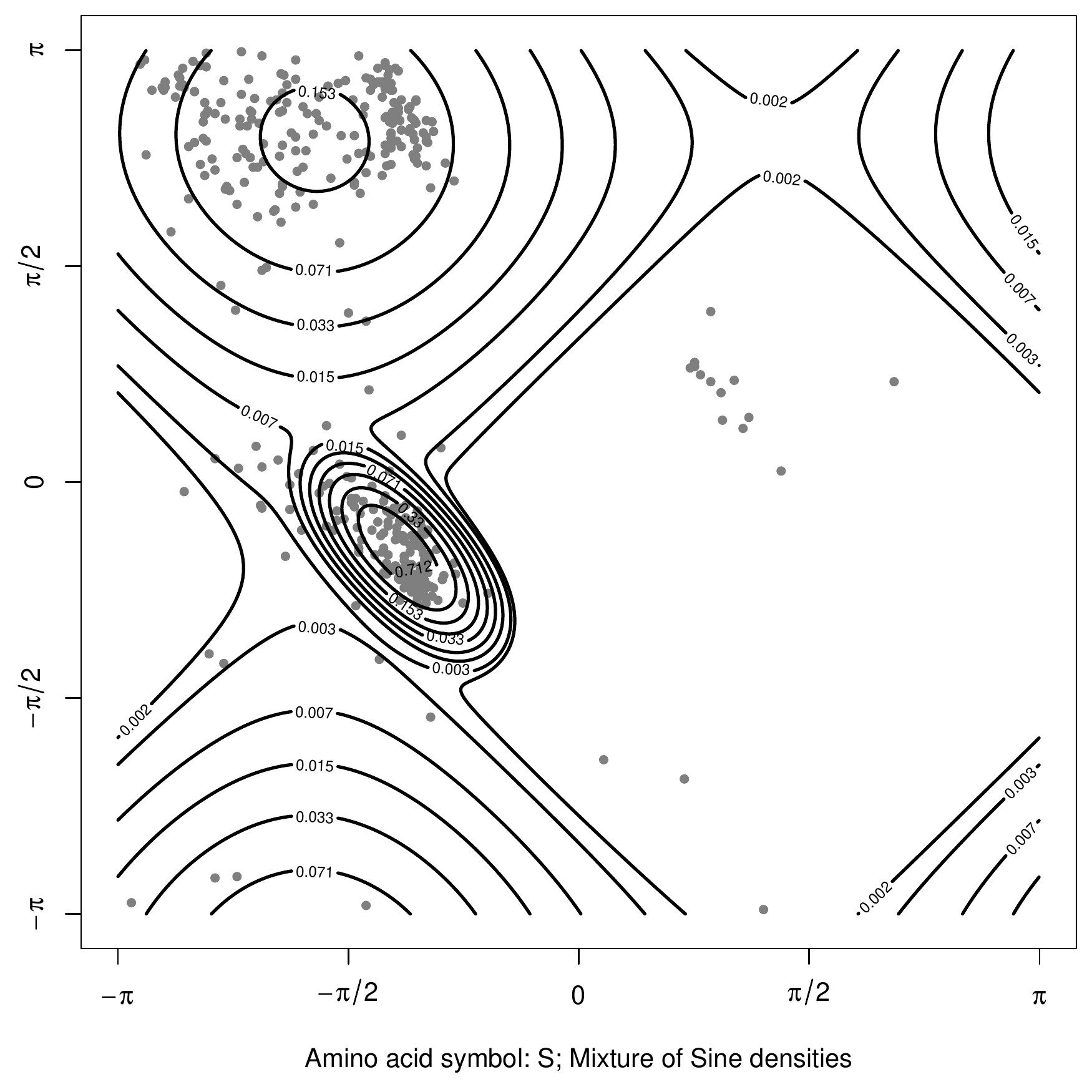}}
\subfloat{
    \includegraphics[height=0.15\textheight]{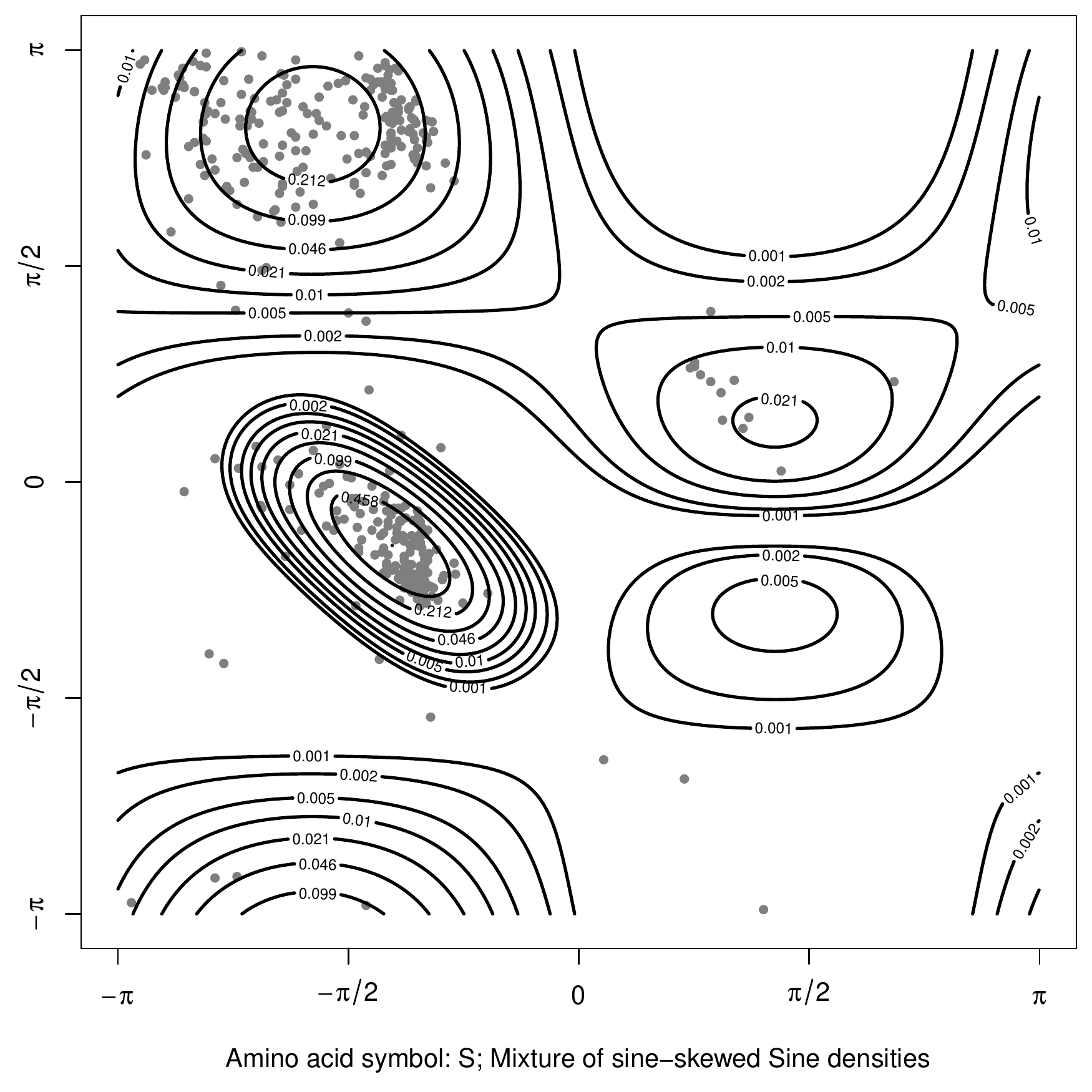}}
		\subfloat{
    \includegraphics[height=0.15\textheight]{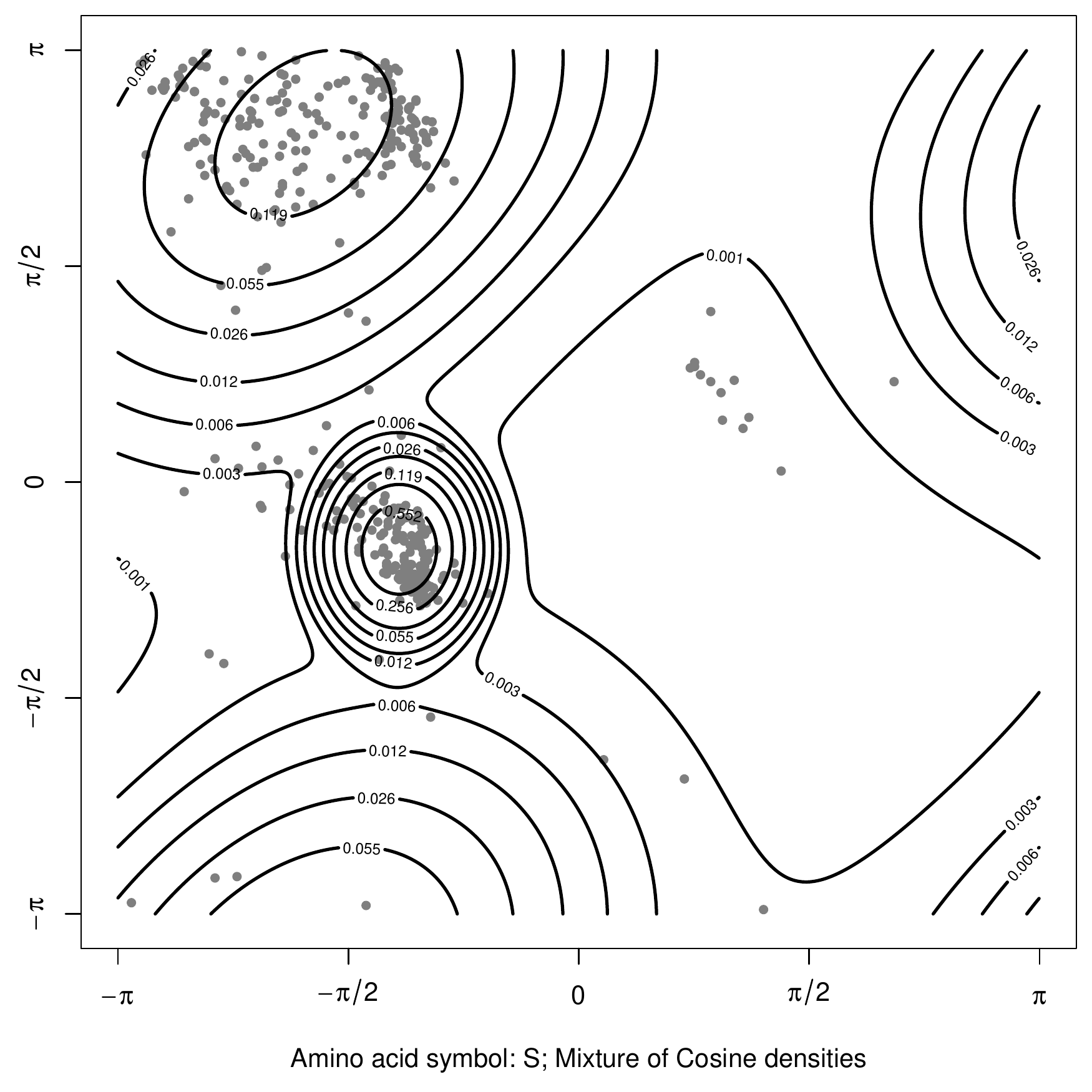}}
\subfloat{
    \includegraphics[height=0.15\textheight]{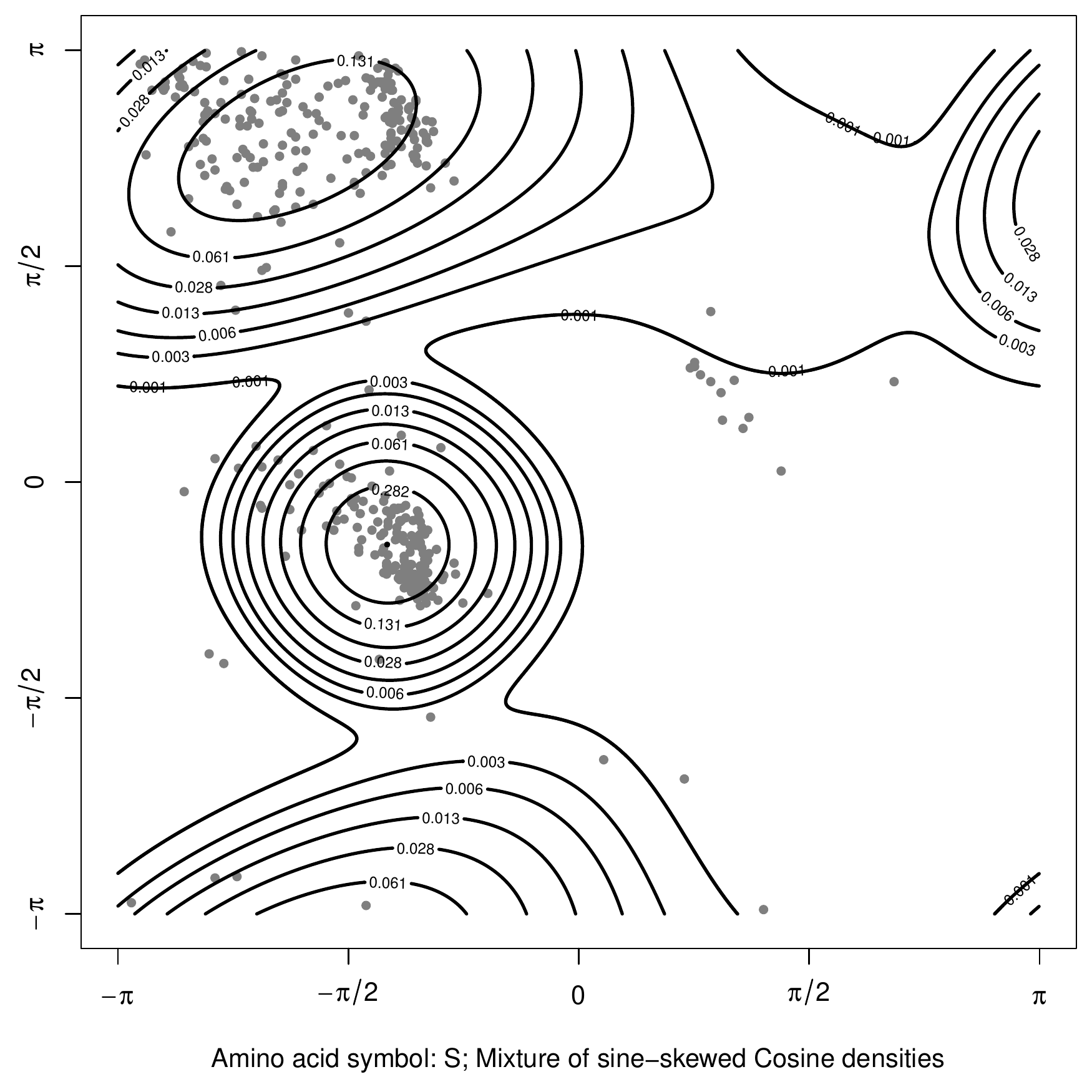}}\\
		\subfloat{
    \includegraphics[height=0.15\textheight]{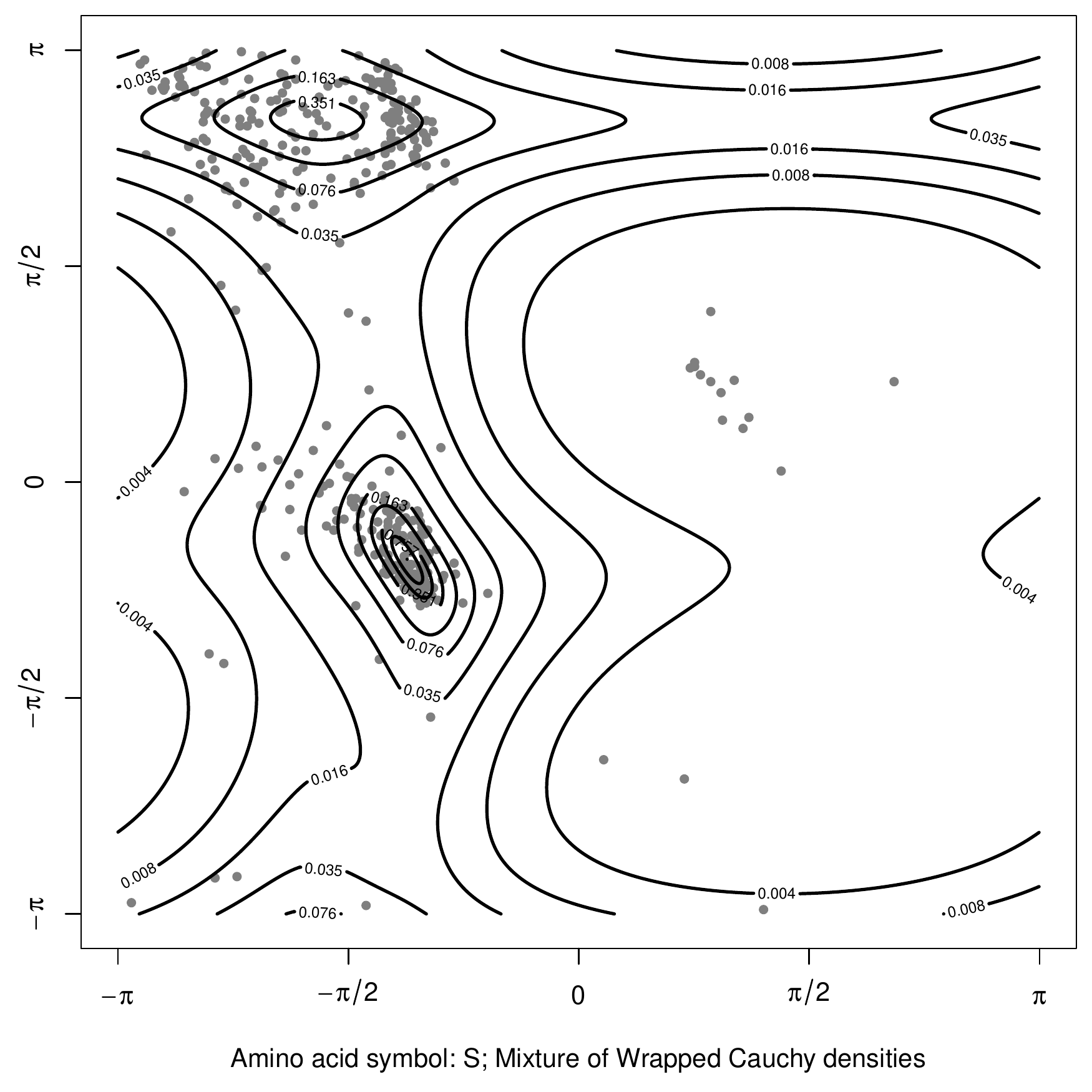}}
		\subfloat{
    \includegraphics[height=0.15\textheight]{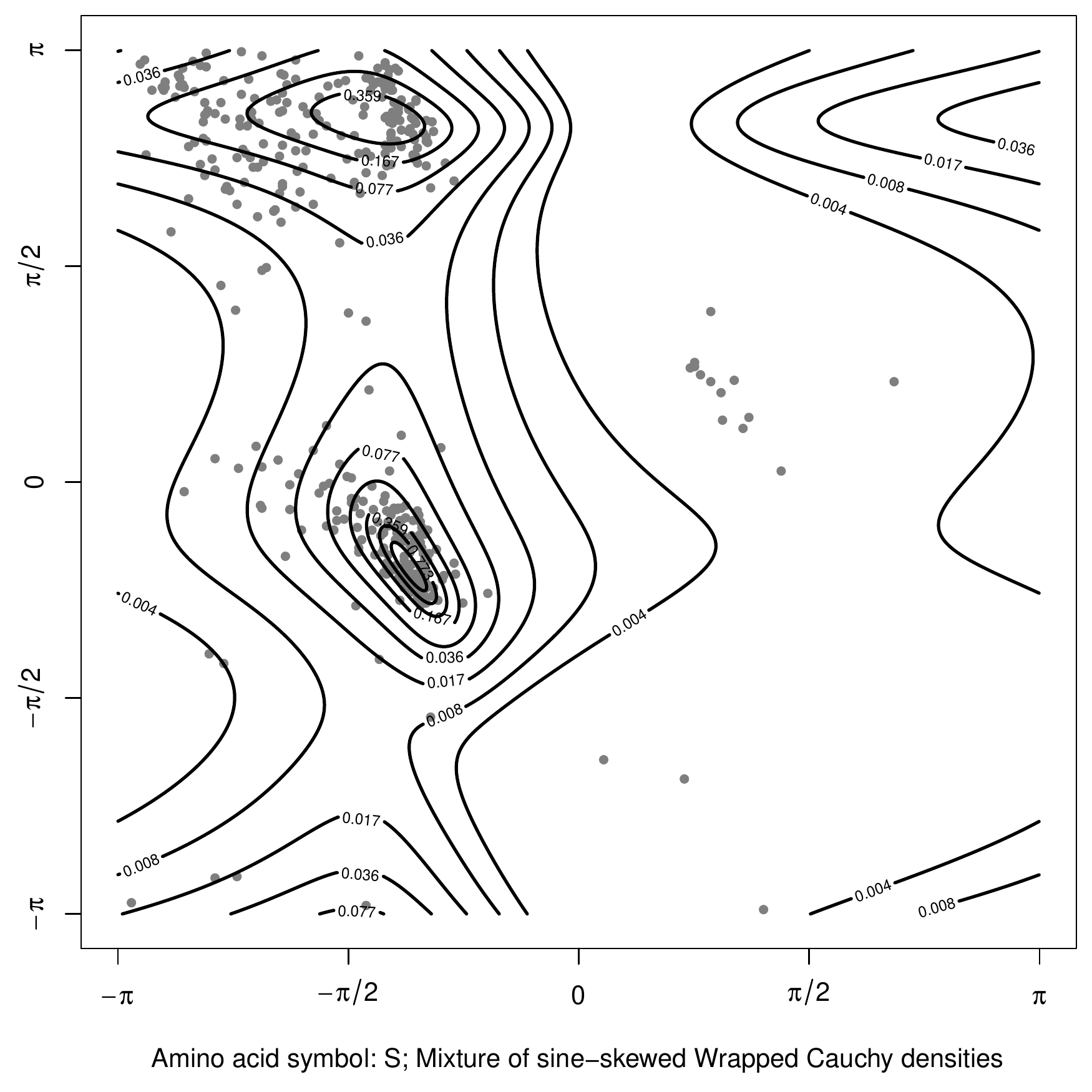}}
\subfloat{
    \includegraphics[height=0.15\textheight]{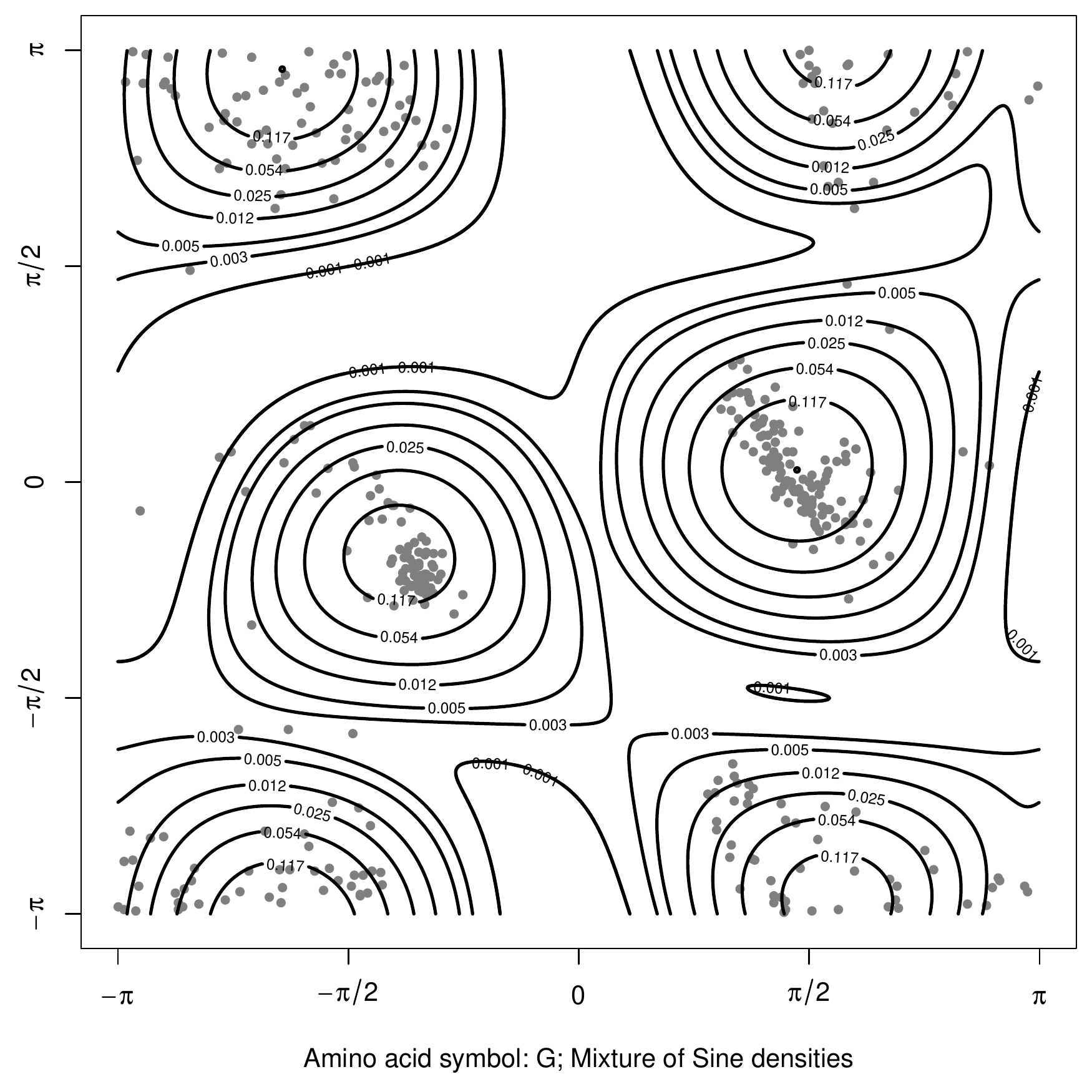}}
\subfloat{
    \includegraphics[height=0.15\textheight]{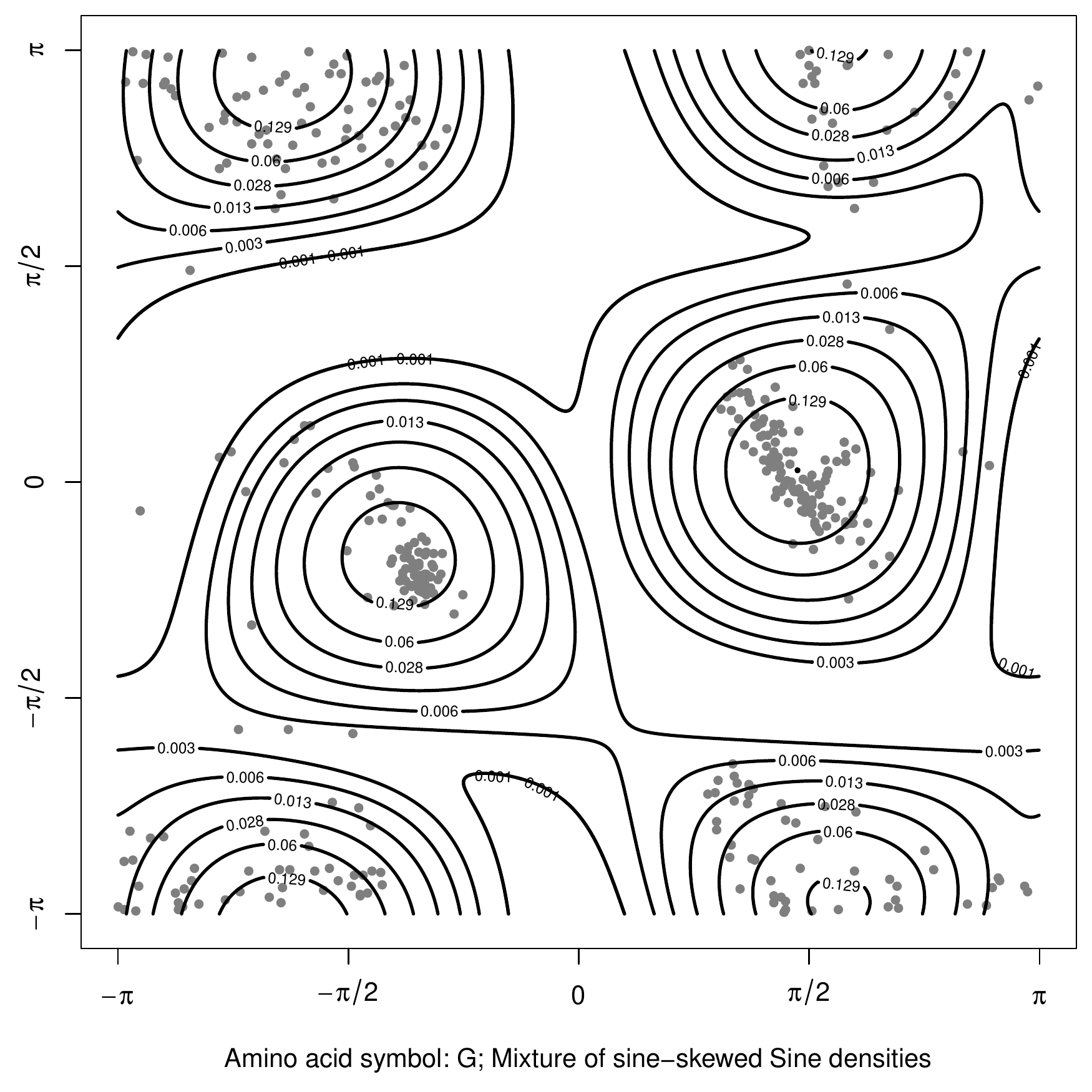}}\\
		\subfloat{
    \includegraphics[height=0.15\textheight]{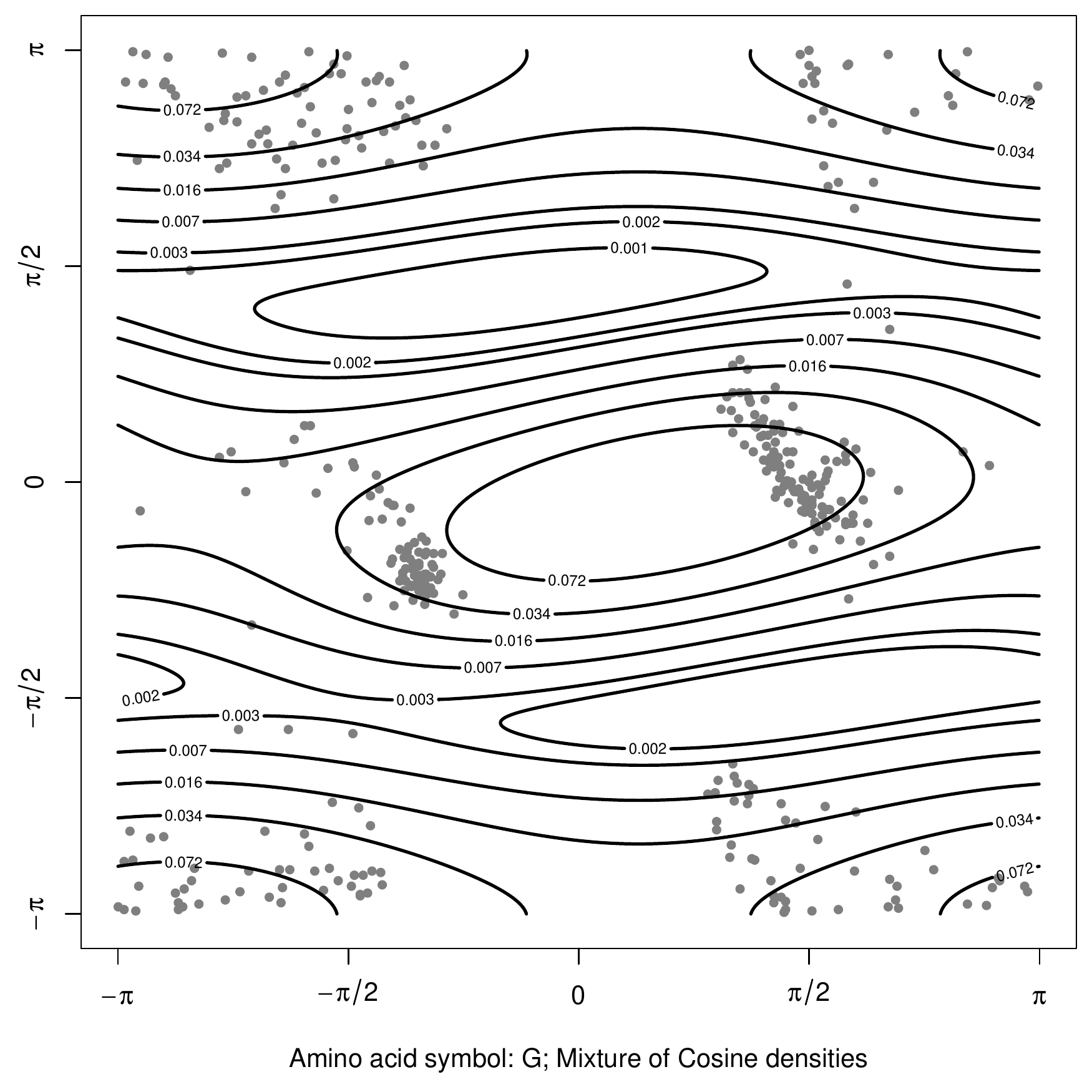}}
\subfloat{
    \includegraphics[height=0.15\textheight]{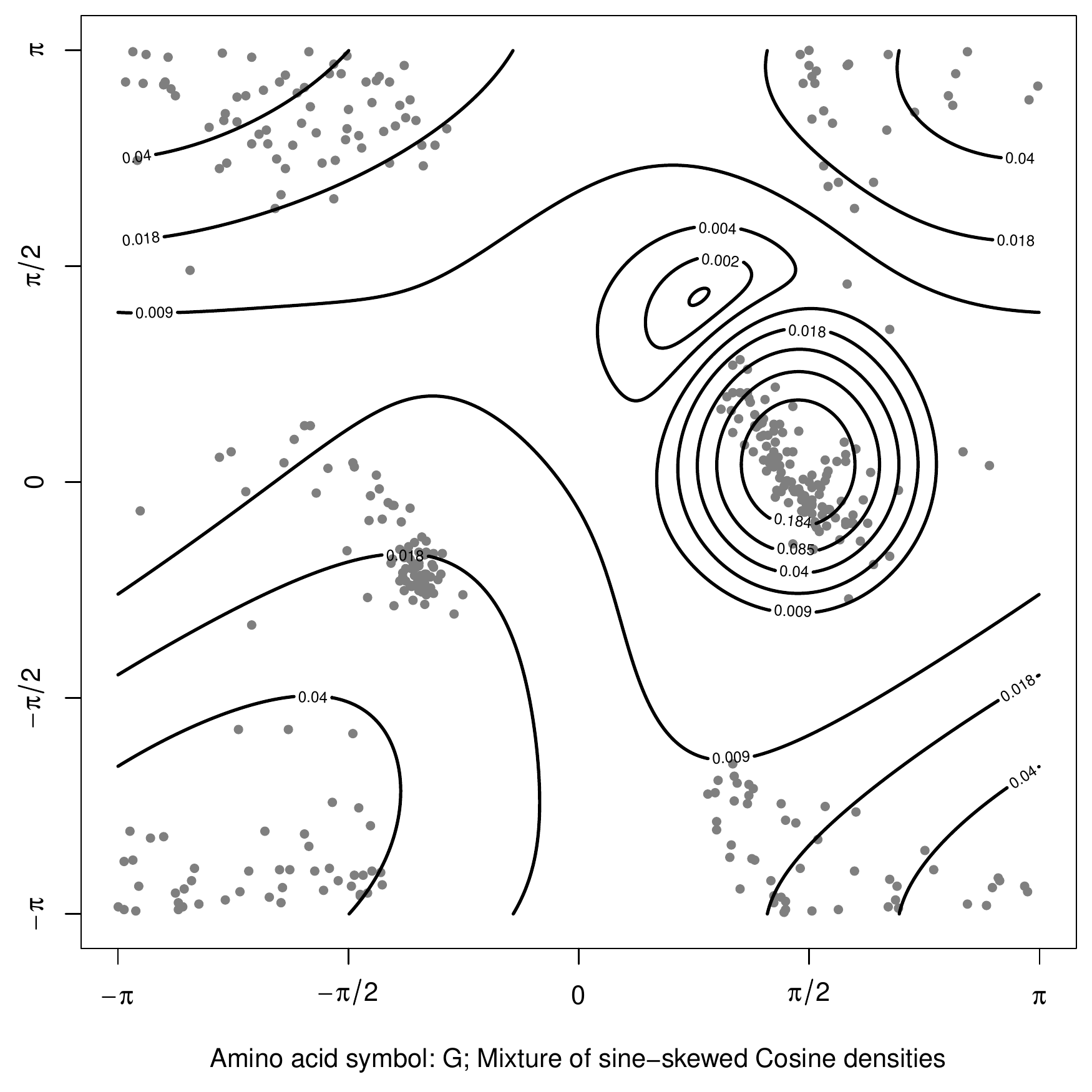}}
		\subfloat{
    \includegraphics[height=0.15\textheight]{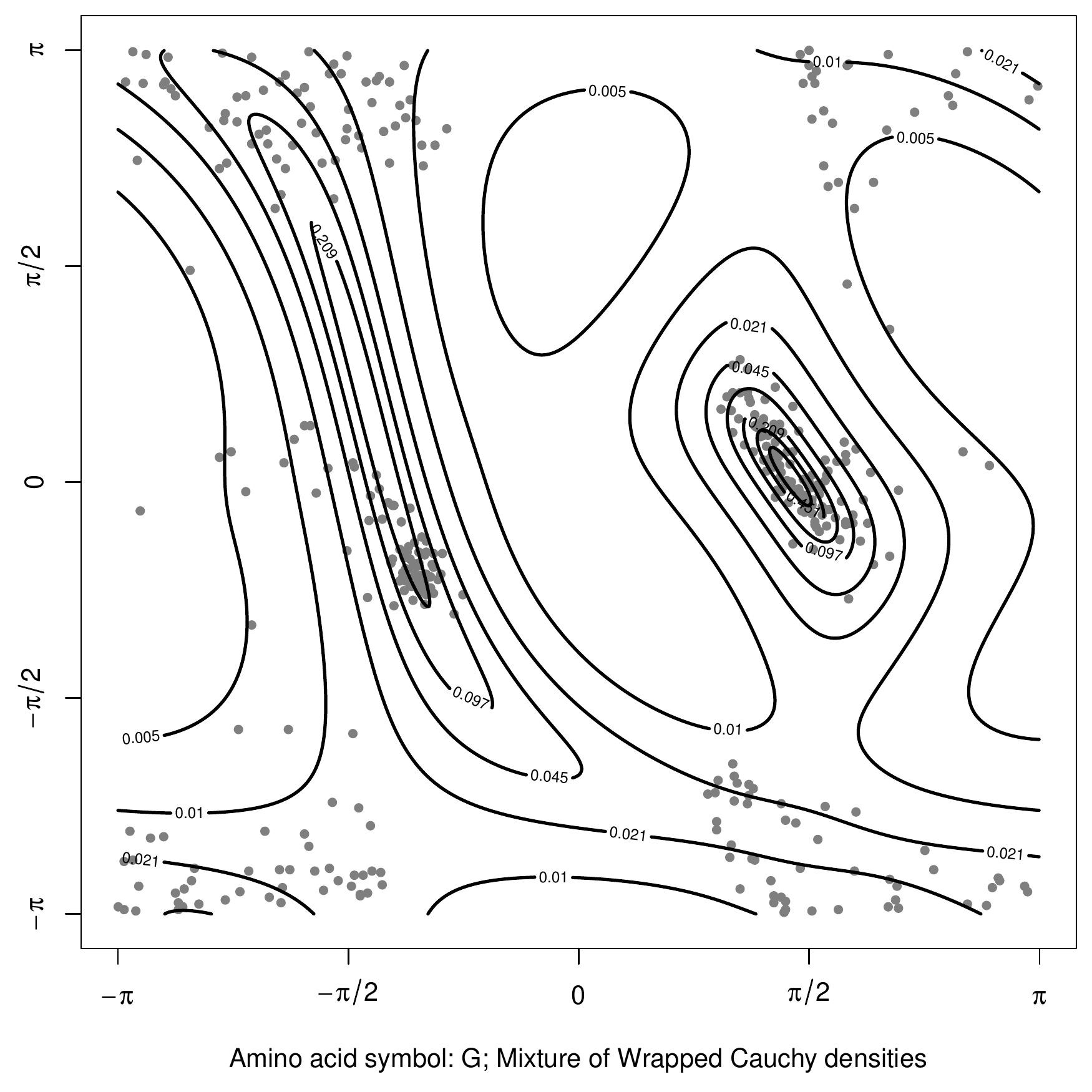}}
		\subfloat{
    \includegraphics[height=0.15\textheight]{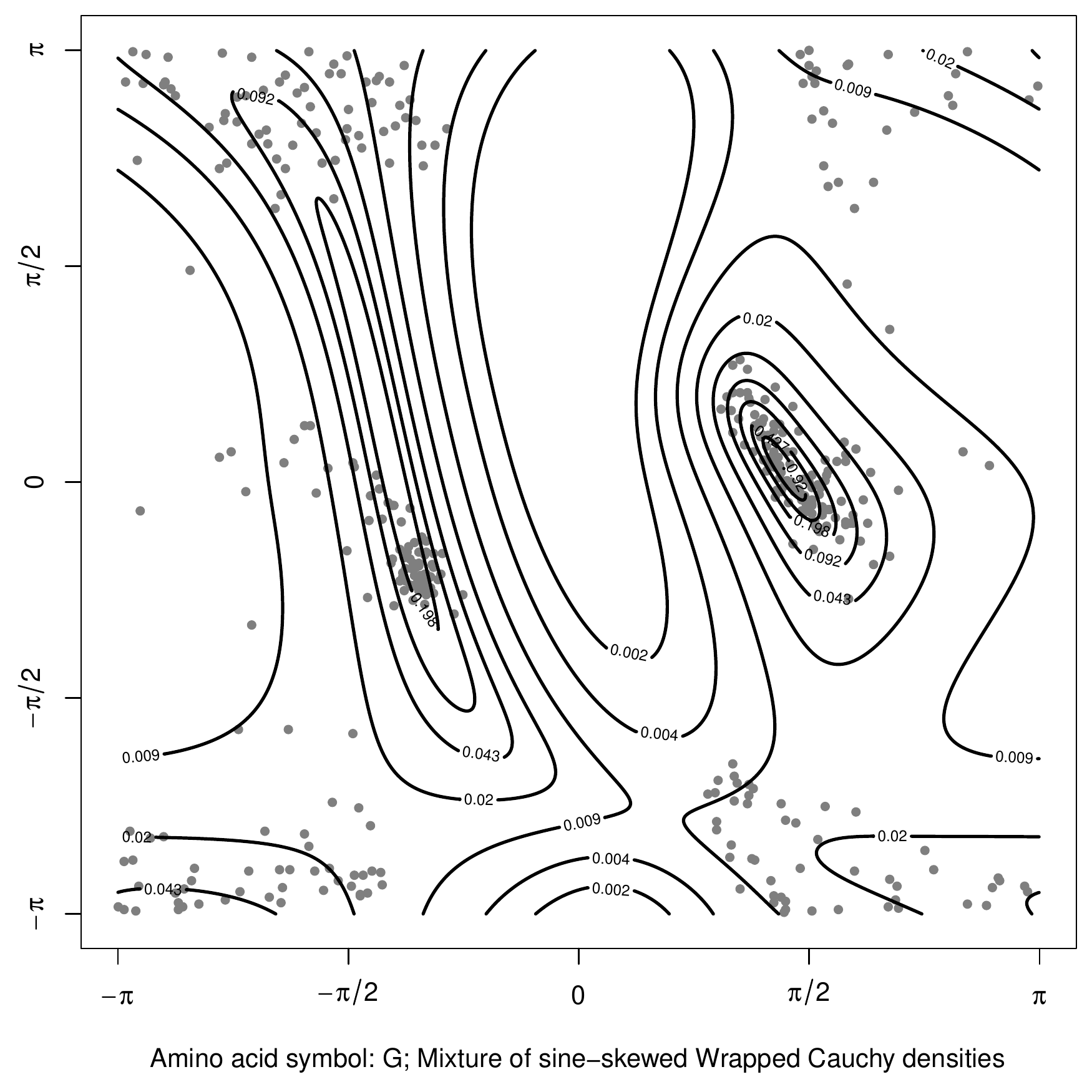}}\\
\subfloat{
    \includegraphics[height=0.15\textheight]{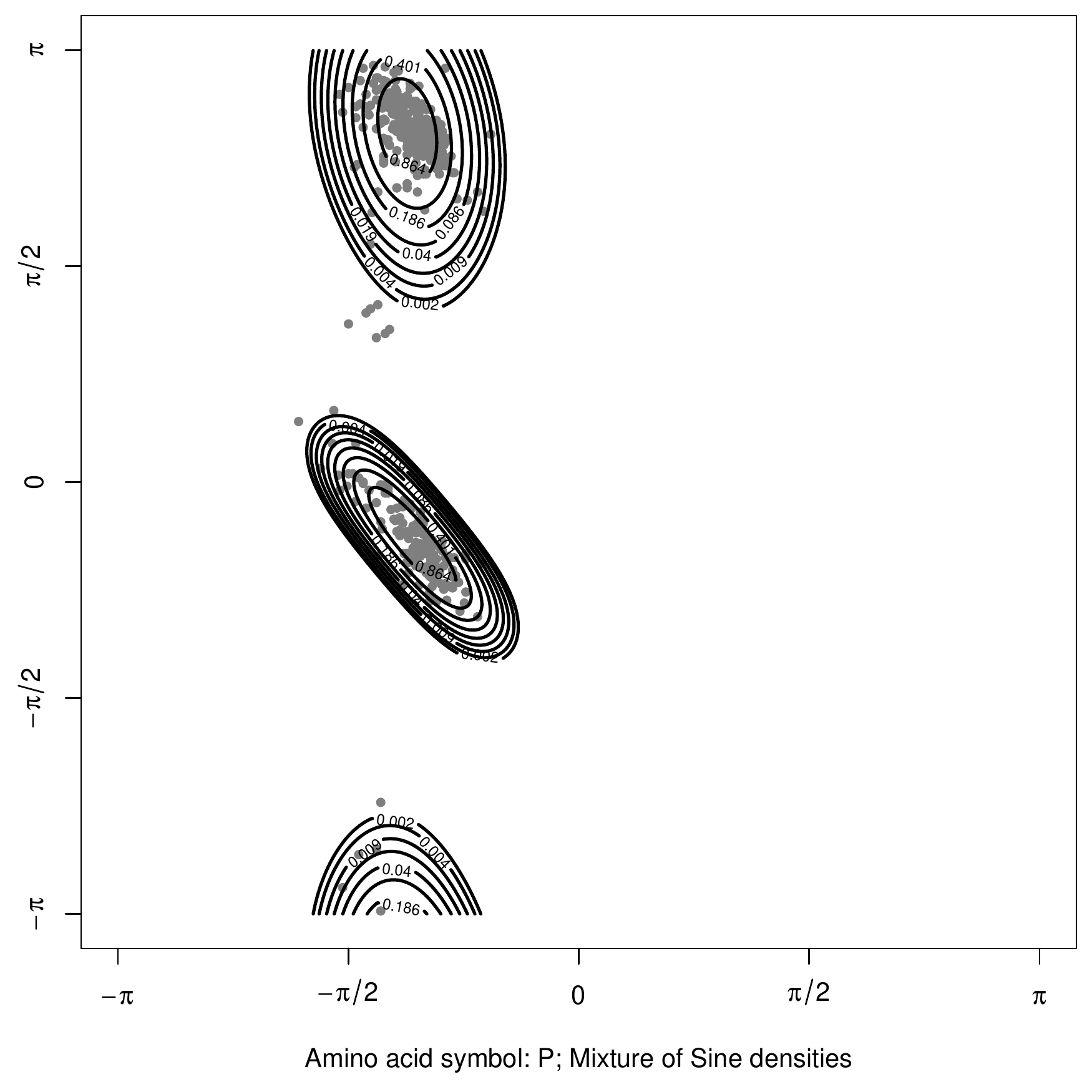}}
\subfloat{
    \includegraphics[height=0.15\textheight]{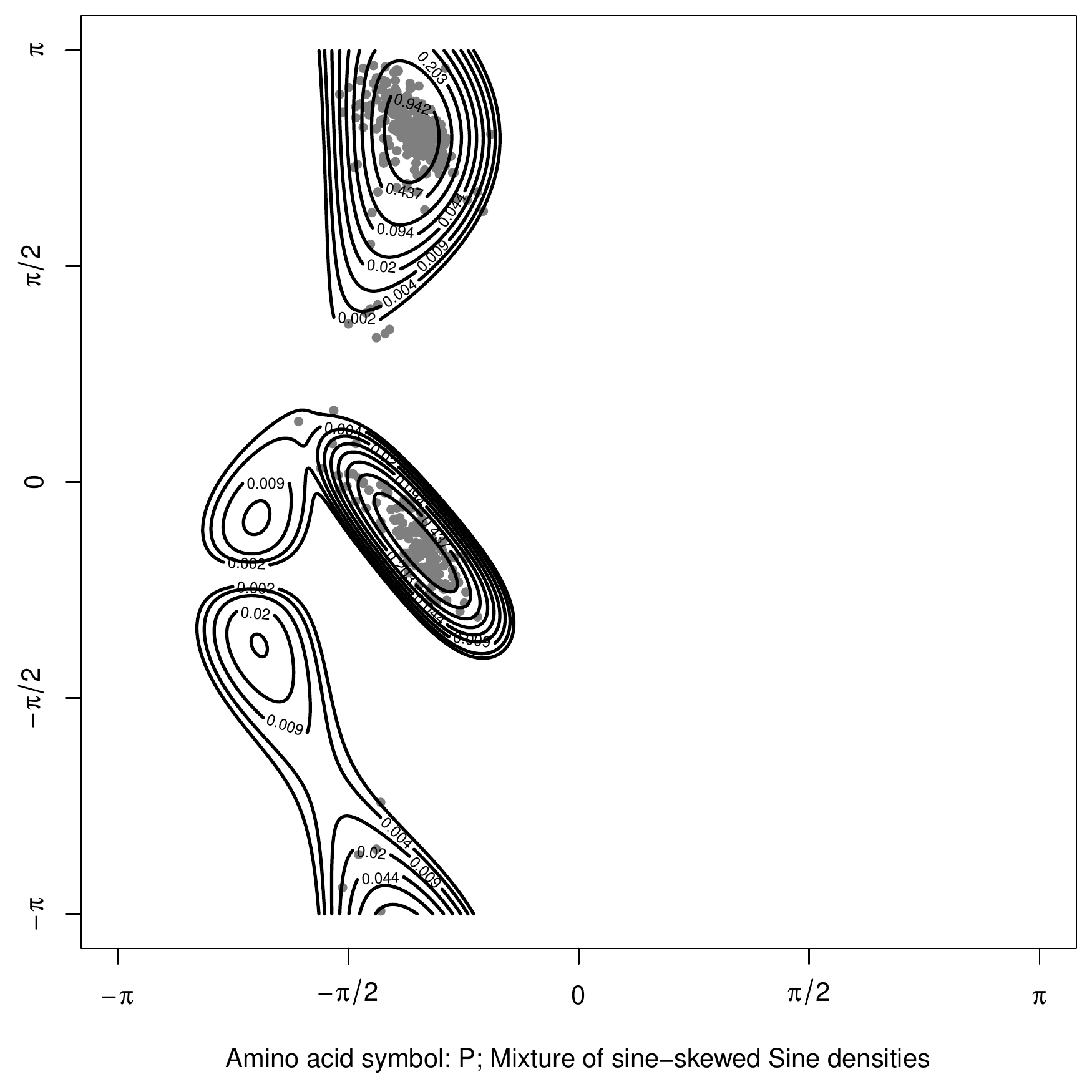}}
		\subfloat{
    \includegraphics[height=0.15\textheight]{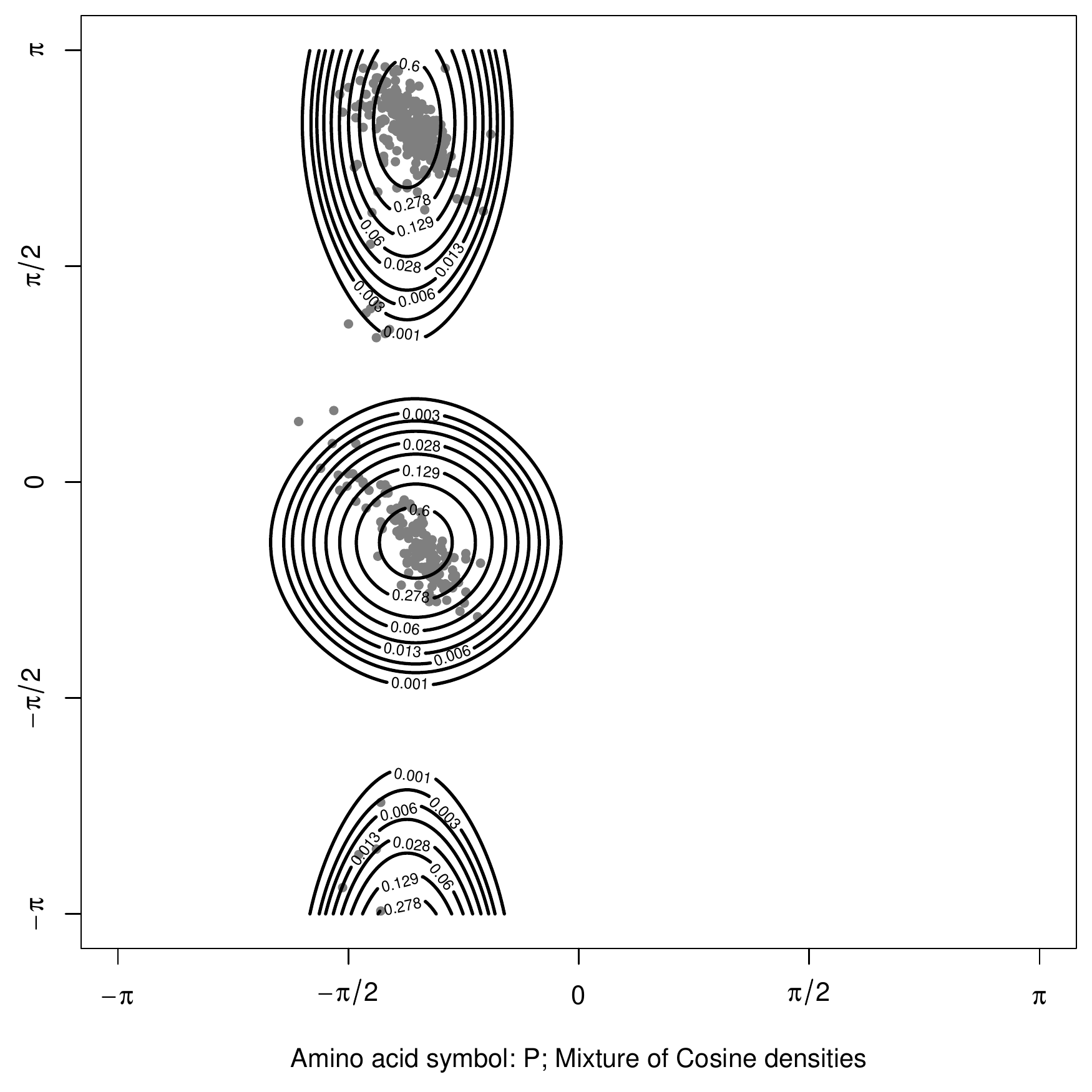}}
\subfloat{
    \includegraphics[height=0.15\textheight]{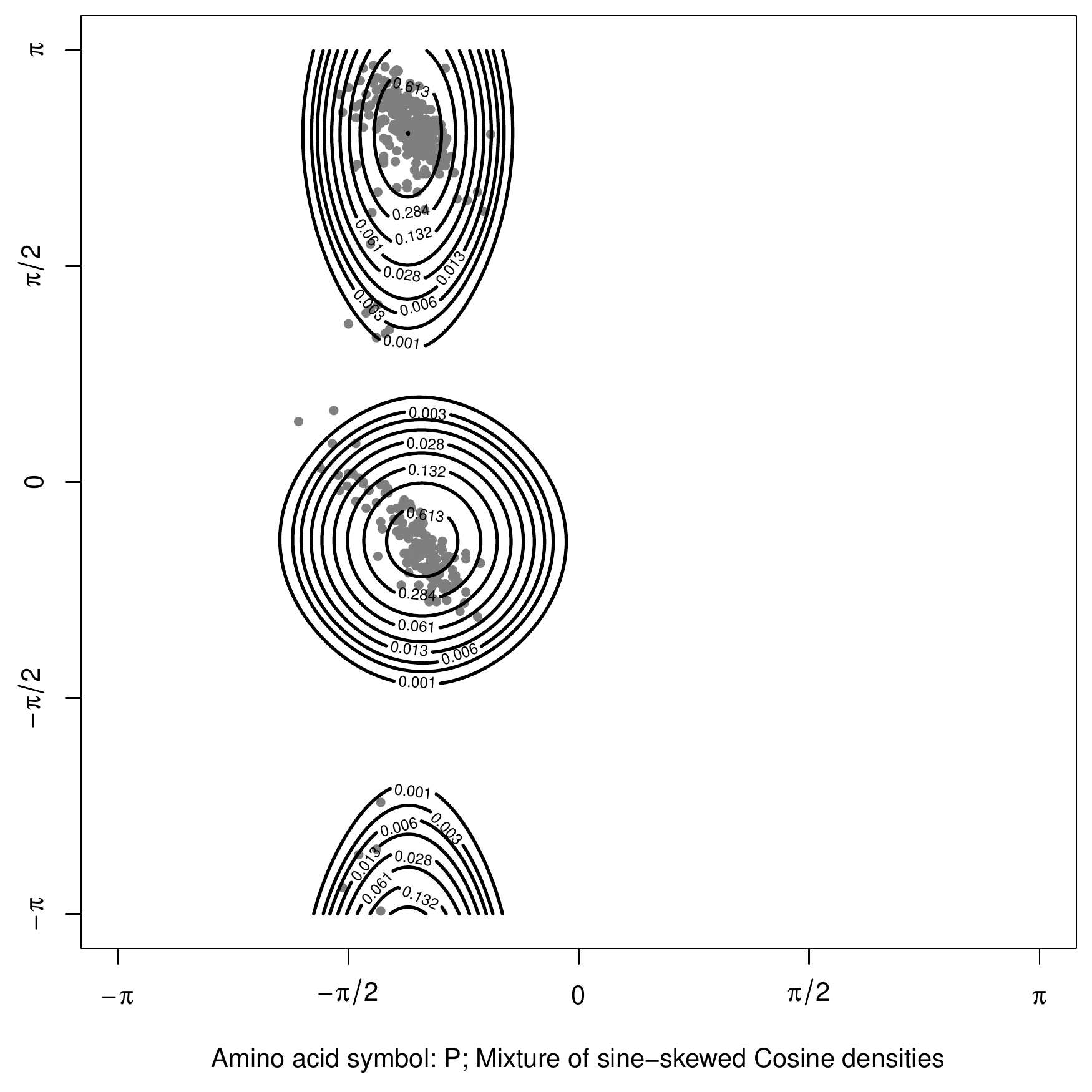}}\\
		\subfloat{
    \includegraphics[height=0.15\textheight]{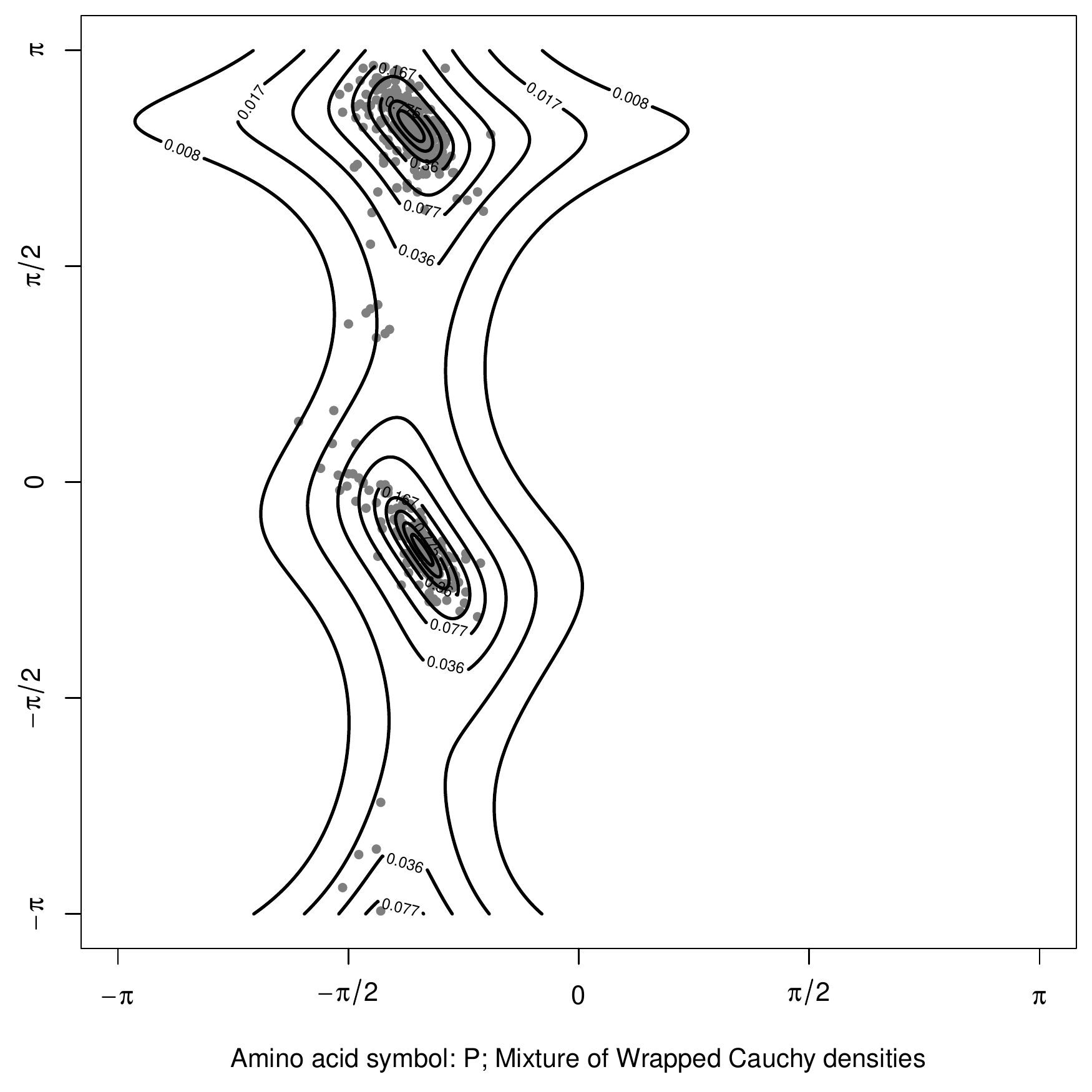}}
		\subfloat{
    \includegraphics[height=0.15\textheight]{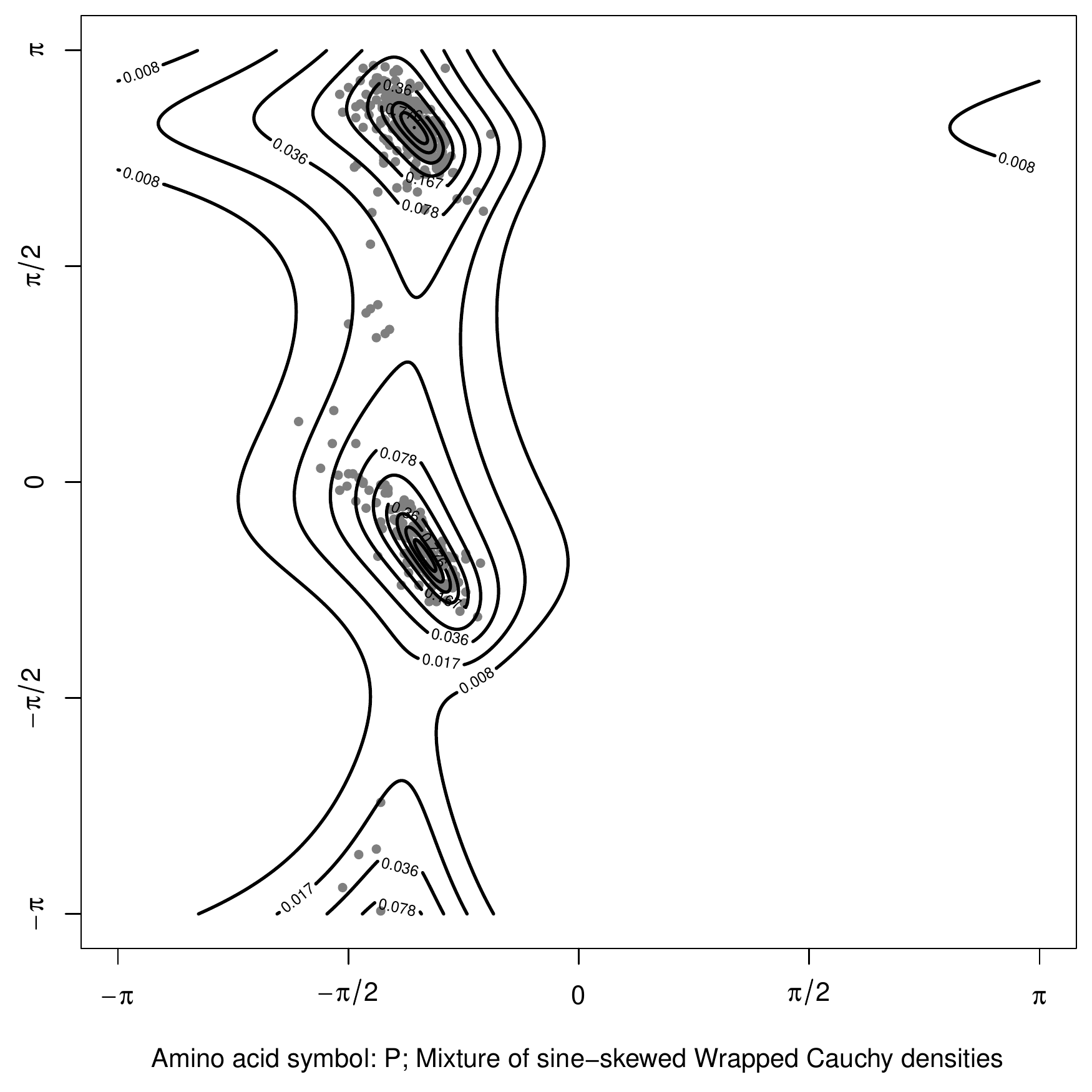}}
		\end{tabular}
		\vspace{0.7cm}
 \caption{Ramachandran plot (gray points) and maximum likelihood fits (black contour) of mixtures of the base models and of mixtures of their sine-skewed versions for the protein data. The studied amino acid types are: serine (S), glycine (G) and proline (P).}
 \label{fig_prot}
\end{figure}

\section{Conclusion}\label{sec:conclu}

In this paper, we have introduced the sine-skewed toroidal distributions, which are obtained by applying our sine-skewing transformation to any existing symmetric distribution on the torus. It is a very general way of allowing distributions to take into account potential skewness occurring in real data sets, and its simple implementation allows future researchers to immediately also build sine-skewed versions of their new distributions. We have thus filled a long-standing gap in the literature that had been identified by various authors such as   \cite{Mardia13}. Our proposal enjoys several nice features: the construction is simple, especially thanks to the absence of need to calculate a normalizing constant, the random number generation is very easy, trigonometric moments are readily obtained from the initial symmetric toroidal distribution, parameter estimation via maximum likelihood works well, and highly versatile shapes can be attained by sine-skewed toroidal distributions. A potential drawback, which is however shared by most distributions on the torus, is that unimodality cannot be always guaranteed.

We have seen in Section~\ref{sec:real} the improved fit to dihedral angle data provided by sine-skewed toroidal distributions in comparison to the popular existing symmetric distributions. As mentioned in the Introduction, toroidal distributions are playing an increasingly important role in protein bioinformatics, in particular in the protein structure prediction problem. The sine-skewed transformation  promises a crucial step forward in this active research area, and it is currently being implemented in the software $\mathtt{R}$ and Pyro to allow a user-friendly usage.

\section*{Acknowledgements}

We thank Prof. Thomas Hamelryck from the University of Copenhagen for the useful discussions and for providing the protein data set that is used in this paper.


\bibliographystyle{chicago}
\bibliography{references_sine_skewed} 

\setlength{\voffset}{0cm}
\setlength{\hoffset}{0cm}

\includepdf[pages=-]{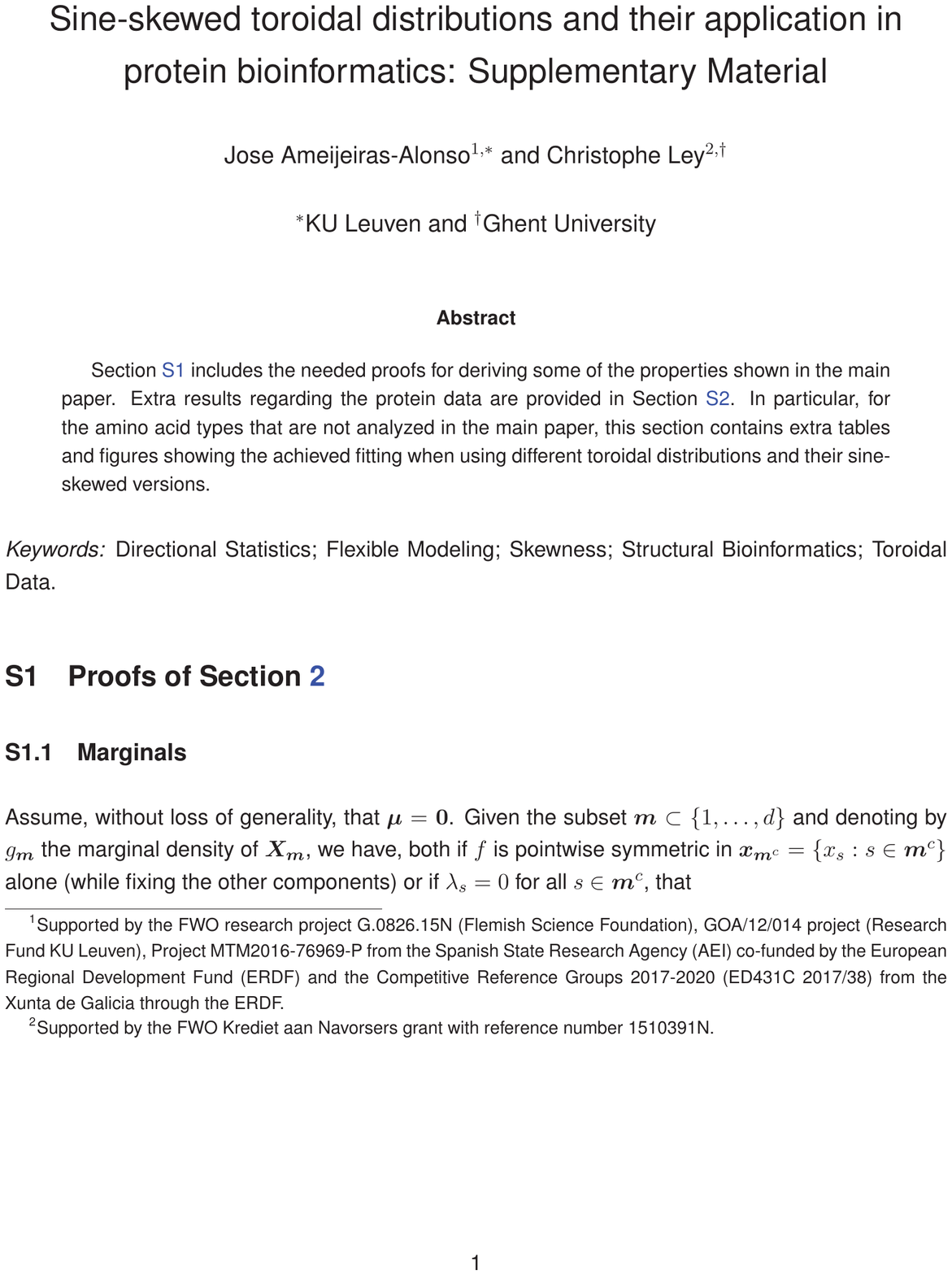}

\setlength{\voffset}{-2.54cm}
\setlength{\hoffset}{-2.54cm}


\end{document}